%% file: Springer_2014.tex
\documentclass[12pt,a4paper]{article}
\usepackage{makeidx}
\usepackage{amssymb}
\makeindex
\newcommand{\seqnoll}{\setcounter{equation}{0}}

\usepackage{euscript}
\usepackage{color}

\usepackage{latexsym}
\usepackage{eqxx}
\usepackage{epsfig}
\usepackage{mathtools}
\makeatletter
\newcommand{\threedot}[1]{%
  {\mathop{#1\hspace{0pt}}\limits^{\vbox to-1.4\ex@{\kern-\tw@\ex@ \hbox {\normalfont .\kern-.1em.\kern-.1em.}\vss}}}}
\makeatother
\newcommand{\bc}{\begin{center}}   
\newcommand{\ec}{\end{center}}
\newcommand{\be}{\begin{equation}}
\newcommand{\ee}{\end{equation}}
\newcommand{\bea}{\begin{eqnarray}}
\newcommand{\eea}{\end{eqnarray}}
\newcommand{\noi}{\noindent}

\newcommand{\sla}[1]{\setbox0=\hbox{$#1$} 
        \dimen0=\dp0 
        \dimen1=\wd0 
        \multiply\dimen0 by 3
        \divide\dimen0 by 4
        \lower\dimen0 \rlap{\hbox to \dimen1{\hss $\,\,/$ \hss}}
        \box0 }
\newcommand{\pres}[2]{\setbox0=\hbox{$\scriptstyle #1$} \dimen0=\dp0  
              \dimen1=\ht0 \divide\dimen1 by 3
              \advance\dimen0 by \dimen1
              \hbox{\lower\dimen0 \hbox{$\scriptstyle #2\!\!$}} #1}

\begin{document}
\newcounter{exenum}
\setcounter{exenum}{1}
\pagebreak
\input{firstpage}

\tableofcontents
\pagebreak
\thispagestyle{empty}
\hphantom{x}

\pagenumbering{arabic}
\input{Introduction}
\input{Gauge_Invariance.tex}
\input{NonRelativistic_Spin.tex}
\input{Magnetic_Monopoles.tex}

\input{Relativistic_Particles.tex}
\input{Yang_Mills_Particles.tex}
\input{Kaluza_Klein_Theory.tex}
\input{Canonical_Quantization.tex}
\input{Pseudo_Classical.tex}
\input{Lagrangians_10.tex}
\newpage
\vspace{-3cm}
\begin{center}
\section{GENERAL REFERENCES}
\end{center}
\noi R. Abraham  and J. S. Marsden, 	"{\it Foundations of Mechanics}"
(Benjamin, Reading, Massachusetts, 1978).
\vspace{2mm}
\\
\noi V. I. Arnold, "{\it Mathematical Methods  of Classical Mechanics}, Graduate
Texts in Mathematics {\bf 60}  (Springer Verlag, New York,  1978).
\vspace{2mm}
\\
\noi Y. Choquet-Bruhat, C. Dewitt-Morette and M. Dillard-Bleick, "{\it Analysis, Manifolds and Physics}"  (North-Holland, Amsterdam, 1977).
\vspace{2mm}
\\
\noi N. S. Craige, P. Goddard and W. Nahm  (Editors), "{\it Monopoles in
Quantum  Field Theory}", Proceedings of the Monopole Meeting, Trieste,  December, 1981 	(World Scientific, Singapore, 1982).
\vspace{2mm}
\\
\noi L. D. Faddeev and A. A. Slavnov, "{\it Gauge Fields - Introduction to Quantum  Theory}" 	(Benjamin/Cummings, London,  1980).
\vspace{2mm}
\\
\noi H. Flanders, "{\it Differential Forms with Applications to the Physical Sciences}" (Academic Press, New York, 1968).
\vspace{2mm}
\\
\noi H. Georgi, "{\it Lie Algebras  in Particle  Physics}"  (Addison-Wesley, New York,  1982).
\vspace{2mm}
\\
\noi R. Gilmore, "{\it Lie Groups,  Lie Algebras  and some of their Applications}"  (Wiley, New York,  1974).
\vspace{2mm}
\\
\noi C. Itzykson and J.-B. Zuber, "{\it Quantum Field Theory}" (McGraw-Hill, New York, 1980).
%
\\
\noi N. Jacobson, "{\it Lie Algebras}"  (Interscience Publishers, New York, 1962).
\vspace{2mm}
\\
\noi S. Kobayashi and K. Nomizu, "{\it Foundations  of Differential Geometry}", Vol.1 and Vol.2 (John Wiley, New York, 1963).
\vspace{2mm}
\\
\noi A. Lichnerowicz,  "{\it Theories Relativistes  de la Graviation et d'Electromagnetisme}"  (Masson et Cie, Paris, 1955).
\vspace{2mm}
\\
\noi D. J. Simms and N. M. J. Woodhouse, "{\it Lectures on Geometric Quantization}", Lecture Notes in Physics (Springer Verlag, Berlin, 1976)
\vspace{2mm}
\\
\noi N. Steenrod, "{\it The Topology of Fibre Bundles}" (Princeton University Press, New Jersey, 1951).
\vspace{2mm}
\\
\noi E. C. G. Sudarshan and N. Mukunda, "{\it Classical Dynamics: A Modern Perspective}" (Wiley, New York, 1974).
\vspace{2mm}
\\
\noi K. Sundermeyer,  "{\it Constrained Dynamics}", Lecture Notes in Physics, 169 (Springer-Verlag, Berlin 1982).
\vspace{2mm}
\\
\noi
W. Thirring, "{\it Classical Dynamical Systems - A Course in Mathematical Physics}", Vol.1 (Springer Verlag, Berlin, 1978).
\vspace{2mm}
\\
\noi
F. W. Warner, "{\it Foundations of Differentiable Manifolds and Lie Groups}" (Scott, Foresman and Co, New York, 1971).
\vspace{2mm}
\\
\noi
S. Weinberg, "{\it Gravitation and Cosmology}" (Wiley, New York, 1972).
 \vspace{2mm}
\\
\noi
C. von Westenholz, "{\it Differential Forms in Mathematical Physics}" (North Holland, Amsterdam, 1978).
\vspace{2mm}
\\
\noi
B. S. Dewitt, "{\it Dynamical Theory of Groups and Fields}" (Blackie and Sons, London and Glasgow, 1965).

\seqnoll
\setcounter{exenum}{1}

\newpage
\printindex
\end{document}

%% file: firstpage.tex
\thispagestyle{empty}
\hphantom{x}
\vspace{-2cm}
\begin{center}
\end{center}
\vspace{1cm}

\begin{center}
%


\rule{\linewidth}{0.1cm}\newline

{ \Huge {\textbf{{Gauge Theories}} }  }
\vspace{0.5cm}

{ \Huge {\textbf{{and}} }  }
\vspace{0.5cm}

{\Huge {\textbf{{Fiber Bundles}} }  }

\vspace{0.5cm}
{\Large {\textbf{{Applications to Particle Dynamics}} }  }

\end{center}

\vspace{0.2cm}
\begin{center}
{\huge by }
\vspace{0.5cm}

{\huge A.P. Balachandran}\raisebox{2ex}{\sl a)} {\huge $\cdot$} {\huge G. Marmo}\raisebox{2ex}{\sl b)}{\huge $\cdot$}
\vspace{0.5cm}

{\huge B.-S. Skagerstam}\raisebox{2ex}{\sl c)} {\huge $\cdot$} {\huge A. Stern}\raisebox{2ex}{\sl d)}

\rule{\linewidth}{0.1cm}\newline

\begin{tabbing}
{\sl \raisebox{1ex}{a)}Physics Department, Syracuse University, Syracuse, New York 13244-1130, U.S.A..}
\vspace*{1mm}
\\
{\sl \raisebox{1ex}{b)}Dipartimento di Fisica, INFN Sezione di Napoli, Universit$\acute{\mbox{a}}$  di Napoli I-80125 Napoli, Italy.}
\vspace*{1mm}
\\
{\sl \raisebox{1ex}{c)}Department of Physics, The Norwegian University of Science and Technology, NTNU,}
\\
\vspace*{1mm}
{\sl \,  N-7491 Trondheim, Norway.}
\\
\vspace*{1mm}
{\sl \raisebox{1ex}{d)}Department of Physics, University of Alabama,
Tuscaloosa, Alabama 35487, U.S.A..}
\\
\end{tabbing}


\vspace{0.5cm} 


%
\end{center}

\newpage

\noi This is an updated version of ''{\sl Gauge Symmetries and Fibre Bundles - Applications to Particle Dynamics}'',  {\sl Lecture Notes in Physics} {\bf 188}, as first published in 1983. For  a related, and a more recent account, see Ref.\cite{bal_1991}.

\newpage

%% file: Introduction.tex
%
\vspace{-3cm}
\begin{center}
\section{INTRODUCTION}
\end{center}
\seqnoll
\setcounter{exenum}{1}
%

{\Huge A} theory defined by an action which is invariant under a time-dependent group of transformations can be called a 
gauge theory. Well known examples of such theories are those defined by the Maxwell and Yang-Mills Lagrangians.  It is widely believed nowadays that the fundamental laws of physics have to be formulated in terms of gauge theories. 

The underlying mathematical structures of gauge theories are known to be geometrical in nature and the local and global features of this geometry have been studied for a long time in mathematics under the name of fibre  bundles. It is now understood that the global properties of gauge theories can have a profound influence on physics. 
For example, instantons and monopoles are both consequences of properties of geometry in the large, and the former can 
lead to, e.g., $CP$ violation, while the latter can lead to such remarkable results as the creation of fermions out of bosons. Some familiarity with global differential geometry and fibre bundles seems therefore very desirable to a physicist who works with gauge theories. One of the purposes of the present work is to introduce the physicist to these disciplines using simple examples. 

There exists a certain amount of literature written by general relativists and particle physicists which attempts to explain the language and techniques of fibre bundles. Generally, however, in these admirable reviews, the concepts are illustrated by field theoretic examples like the gravitational and the Yang-Mills systems. This practice tends to create the impression that the subtleties of gauge invariance can be understood only through the medium of complicated field theories. Such an impression, however, is false and simple systems with gauge invariance occur in plentiful quantities in the mechanics of point  particles and extended objects. Further, it is often the case that the large scale properties of geometry play an essential role in determining the physics of these systems. They are thus ideal to commence studies of gauge theories 
from a geometrical point of view. Besides, such systems have an intrinsic physical interest as they deal with particles with spin, interacting charges and monopoles, particles in Yang-Mills fields, etc.. We shall present an exposition of these systems and use them to introduce 
the reader to the mathematical concepts which underlie gauge theories. Many of these examples are known to exponents of geometric quantization, but we suspect that, due in part to mathematical difficulties, the wide community of physicists is not very familiar with their publications. We admit that our own acquaintance with these publications is slight. If we are amiss in giving proper credit, the reason is ignorance and not deliberate intent. 

The matter is organized as follows. After a brief introduction to the concept of gauge invariance and its relationship to determinism in Section \ref{section_gaugeinvariance}, we introduce in Chapters 3 and 4 the notion of fibre bundles in the context of a discussion on spinning point particles and Dirac monopoles. The fibre bundle language provides for a singularity-free global description of the interaction between a magnetic monopole and an electrically charged test particle. Chapter 3 deals with a non-relativistic treatment of the spinning particle. The non-trivial extension to relativistic 
spinning particles is dealt with in Chapter 5. The free particle system as well as interactions with external electro-magnetic and gravitational fields are discussed in detail. In Chapter 5 we also elaborate on a remarkable relationship between the charge-monopole system and the 
system of a massless particle with spin. The classical description of Yang-Mills particles with internal degrees of freedom, such as isospinor colour, is given in Chapter 6. We apply the above in a discussion of the classical scattering of particles off a 
't Hooft-Polyakov monopole. 
In Chapter 7 we elaborate on a Kaluza-Klein description of particles with internal degrees of freedom. The canonical 
formalism and the quantization of most of the preceding systems are discussed in Chapter 8. The dynamical systems given in Chapters 3-7 are formulated on group manifolds. The procedure for obtaining the extension to super-group manifolds is briefly discussed in Chapter 9. In Chapter  10, we show that if a system admits only local Lagrangians for a configuration space $Q$, then under certain conditions, 
it admits a global Lagrangian when $Q$ is enlarged to a suitable $U(1)$ bundle over $Q$. Conditions under which a symplectic form is derivable from a Lagrangian are also found.

The list of references cited in the text is, of course, not complete, but it is instead intended to be a guide to  the extensive literature in the field.

\noi 
\newpage

%% file: Gauge_Invariance.tex
%
\vspace{-3cm}
\begin{center}
\section{THE MEANING OF GAUGE INVARIANCE}
\label{section_gaugeinvariance}
\end{center}
\seqnoll
\setcounter{exenum}{1}

{\Huge B}elow we will deal with systems which exhibit a gauge symmetry.
It is thus useful to clarify the distinction between ordinary symmetries
and gauge symmetries at the beginning.

\subsection{The Action}

The action $S$ is a functional of fields with values in a suitable
range space. The domain of the fields is a suitable parameter space.
Thus for a non-relativistic particle, the range space may be $\mathbf{R}^3$,
a point of which denotes the coordinate of the particle. The parameter space is $\mathbf{R}$, a point of which denotes an instant of time. The fields
are functions from $\mathbf{R}$ to $\mathbf{R}^3$:
\begin{equation}
\label{eq:2_1}
{\cal F}(\mathbf{R},\mathbf{R}^3) = q~,~ q=(q_1,q_2,q_3)~,~q(t) \in \mathbf{R}^3\,\, .
\end{equation}
\noi
Thus each field $q$ assigns a point $q(t)$ in $\mathbf{R}^3$ to each instant of time
$t$.

For a real scalar field theory in Minkowski space $M^4$, the parameter
space is $M^4$, the range space is $\mathbf{R}$ and the set of fields
${\cal F}(\mathbf{R}^4,\mathbf{R})$ is the set of functions from $\mathbf{R}^4$
to $\mathbf{R}$.

Let us denote the parameter space by $D$, the range space by $R$ and
the set of fields by ${\cal F}(D,R)$. Then the action $S$ is a functional on
${\cal F}(D,R)$. It assigns to each $f \in{\cal F}(D,R)$ a number $S[f]$. For instance, in
the non-relativistic example cited above,
\begin{equation}
\label{eq:2_2}
 S[f] = \frac{m}{2}\int dt \frac{dq_i(t)}{dt} \frac{dq_i(t)}{dt} \,\, .
\end{equation}
\noi
The action also depends on the limits of the time integration. Since
these limits are not important for us, they have here been ignored.
If necessary, they can be introduced by restricting $D$ suitably. In
this case, for example, instead of $\mathbf{R}$, we can choose for $D$ the interval $t_1 \leq t \leq t_2 $.

The concept of a {\underline {global symmetry group}}  $G = \{ g \}$ may be defined as
follows: Suppose $G$ is a group with a specified  action  $r \rightarrow  gr$ on $ R\equiv  \{r \}$.
Then, $G$ has a natural action $f \rightarrow gf$ on ${\cal F}(D,R)$, where $(gf)(t) 	=  gf(t)$. This group of transformations on ${\cal F}(D,R)$ is the global group associated with $G$. We  denote it by the same symbol $G$. We say further that $G$ is a \underline{global symmetry group} if
\begin{equation}
\label{eq:2_3}
S[f] = 	S[gf] 	\,\, ,
\end{equation}
up to surface terms. For reasons of simplicity we shall assume hereafter that  $G$ is a Lie group.

As an example, consider the non-relativistic 	free particle with $D =  \{  t |   - \infty < t < \infty \}$, $R=\mathbf{R}^3$ and $G = SO(3)$. The rotation  group has a standard action on $\mathbf{R}^3$. It can be "lifted" to the action $q \rightarrow gq$ on ${\cal F}(D,R)$, where
\begin{equation}
\label{eq:2_4}
[gq] (t) =  g q(t)~~[\equiv (g_{ij}q_j(t)]) 	\,\, .
\end{equation}
\noi	                                                            
Thus in the usual language, $g$ is a global rotation. Further, $SO(3)$ is a global symmetry group since for the action (\ref{eq:2_4})
\begin{equation}
\label{eq:2_5}
S[q] = S[gq] 	\,\, .
\end{equation}

In contrast, the \underline{gauge group  ${\cal G}$ associated with  $G$} is defined to be the set of all functions from $D$ to $G$, i.e. ${\cal G}={\cal F}(D,G)=\{h\}$, where for $d \in D$, $d \stackrel{h}{\rightarrow}  h(d) \in G$. The group multiplication in ${\cal G}$ is defined by 
$(hh')(d)=h(d)h'(d)$.	This group, as well, has a natural action on ${\cal F}(D,R)$, i.e.,  $(hf)(d)=h(d)f(d)$. If $S$ is invariant under ${\cal G}$ (up to surface terms), i.e., $S[f]=S[hf]$ $+$ possible surface terms, then the gauge group is a \underline{gauge symmetry group}.

It is possible that the sort of boundary conditions we impose on the set of functions in the gauge group can have serious consequence
for the theory (see, e.g., Ref.\cite{bal_76}). If we do not impose any particular boundary conditions so that the boundary conditions are ''free", ${\cal G}$ will contain constant functions and the associated global group  $G$ may be thought of as the subgroup of ${\cal G}$ of these constant functions.

Let ${\cal G}$ be a gauge symmetry group and $\Gamma$ be a global symmetry group not associated with ${\cal G}$. Now recall that the parameter 	space contains a parameter which we identify as time $t$. The profound difference between the gauge symmetry group ${\cal G}$ and $\Gamma$  is due to the fact that \underline{${\cal G}$ contains time-dependent symmetries unlike $\Gamma$}. It affects the deterministic  aspects of the theory and also has its impact on Noether's derivation of conservation laws.  These twin aspects are manifested as constraints in the Hamiltonian
frame work Ref.\cite{dirac_64}. We can illustrate these remarks as follows:

\underline{a) Determinism}

 A trajectory, in our language, is a function ${\bar f} \in {\cal F}(D,R)$ such that
\begin{equation}
\label{eq:2_6}
\delta S[{\bar f}] = 0 	\,\, .
\end{equation}
Suppose ${\bar f}$ is a possible trajectory for a specified set of initial conditions $d^k{\bar f}/dt^k |_{t=0}\, ,\, k=1,2,...,n$.
Since ${\cal G}$ is a gauge symmetry group, $h{\bar f}$ is also a trajectory. Further since the time dependence of $h$ is at our disposal, we can \underline{choose} $h$ such that
\begin{equation}
\label{eq:2_7}
\left. \frac{d^k(h{\bar f})(t)}{dt^k} \right| _{t=0} = \left. \frac{d^k{\bar f}(t)}{dt^k} \right| _{t=0}\, k=0,1,...,n\,\, .
\end{equation}
This does not constrain $h$ to be trivial for \underline{all} time. Here we assume, of course, that ${\cal G}$ acts non-trivially on fields. The conclusion is that there are several possible trajectories for specified initial conditions. In this sense, the theory does not determine the future from the present if the state of the system is given by the values of ${\bar f}$ and its derivatives at a given time.

In the customary formulation, determinism is restored by considering only those functions which are invariant under ${\cal G}$. These 	gauge invariant functions and their derivatives at a given time are then \underline{defined} to constitute the observables of the theory. Such a definition of observables seems to have little direct bearing on whether they are accessible to experimental observation. It is a definition which is \underline{internal} to the theory and dictated by requirements of determinism.

In a Hamiltonian formulation with no constraints, the specification of Cauchy data, i.e., a point of phase space, allows us to uniquely specify the future state of the system, at least for sufficiently small times. The existence of a gauge symmetry group for the action
$S$ thus means that $S$ should lead to a constrained Hamiltonian dynamics. An orderly way to treat such a dynamics is due to Dirac \cite{dirac_64}. We will have occasion to use it later.

 \underline{b) Conservation Laws}

The infinitesimal variation of $S$ under a gauge transformation is characterized by arbitrary functions $\epsilon_{\alpha}$. 
If ${\cal G}$ is a symmetry group, Noether's argument shows that there is a charge formally written in the form
\begin{equation}
\label{eq:2_8}
Q = \int_{{\bar D}}dt\epsilon_{\alpha}Q_{\alpha}\, \, ,
\end{equation}
which is a constant of motion 
\begin{equation}
\label{eq:2_9}
\frac{dQ}{dt} = 0 \,\, .
\end{equation}
Here ${\bar D}$ is a fixed time slice of $D$. Since the $\epsilon_{\alpha}$'s are arbitrary functions, we can conclude that  \cite{bergmann_1983}
\begin{equation}
\label{eq:2_10}
Q_{\alpha} = 0 \,\, .
\end{equation}
Thus the generators $Q_{\alpha}$ of the gauge group vanish identically.

In electro-magnetism, the analogues of Eq.(\ref{eq:2_10}) are the Gauss law
\begin{equation}
\label{eq:2_11}
\nabla \cdot {\bf E} + j_0 = 0 \,\, .
\end{equation}
and the vanishing of the canonical momentum $\Pi ^0$  conjugate to $A^0$. The
non-Abelian generalizations  of these equations  are well known \cite{abers_1973}.

In the Hamiltonian framework,  the equations $Q_{\alpha} =  0$ become first
class constraints. Quantization of the system then often becomes highly non-trivial in their presence.

\subsection{The Lagrangian}

The configuration  space is usually identified  with ${\cal F}({\bar D},R)$, where ${\bar D}$ now is a fixed time-slice of $D$. It is clear however that for precision we should write ${\bar D}_t$ for the time slice at time $t$. The customary hypothesis is that ${\bar D}_t$ for different t are diffeomorphic  and that there is a natural identification  of points of ${\bar D}_t$  for different times. Under
these circumstances, (which we now assume, we are justified in writing ${\bar D}$. 

As an example, consider a field theory in Minkowski space $M^4$.
Slices at different times $t$ give different three dimensional sub­spaces $\mathbf{R}^3_t \subseteq \mathbf{R}^3$. Without further considerations,  there is no natural identification of points of these spaces, that is, there is as yet no obvious meaning to the identity of spatial points for observations at
different times. What is done in practice is as follows: On $M^4$, there is an action of the time translation group $\{ U_{\tau}\,|\,  -\infty < \tau < \infty \}$.  The latter maps $\mathbf{R}^3_t$ to $\mathbf{R}^3_{t+\tau}$ in a smooth, invertible way. We then identify all points in $\mathbf{R}^3_t$ and  $\mathbf{R}^3_{t+\tau}$ which are carried into each other by time translation $U_{\tau}$. In terms of the conventional coordinates  $({\bf x},t)$,
\begin{equation}
\label{eq:2_12} 
U_{\tau} ({\bf x},t) = ({\bf x},t+\tau)\, \, ,
 \end{equation}
and we think of ${\bf x}$ as referring to the \underline{same} three dimensional point for all times.

A field $f \in {\cal F}({\bar D},R)$ restricted to a given time $t$ is a function on $\bar{D}_t$. Since we have an identification of points of $\bar{D}_t$ for different $t$, the field $f$ can be regarded as a one-dimensional family of functions $f_t \in {\cal F}(\bar{D},R)$ parameterized by time. We have thus established a correspondence
\begin{equation}
\label{eq:2_13} 
{\cal F}(D,R) \rightarrow {\cal F}(\mathbf{R} ,{\cal F}(\bar{D},R))\, \, ,
 \end{equation}
between functions appropriate to the action principle and curves in the configuration space ${\cal F}(\bar{D},R)$.

The Lagrangian is a function, or more precisely  a functional, of "coordinates and velocities". That is, it is a function of a point    $\alpha \in {\cal F}(D,R)$ on the configuration space and of the tangent, denoted by  $\dot{\alpha}$,  to this space at this point. This new space, a point of which is a point and a tangent 	at the point of the configuration space,  is the tangent bundle $\mbox{T} F({\bar D},R)$ on the configuration space.

When the action is reconstructed from the Lagrangian by the formula
\begin{equation}
\label{eq:2_14} 
S 	=	\int  dt L[\alpha(t), \dot{\alpha}(t)] \, \, ,
 \end{equation}
we are integrating $L$ along a curve 	in the tangent bundle. This curve is not arbitrary since we require that $\dot{\alpha}(t) =  d\alpha(t)/dt $. Such a curve in the tangent bundle is the "lift of a curve" from the configuration space. It is defined by a "second order" vector field in the tangent bundle. With this restriction on curves, a curve on the tangent bundle is uniquely determined by a curve $\alpha_t \in {\cal F}(\bar{D},R)$. Since such a curve in turn defines a function in ${\cal F}(D,R)$, we recover the original interpretation of the action as a functional on ${\cal F}(\bar{D},R)$.

We need to investigate the action of the gauge group on the tangent bundle. It turns out that in its action on the tangent bundle, the gauge group, in its simplest version, is associated to the global group
%
%
%
%
\begin{equation}
\label{eq:2_15}
G \ltimes  \underline{G}  =  \{\, (h,l) | h \in G, 	l \in \underline{G} \,\} 	\,\, .
\end{equation}
where $G$ is the associated global group, and $\underline{G}$ is its Lie algebra and the group multiplication  is
\begin{equation}
\label{eq:2_16}
(h' ,l') (h,l) =	(h'h,  l' +  adh'l)	\,\, .
\end{equation}
Here $ad$ is the adjoint representation of $G$  on $\underline{G}$. In the notation
common in physics literature
\begin{equation}
\label{eq:2_17}
adh'l =  h'l h'^{-1}\,\, . 	
\end{equation}
Thus $G  \ltimes \underline{G}$  is the semi-direct product of $G$  with $\underline{G}$. This result has
been discussed before by Sudarshan and Mukunda  \cite{dirac_64}.

 We denote the gauge group associated to $G$ at a given time by $\hat{{\cal G}}$.
The elements of  $\hat{{\cal G}}$ are functions ${\cal F}(\bar{D},G) =  \{h\}$ 	with group multiplication defined by
\begin{equation}
\label{eq:2_18}
(hh')(\bar{d}) =  h(\bar{d})h'(\bar{d}) \,\, , \, \, \bar{d}\in \bar{D} 	 \, \, .
\end{equation}
 The Lie algebra $\underline{G}$ is a group under addition and the associated gauge
group at a given time is denoted by $\underline{\hat{{\cal G}}}$. Finally the gauge group associated to $G  \ltimes \underline{G}$ at a given time is denoted by $\hat{{\cal G}}  \ltimes \hat{\underline{{\cal G}}}$.

These remarks can be established by examining the way the action of the gauge group "projects down" to an action on coordinates and velocities. A function $f \in {\cal F}(D,R)$ is transformed to $hf$. Thus the curve at $\alpha_t \in {\cal F}(\bar{D},R)$ is transformed into $(h\alpha)_t$,  where $h$ is time-dependent. Thus a point of the tangent bundle is transformed according to
\begin{equation}
\label{eq:2_19}
(\alpha, \dot{\alpha}) \rightarrow  (h\alpha, \frac{d(h\alpha)}{dt}) = (h\alpha, h\dot{\alpha} + l(h\alpha))	 \, \, ,
\end{equation}
where $h\in \hat{{\cal G}}$, $l =  \dot{h}h^{-1} \in \underline{\hat{{\cal G}}}$.  In Eq.(\ref{eq:2_19}), the time-dependence of $h$ and
 $l$ have disappeared since we are examining the action at a point of
$\mbox{T} {\cal F}(\bar{D},R)$. In writing 	Eq.(\ref{eq:2_19}), we have also assumed that the action of the gauge group is local in time, that is
\begin{equation}
\label{eq:2_20}
(h\alpha )_t =  h(t)\alpha (t)	 \, \, .
\end{equation}
If  $(h\alpha)_t$ depends on $h(t)$ as well as (say) its derivatives $d^kh(t)/dt^k$,  Eq.(\ref{eq:2_19}) will have to be modified. For Yang-Mills theories, this actual happens (see  below). We prefer  to illustrate the idea without this complication. With this assumption we   can write
\begin{equation}
\label{eq:2_21}
(h,l) \in  \hat{{\cal G}}  \ltimes \hat{\underline{{\cal G}}}  \,, \,  (h,l)(\alpha,\dot{\alpha})=(h\alpha,h\dot{\alpha} + l(h\alpha))	 \, \, .
\end{equation}
The group multiplication  Eq.(\ref{eq:2_16}) follows from
\begin{eqnarray}
\label{eq:2_22}
(h',l')(h\alpha, h\dot{\alpha} + l(h\alpha))= (hh'\alpha,hh'\dot{\alpha}+(h'lh'^{-1})(h'h\alpha) + l'(h'h\alpha) )	\nonumber \\
=(hh'\alpha,h'h\dot{\alpha} +(l' + adh'l)(h'h\alpha)) = (h'h,l'+adh'l)(\alpha,\dot{\alpha})\, \, .
\end{eqnarray}

The preceding considerations are easily illustrated by Yang­Mills theory where the vector potential $A_{\mu}$ 	has values in the Lie
algebra $\underline{G}$ of the gauge group $G$ and transforms as follows:
\begin{equation}
\label{eq:2_23}
A_{\mu} \rightarrow hA_{\mu}h^{-1} + h\partial_{\mu}h^{-1}	 \, \, .
\end{equation}
Thus at a fixed time
\begin{equation}
\label{eq:2_24}
(h,l) A_i = hA_ih^{-1} \, \, ,
\end{equation}
and 
\begin{equation}
\label{eq:2_25}
(h,l) A_0 = hA_0h^{-1} -l\, \, ,
\end{equation}
where
\begin{equation}
\label{eq:2_26}
l=\dot{h}h^{-1}\, \, .
\end{equation}
The group multiplication  law  Eq.(\ref{eq:2_16}) follows by considering the application of $(h',l'$) to the left-hand side of Eqs. (\ref{eq:2_24}) and (\ref{eq:2_25}).

The transformation  Eq.(\ref{eq:2_25}) on the configuration space variable $A_0$ 	is not local in time since (\ref{eq:2_26}) involves $dh/dt$. Nonetheless,
the group multiplication  Eq.(\ref{eq:2_16}) is unaffected.

The space on which the group is supposed to act, however, is not the space of $A_{\mu}$, but of $(A_{\mu}, \dot{A}_{\mu})$. If we consider the subspace $(A_{i}, \dot{A}_{i})$, since (2.24) does not involve $\dot{h}$, we find the group $\hat{G}  \ltimes \hat{\underline{G}}$. However, the argument has to be modified if $\dot{A}_0$ is considered since its transformation involves $\dot{l}$. An element of the gauge group is
now a triple $(h,l,\dot{l})$ with the action
\begin{equation}
\label{eq:2_27}
(h,l,\dot{l})(A_{0}, \dot{A}_{0})= (hA_{0}h^{-1} -l, h\dot{A}_{0}h^{-1}+ [l,hA_{0}h^{-1}]- \dot{l})\, \, ,
\end{equation}
and the multiplication law
\begin{equation}
\label{eq:2_28}
(h_1,l_1,\dot{l}_1)(h_2,l_2,\dot{l}_2)= (h_1h_2, l_1+h_1l_2h_1^{-1},\dot{l}_1+[l_1,h_1h_2h_1^{-1}]+ h_1\dot{l}_2h_1^{-1})\, \, .
\end{equation}
The action of $(h,l,\dot{l})$ on $(A_i, \dot{A}_i)$ is obtained from taking the derivative of Eq.(\ref{eq:2_24}). In this action, $\dot{l}$ is passive. The general gauge group $G_L$ at the Lagrangian level can thus in general involve $\dot{l}, \ddot{l}, \threedot{l}, ... \, $.

The subgroup of \underline{constant} functions from $\bar{D}$ to $G$ is what is called the global symmetry group. Since it is isomorphic to $G$, we can denote it by the same symbol $G$. It is a subgroup of $\hat{{\cal G}}$ if there are no boundary conditions on functions in $\hat{{\cal G}}$, that is if all constant functions are allowed in $\hat{{\cal G}}$. Thus, with free boundary conditions, we can conclude the following: Since observables are invariant under $\hat{{\cal G}}$, they are invariant under the global group $G$. That is, all observables are globally neutral.

\subsection{The Hamiltonian} 

The Hamiltonian framework provides an algebraic formulation of the classical theory in terms of Poisson	brackets (PB's). It is an essential step in the quantization of the classical theory according to  Ref.\cite{dirac_47}.

In this section, we qualitatively describe the preliminaries to Dirac's procedure for setting up the canonical formalism in the presence of constraints. Concrete examples will be worked out in the
subsequent chapters.
In the canonical formalism, we start with defining a "cotangent bundle"	$T^{\star}{\cal F}(\bar{D},R)$ on the• configuration space ${\cal F}(\bar{D},R)$ and PB's between functions on this bundle. This construction is carried	out whether or not there are constraints present in the theory. A
point in this bundle is labeled by $(\alpha,p)$ where $ \alpha \in {\cal F}(\bar{D},R)$ is a point of the configuration space and $p$ is the ``conjugate momentum variable''. It is also a function on $\bar{D}$. The PB's involving $\alpha$ and
$p$ are conventional.

If we are given a Lagrangian $L$ , then it defines a map $T {\cal F} (\bar{D},R) \rightarrow T^{\star} {\cal F}(\bar{D},R)$ by the formula
\begin{equation}
\label{eq:2_29}
(\alpha, \dot{\alpha}) \rightarrow (\alpha, \frac{\partial L}{\partial \dot{\alpha}})\, \, .
\end{equation}
The Lagrangian is 	non-singular	if this map is one-to-one	onto $T^{\star} {\cal F}(\bar{D},R)$.
In that case, when $\alpha$ and $\dot{\alpha}$ range over the allowed values,
 all of 	$T^{\star} {\cal F}(\bar{D},R)$  is recovered and every point of  $T^{\star} {\cal F}(\bar{D},R)$  specifies a state of the system.

It is then an elementary result that the time-evolution on  $T^{\star} {\cal F}(\bar{D},R)$
can be generated by the formula 
\begin{equation}
\label{eq:2_30}
\dot{x} = \{x, H \} ~~,~~ x \in T^{\star} {\cal F}(\bar{D},R) \, \, .
\end{equation}
 where $H$ is the Hamiltonian for the system under consideration and is constructed by the Legendre transform from $L$ and $\{ \cdot,\cdot \}$ is the Poisson bracket.

As we remarked earlier, in gauge theories, the image of the map
$(\alpha, \dot{\alpha}) \rightarrow (\alpha, \partial L/\partial \dot{\alpha})$
is not all of $T^{\star} {\cal F}(\bar{D},R)$, but only a sub-manifold  ${\cal M}$ of this space. That is, there are functions $\phi_n , n=1,2, ...$, on $T^{\star} {\cal F}(\bar{D},R)$ such that $\phi_n$ is zero on  ${\cal M}$:
\begin{equation}
\label{eq:2_31}
\phi_n (\alpha, \frac{\partial L}{\partial \dot{\alpha}}) \equiv 0 \, \, .
\end{equation}
We note that not all functions on $T^{\star} {\cal F}(\bar{D},R)$ need to have zero PB's with $\phi_n$
on ${\cal M}$, i.e., $\{ f,\phi_n \}$ need not vanish on ${\cal M}$ for all functions $f$ .
For instance, in electro-dynamics the electric field $\bf E $ is conjugate to the
potential $\bf A $  and Gauss' law
\begin{equation}
\label{eq:2_32}
\nabla \cdot {\bf E}+ j_0 = \phi_1 \, \, .
\end{equation}
is a (secondary) constraint. But obviously,
\begin{equation}
\label{eq:2_33}
\{A_i({\bf x}), \phi_1({\bf y})\} \neq 0 \, \, .
\end{equation}
on ${\cal M}$.  Note that we must \underline{first} evaluate the PB's and \underline{then} substitute 	$\phi_n = 0$. Due to the existence of such $f$, we cannot set 	$\phi_n = 0$ before evaluating PB's. Thus we cannot  eliminate redundant degrees of freedom using 	Eq.(\ref{eq:2_31}) without trouble from the Poisson bracket algebra. 	

A systematic method to treat the constraints is due to Dirac. References \cite{dirac_64} contain a detailed exposition of the method. In Chapter \ref{canonical_formalism} we will have occasion to illustrate the method in specific examples.

\noi 
\newpage

%% file: NonRelativistic_Spin.tex
%
\vspace{-3cm}
\begin{center}
\section{NON-RELATIVISTIC PARTICLES WITH SPIN}
\label{section_nonrelativistic_spin}
\end{center}
\seqnoll
\setcounter{exenum}{1}

{\Huge A} classical non-relativistic particle with spin is an example of an elementary system where the utility of the fibre bundle formalism
can be illustrated. The Hamiltonian description of such systems is well known (see, e.g., Ref.\cite{dirac_64}) and is recalled below. The construction of a Lagrangian description however is not quite straightforward.
One such construction involves the use of non-trivial fibre bundles. Below we will only discuss particles with zero electric charge. In
Chapter 5 we return to a relativistic description of charged particles.

\subsection{The Hamiltonian Description}
\label{section_nonrelativistic_spin_1}
Let ${\bf x}= (x_1,x_2,x_3)$, ${\bf p}= (p_1,p_2,p_3)$, and ${\bf S}= (S_1,S_2,S_3)$ denote the coordinate, the momentum and the spin of the particle. Here we therefore impose the constraint
\begin{equation}
\label{eq:3_1}
S^2 \equiv S_iS_i = \lambda^2 \, \, ,
\end{equation}
where $\lambda $ is a constant.
\noi
The Poisson brackets are
\begin{eqnarray}
\label{eq:3_2}
\{x_i,x_j\} &=& \{p_i,p_j\} =0 \, \, ,\\
\label{eq:3_3}
\{x_i,p_j\} &=& \delta_{ij} \, \, ,\\
\label{eq:3_4}
\{S_i,S_j\} &=& \epsilon_{ijk}S_k \, \, .
\end{eqnarray}
%
%
%
%
%
%
%
%

If the particle is free, the Hamiltonian of the system is
\begin{equation}
\label{eq:3_5}
H_0 = \frac{{\bf p}^2}{2m} \, \, ,
\end{equation}
where $m$ is the mass of the particle. If there is an external, not necessarily homogeneous,  magnetic field ${\bf B} = (B_1 ,B_2 ,B_3)$ present, and the particle has a magnetic moment $\mu $,  the Hamiltonian has the following form:
\begin{equation}
\label{eq:3_6}
H= H_0 + \mu {\bf S}\cdot{\bf B} \, \, .
\end{equation}
The equations of motion for the free particle and the interacting particle are, respectively
\begin{eqnarray}
\label{eq:3_7}
\dot{x}_i &=& \frac{p_i}{m} \, \, ,\\
\label{eq:3_8}
\dot{p}_i &=& 0 \, \, ,\\
\label{eq:3_9}
\dot{S}_i &=& 0\, \, ,
\end{eqnarray}
%
%
%
%
%
%
and
\begin{eqnarray}
\label{eq:3_10}
\dot{x}_i &=& \frac{p_i}{m} \, \, ,\\
\label{eq:3_11}
\dot{p}_i &=& -\mu S_j\partial_iB_j \, \, ,\\
\label{eq:3_12}
\dot{S}_i &=& \mu\epsilon_{ijk}B_jS_k\, \, .
\end{eqnarray}
Here we make use of the notation $\partial_i \equiv \partial/\partial x_i$.
%
%
%
%

\subsection{The Lagrangian Description}
\label{section_nonrelativistic_spin_2}

If we know the Hamiltonian description, it is often possible to find the Lagrangian of the system by a Legendre transformation.
We can perform the Legendre transformation provided we can find coordinates for the phase space which are canonical. By this
we mean the following. Let $Q$ denote the configuration space of the
system under consideration. The phase space $T^* Q$ , in our case, is
eight-dimensional. A canonical system of coordinates for this
space is by definition of the form
\begin{equation}
\label{eq:3_13}
T^* Q = {\Big \{}(Q_1,Q_2,Q_3,Q_4,P_1,P_2,P_3,P_4){\Big\}}  \, \, ,
\end{equation}
where
\begin{equation}
\label{eq:3_14}
\{Q_i, Q_j\}= \{P_i, P_j\}= 0  \, \, ,
\end{equation}
and 
\begin{equation}
\label{eq:3_15}
\{Q_i, P_j\}= \delta_{ij}  \, \, .
\end{equation}
For our system we can, of course, set
\begin{equation}
\label{eq:3_16}
Q_i = x_i\, \,  , \, \,  P_i = p_i  \, \, , \, \,  i=1,2,3  \, \, .
\end{equation}
It remains to find $Q_4$ and $P_4$. They will depend on ${\bf S}$ and perhaps
${\bf x}$ and ${\bf p}$. One can show, however, that there exists no such coordinates 
$Q_4$ and $P_4$ which are smooth functions of ${\bf S}$.  From the
constraint Eq.(\ref{eq:3_1}), ${\bf S}$ spans a $2$-dimensional sphere. It is well-known
that a $2$-dimensional sphere cannot be globally coordinatized by
a set of coordinates $(Q_4,P_4)$ (see, e.g., Ref.\cite{Flanders_63}). Any choice of $Q_4$ and $P_4$ will
therefore show a singularity for at least one ${\bf S}$. This
singularity is the analogue of the Dirac string \cite{Dirac_31,Dirac_48} in the theory
of magnetic monopoles. We refer to Section \ref{sec:3_3} for a proof of this
result.

Thus we cannot find a global Lagrangian by a Legendre transformation when we
have a constraint like Eq.(\ref{eq:3_1}). For a local
Lagrangian description we refer to Ref.\cite{Horovaty_79}. Although it is not possible to find a global
Lagrangian by a Legendre transformation, the above system does admit a global
Lagrangian description  by enlarging the configuration space. We shall now construct it and
point out some of its novel features. The canonical formalism for this Lagrangian is the
one discussed above. We will discuss the derivation of this formalism in Chapter \ref{canonical_formalism}.

Let $\Gamma = \{s\}$ denote the usual spin $1/2$ representation of the rotation
group (see, e.g., Ref.\cite{Talman_68}). Thus we have
\begin{equation}
\label{eq:3_17}
s^{\dag}s = {\bf 1}~,~~\det s = 1~.
\end{equation}
%
%
The configuration space $Q$ for the Lagrangian will be the product
space ${\bf R}^{3} \times \Gamma$. The points of ${\bf R}^{3}$ as usual correspond to the
position coordinates of the particle while a point $s \in \Gamma$ is
related to the spin degrees of freedom $S_{i}$ through
\begin{equation}
\label{eq:3_18}
S_{i} \sigma_{i} = \lambda s \sigma_{3} s^{-1}~~~.
\end{equation}
Here $\sigma_{i}~,~~i =1,2,3$, are the three Pauli matrices. As a consequence
of this definition, the constraint Eq.(\ref{eq:3_1}) is fulfilled as an identity. 

The Lagrangian of the free spinning particle then is
\begin{equation}
\label{eq:3_19}
L_{0} = \frac{1}{2} m \dot{{\bf x}}^{2} + i\lambda  ~\mbox{Tr}(\sigma_{3} s^{-1}
\dot{s})~~~,
\end{equation}
where the dot indicates differentiation with respect to time.

We now verify that $L_{0}$ gives the correct equations of motion. Variation
of the coordinate ${\bf x}$ leads in a known way to
\begin{equation}
\label{eq:3_20}
m \ddot{x}_{i} = 0~,~~i = 1,2,3~~.
\end{equation}
Variation of $s$ can be performed as follows. The most general variation of $s$
can be written in the form
\begin{eqnarray}
\label{eq:3_21}
\delta s = i~\varepsilon \cdot \sigma~s~~~, 
\end{eqnarray}
where
\begin{eqnarray}
\label{eq:3_22}
\varepsilon \cdot \sigma = \varepsilon_{i} \sigma_{i}~~~.
\end{eqnarray}
%
%
This is so because $i \varepsilon \cdot \sigma$ is a generic element of the Lie
algebra of $\Gamma$ and the general variation of $s$ is induced by such an
element. Equations (\ref{eq:3_17}) and (\ref{eq:3_21}) imply
\begin{equation}
\label{eq:3_23}
\delta s^{-1} = -is^{-1} \varepsilon \cdot \sigma~~.
\end{equation}
Hence, for the variation Eq.(\ref{eq:3_21}),
\begin{equation}
\label{eq:3_24}
\delta L_{0} = - \lambda ~\mbox{Tr}~(\sigma_{3} s^{-1} \dot{\varepsilon} \cdot
\sigma s) = - 2 S_{i} \dot{\varepsilon}_{i}~~~.
\end{equation}
After a trivial partial integration, this yields the required equation of motion
\begin{equation}
\label{eq:3_25}
\dot{S}_{i} = 0~~.
\end{equation}

If the particle has a magnetic moment $\mu$, the Lagrangian in the presence
of an external magnetic field ${\bf B}$ is
\begin{equation}
\label{eq:3_26}
L = L_{0} - \frac{\mu}{2} ~\mbox{Tr}({\cal S} {\cal B}) \equiv L_{0} - \mu S_{i}
B_{i}~~~,
\end{equation}
where
\begin{equation}
\label{eq:3_27}
{\cal S} \equiv S_{i} \sigma_{i} = \lambda s \sigma_{3} s^{-1}
\end{equation}
and
\begin{equation}
\label{eq:3_28}
{\cal B} = B_{i} \sigma_{i}~~~.
\end{equation}
In Eq.(\ref{eq:3_26}), during variations, we should regard $S_{i}$ as a function of $s$.
Now the variation of ${\bf x}$ gives
\begin{equation}
\label{eq:3_29}
\delta L = m \dot{x}_{i} \delta \dot{x}_{i} - \mu S_{j} \partial_{i} B_{j}
\delta x_{i}
\end{equation}
where
\begin{equation}
\label{eq:3_30}
\partial_{i} B_{j}\equiv \frac{\partial B_{j}}{\partial x_{i}}~~~~.
\end{equation}
Hence
\begin{equation}
\label{eq:3_31}
m \ddot{x}_{i} = - \mu S_{j} \partial_{i} B_{j}~~~~.
\end{equation}
The variation Eq.(\ref{eq:3_21}) of $s$ gives in this case
\begin{equation}
\label{eq:3_32}
\delta L_B = - ~\mbox{Tr}({\cal S} \sigma \cdot \dot{\varepsilon}) - \frac{i \mu}{2}
~\mbox{Tr}([{\cal S},{\cal B}] \varepsilon \cdot \sigma)\,\, ,
\end{equation}
where we have used the cyclic property of the trace operation, i.e.,
\begin{equation}
\label{eq:3_33}
~\mbox{Tr}(A[B,C]) = ~\mbox{Tr}(B[C,A])~~.
\end{equation}
Thus
\begin{equation}
\label{eq:3_34}
\dot{S}_{i} = \mu \varepsilon_{ijk} B_{j} S_{k}~~~.
\end{equation}
Equations (\ref{eq:3_31}) and (\ref{eq:3_34}) are the same as those given by the Hamiltonian
discussed above.

\subsection{Gauge Properties of $L_0$ and $L$
\label{sec:3_3}}

The Lagrangian $L_A$, $A=0,1$, where $L_1 \equiv L$, exhibits gauge
invariance under a gauge group ${\cal G}$ which we now discuss in some detail.
 
Let $U(1) = \{ \exp (i \sigma_{3} \theta/2) \}$ and consider the transformation
\begin{equation}
\label{eq:3_35}
s \rightarrow s \exp (i \sigma_{3} \theta/2)~~~,
\end{equation}
where $\theta$ in general is time-dependent. Under this transformation, $L_{A}$
changes only by the time derivative of a function, that is,
\begin{equation}
\label{eq:3_36}
L_{A} \rightarrow L_{A} + \lambda \dot{\theta}~~~.
\end{equation}

We distinguish this invariance property of a Lagrangian function from the
conventional one where the last term in Eq.(\ref{eq:3_36}) is absent by saying that
$L_{A}$ is \underline{weakly invariant} under the gauge transformation Eq.(\ref{eq:3_35}).
This weak invariance of $L_{A}$ clearly suggests that the equations of motion
involve only variables invariant under the gauge transformation Eq.(\ref{eq:3_35}).
For dynamical variables, ``invariance'' under the transformation Eq.(\ref{eq:3_35}), of
course, has the conventional meaning. We may note here that the equations
of motion Eqs.(\ref{eq:3_31}) and (\ref{eq:3_34}) in fact only contain the gauge invariant
dynamical variables $x_{i}$ and $S_{i}$.

Since $L_A$ changes under the gauge transformation Eq.(\ref{eq:3_35}), it
is not possible to write it as it stands in terms of gauge invariant
quantities only. We can instead attempt to eliminate the additional
gauge degree of freedom in  $L_A$ by fixing the gauge. This
means the following:
We can now show \cite{Bal_78} that any gauge invariant
 is a function of $S_{i}$ and, of course, of $x_{i}$.
Gauge fixing means that for each ${\bf S} = (S_{1}, S_{2}, S_{3})$, we try to
find an element $s({\bf S}) \in SU(2)$ such that
\begin{equation}
\label{eq:3_37}
S_{i} \sigma_{i} = \lambda s({\bf S}) \sigma_{3} s({\bf S})^{-1}~~~.
\end{equation}
If such an $s({\bf S})$ existed, we could substitute $s({\bf S})$ for $s$ in the
Lagrangian $L_A$ and thereby eliminate the gauge degree of freedom. We can
show, however, that there exists no such choice of $s({\bf S})$ which is
continuous for all ${\bf S}$. The reason for this is as follows. The vectors
${\bf S}$ which satisfy the normalization conditions $S_{i}S_{i} = \lambda^{2}$
span the two sphere $S^{2}$. The existence of a smooth $s({\bf S})$ with the
property Eq.(\ref{eq:3_37}) means that
\begin{equation}
\label{eq:3_38}
SU(2) = S^{2}\times U(1)~~,
\end{equation}
since any point in $\Gamma$ can then written in the form $s({\bf S}) \exp
(i \sigma_{3} \theta/2)$. But $SU(2)$ is simply connected while $U(1)$ on the
right hand side of Eq.(\ref{eq:3_38}) is infinitely connected, and so the right hand side
of Eq.(\ref{eq:3_38}) is infinitely connected. Here we recall that $U(1)$ is topologically
identical to the circle $S^{1}$. Hence (\ref{eq:3_38}) and a smooth $s({\bf S})$ do not exist.
Thus we have the  remarkable situation that a Lagrangian for
a non-relativistic spinning system exists only if the space of
coordinates and spin variables is non-trivially enlarged to include
a $U(1)$ gauge degree of freedom (at least in our approach).

It is often stated in the literature that $U(1)$ gauge invariance
implies electro-magnetism. But the $U(1)$ gauge invariance of the
Lagrangian $L_A$ seems to have little to do with electro-magnetism. In
the sections which follow, we will encounter other weakly gauge
invariant Lagrangians in contexts which seem equally remote from
Abelian or non-Abelian gauge fields. Thus the assertions in the
literature seem to require qualifications.

\subsection{Principal Fibre Bundles}

The Lagrangians $L_A$ are associated with what in differential
geometry is called a principal fibre bundle structure. We now discuss this bundle structure .

As we have seen above, the configuration space appropriate to the Lagrangian
$L_A$ is the group space $SU(2) = \{s \}$. More accurately, it is ${\bf R}^{3}
\times SU(2)$. But ${\bf R}^{3}$ being, in this case, the set of positions of the
particle under consideration, is not relevant in the present context and will
be simply ignored. On the space $SU(2)$, there is the action of the group $U(1)$, i.e. there is an action
\begin{equation}
\label{eq:3_39}
s \rightarrow s \exp (i \sigma_{3} \theta/2)
\end{equation}
of $U(1)$. Under this action, $L_A$ is weakly invariant for time dependent
$\theta$'s. If we now define the projection map $\pi$ by
\begin{equation}
\label{eq:3_40}
\pi : SU(2) \rightarrow S^{2}~~~,
\end{equation}
where
\begin{equation}
\label{eq:3_41}
\pi : s \rightarrow \lambda s \sigma_{3} s^{-1} \equiv S_{i} \sigma_{i}~~,
\end{equation}
weak invariance of $L_{A}$ implies that the equations of motion depend only
on the \underline{base manifold} $S^{2} = \{ {\bf S} \}$.

Thus we have the following mathematical structure:
\begin{enumerate}
\item \hspace{-2mm}) A manifold $SU(2)$ which topologically is the same as the three sphere
$S^{3}$,
\item \hspace{-2mm}) the action of a \underline{structure group} $U(1)$ on the manifold
$SU(2)$,
\item \hspace{-2mm}) the projection map $\pi $ from $SU(2)$ to the base manifold $S^{2}$. Further, 
\item \hspace{-2mm})  the $U(1)$ action is \underline{free}, that is, $sg=s$ for $g \in U(1)$
implies that $g$ equals the identity element $e$ of the structure group $U(1)$.
\end{enumerate}
Note that the projection $\pi $ maps all the right cosets
\begin{equation}
\label{eq:3_42}
s U(1) \equiv s\cdot\{ \exp (i \sigma_{3} \theta/2) \}
\end{equation}
to a single point on the base space $S^{2}$. This right coset is just the
orbit of $s$ under the action of the $U(1)$ group. It is also easy to check
that distinct orbits have distinct images on $S^{2}$ and that the mapping
is onto $S^{2}$. That is, the space $SU(2)/U(1)$ of the right cosets can be
identified with the base space $S^{2}$. Thus, if we define an equivalence
relation $\sim$ by the statement
\begin{equation}
\label{eq:3_43}
s_{1} \sim s_{2}~\mbox{if}~s_{1}g = s_{2}~\mbox{for some}~g \in U(1)~,
\end{equation}
then $\pi$ is just the map from $SU(2)$ to the space of equivalence classes
generated by the relation $\sim$, that is, to the space $SU(2) / U(1)$.

The preceding features define a principal fibre bundle, as denoted by $U(1)\rightarrow S^3\rightarrow S^2$, with the bundle space
$S^{3} \equiv SU(2)$ as a manifold, structure group $U(1)$ and base
space $S^{2}$. It is a well-known structure in mathematics - the Hopf fibration
of the two sphere $S^{2}$ (see, e.g., Ref.\cite{Steenrod_51}).

We now give the general definition of a principle fibre bundle $G\rightarrow E\rightarrow B$.
For details, see for, e.g., Daniel and Viallet, Ref.\cite{bal_76} and Ref.\cite{Thomas_80}.
It consists of a bundle space $E$, a structure
group $G$, a base space $B$ and a projection map $\Pi$ from $E$ onto $B$. The
group $G = \{g\}$ has an action on the bundle space $E$:
\begin{equation}
\label{eq:3_44}
E \ni p \rightarrow pg \in E~~.
\end{equation}
This action is required to be \underline{free}, that is,
\begin{equation}
\label{eq:3_45}
~pg = p\, , \mbox{for any}\, p\, ,\, \mbox{implies that}~g~\mbox{equals
the identity}~e~ \mbox{of}~G~.
\end{equation}
The projection $\Pi$ is just the identification of points 
related by the $G$-action. Thus
\begin{equation}
\label{eq:3_46}
\Pi (p) = \Pi (pg)\, \, ,
\end{equation}
while
\begin{equation}
\label{eq:3_47}
\Pi (p) = \Pi (q) 
\end{equation}
implies that
\begin{equation}
\label{eq:3_48}
q = pg
\end{equation}
for some $g \in G$. We can think of $B$ as the space of $G$-orbits in $E$.

A global section is a map
\begin{equation}
\label{eq:3_49}
\varphi:B \rightarrow E
\end{equation}
such that
\begin{equation}
\label{eq:3_50}
\pi \circ \varphi =~\mbox{identity map on}~B~.
\end{equation}
Thus for $b \in B$, $\varphi(b)$ is in $E$ and
\begin{equation}
\label{eq:3_51}
\Pi (\varphi (b)) = b~\mbox{for all}~b~\mbox{in}~B~.
\end{equation}
A local section is defined analogously by restricting the domain of definition
of the map $B \rightarrow E$ to an open set in $B$. For suitable open sets in
$B$, a local section always exists. In fact, there is always a covering
$\{V_{\alpha}\}$ of $B$ by open sets $V_{\alpha}$ where
$\cup\!\!\!\raisebox{-8pt}
{{$\scriptstyle \alpha$}}$
$V_{\alpha} = B$, such that each $V_{\alpha}$ admits a local
section. 

The principal fibre bundle $E$ is said to be trivial if $E = B \times G$. A
principal fibre bundle is trivial {\em if and only if} it admits a global section.
Note that a point $p$ in a trivial bundle is of the form $p=(b,g)$, where
$b \in B$ and $g \in G$, while the group acts on $E$ as follows:
\begin{equation}
\label{eq:3_52}
(b,g) \rightarrow (b,gg')~,~g' \in G~~~.
\end{equation}
Thus the projection map is just
\begin{equation}
\label{eq:3_53}
\Pi (b,g) = b~~~.
\end{equation}

\subsection{Gauge Fixing}

In the conventional treatment of gauge theories (see, e.g., Ref.\cite{abers_1973}) there is a procedure called \underline{gauge fixing}. It may be explained in the  following way. Suppose the
configuration space of the Lagrangian  is $\{ \xi \}$. Here $\xi$ can be
a multi-component, as well as a space-time dependent field. In the latter case, the considerations which follow are only formal. Suppose the gauge group ${\cal G}$ is described by the set $\{ g \}$,  a time-dependent,  and also possibly 
space-dependent, group, and has the action
\begin{equation}
\label{eq:3_54}
\xi \rightarrow \xi g
\end{equation}
on $\{ \xi \}$. Fixing the gauge consists of picking exactly one point from
each orbit $\{ \xi  g \}$. This is accomplished by imposing a condition of the
form
\begin{equation}
\label{eq:3_55}
\chi (\xi) = 0
\end{equation}
on $\xi $. Here $\chi $, of course, can be multi-component, $\chi = (\chi_{1},
\chi_{2}, \ldots , \chi_{n})$. Thus Eq.(\ref{eq:3_55}) can actually be many conditions.
The equation (\ref{eq:3_55}) defines a surface $M$. From the previous remarks, it is clear
that the surface $M$ must be such that each orbit cuts $M$ once and exactly
once.

If the action (\ref{eq:3_54}) is free, the previous discussion shows that $M$ is a
global section in a principal fibre bundle. In this case, $M$ exists if and
only if $\{ \xi \}$ is a trivial bundle. Global gauge fixing is possible only
in such a case.

In general, the action of the gauge group ${\cal G}$ on $\{ \xi \}$ can be quite
involved. Thus: 

a) The action of ${\cal G}$ may not be free. Then the orbit $\xi {\cal G}$
is not diffeomorphic to ${\cal G}$ since some elements of ${\cal G}$ leave $\xi$ fixed,
that is, some degrees of freedom of ${\cal G}$ disappear in the map
\begin{equation}
\label{eq:3_56}
g \rightarrow \xi g~~~.
\end{equation}

b) The little group, also called the stability group or the isotropy group, ${\cal G}_{\xi}$ of
$\xi$ is the set
\begin{equation}
\label{eq:3_57}
{\cal G}_{\xi} = \{ g \in {\cal G} | g = \xi g \}~~.
\end{equation}
It may happen that two distinct points $\xi$ and $\xi'$ have little groups
${\cal G}_{\xi}$ and ${\cal G}_{\xi'}$ which are not conjugate in ${\cal G}$. 
That is, there exist
no element $ \bar{g} \in {\cal G}$, such that
\begin{equation}
\label{eq:3_58}
\bar{g}{\cal G}_{\xi} \bar{g}^{-1} \equiv \{ \bar{g}g \bar{g}^{-1} | g \in {\cal G}_{\xi} \}
= {\cal G}_{\xi'}~~~.
\end{equation}
In fact, ${\cal G}_{\xi}$ and ${\cal G}_{\xi'}$ may not even be isomorphic. An example is the
action of the connected Lorentz group $L_{+}^{\uparrow}$ on the Minkowski space
$M^{4}$. In this case, if for instance $x \in M^{4}$ is time-like, the little group is
$SO(3)$, while if $x$ is space-like, the little group is the connected $2 + 1$
Lorentz group. In case b), the different orbits are not diffeomorphic.

If the orbits $\xi {\cal G}$ for different $\xi$ are diffeomorphic, we have a
\underline{fibration} of the space $\{ \xi \}$ by the group ${\cal G}$. If there are
topologically distinct orbits, we have a \underline{singular fibration} of the space
$\{ \xi \}$ by the group ${\cal G}$.
 
In Yang-Mills field theory, there are some results which show the non-existence
of a global gauge condition \cite{Gibov_77,Singer_78}, that is, of a global surface $M$ with the
properties discussed above. These results are usually proved either when
the Euclidean space-time is compactified to the four sphere $S^{4}$ or its time-slices are compactified to three spheres $S^{3}$. The physical meaning of such a compactification is obscure to us \cite{bal_76}.

It may be noted that in principle, it is unnecessary to fix a gauge. The orbits
of ${\cal G}$  in $\{ \xi \}$ are well defined. We can work on the space of these
orbits. That is, ${\cal G}$ defines an equivalence relation $\sim$ on $\{ \xi \}$,
$\xi$ and $\xi'$ being equivalent if they are connected by the ${\cal G}$-action
that is,
\begin{equation}
\label{eq:3_59}
\xi \sim \xi' \Longleftrightarrow \xi' = \xi g~\mbox{for some}~g \in {\cal G}~.
\end{equation}

The space of orbits is just the space $\{ \xi \}$ with ${\cal G}$-equivalent points
identified, that is, $\{ \xi \}/\sim$. Thus for the spinning particle system discussed
above, it is unnecessary to fix a gauge. In fact, a global gauge does not exist
for this system
since the bundle is non-trivial. For each fixed time, the space $\{ \xi \}$ in
this case is the three sphere $S^{3}$, the group $G$ which is gauged is $U(1)$ and the space
of orbits $S^{3} / \sim$ is $S^{2}$. This example also shows that even if a
global gauge does not exist, the space of orbits, or the space of gauge
invariant variables, can still be well defined.

However, the sort of systems (like the spinning particle) we discuss in the present work
are rather exceptional. Here we can readily identify the space of gauge
invariant variables in a concrete way. In field theoretical problems, this
usually turns out to be difficult to do. The practice in these problems is to
fix the gauge by some convenient procedure. We have seen that a global gauge
fixing is not always possible. Such a circumstance can cause difficulties in
such problems during quantization.
 
Recently a perturbation theory for gauge fields without gauge fixing has been developed \cite{Parisi_81} based on the Langevin equation of non-equilibrium statistical mechanics. We will not, however, enter into its discussion here.
\noi 
\newpage

%% file: Magnetic_Monopoles.tex
%
%
\vspace{-3cm}
\begin{center}
\section{MAGNETIC MONOPOLES}
\label{section_magnetic_monopoles}
\end{center}
\seqnoll
\setcounter{exenum}{1}

{\Huge I}n this chapter, we discuss the classical formalism for the description of
a non-relativistic  charged particle in the field of a point-like Dirac
magnetic monopole   Ref.\cite{Dirac_31,Dirac_48,Olive_78}. This system as well illustrates the utility of the
fibre bundle formalism in an elementary context. See in this context also Refs.\cite{Friedman_79,Horvathy_80, Yang_75}.

\subsection{Equations of Motion}
\label{section_magnetic_monopoles_1}
 Let ${\bf x} = (x_1,x_2,x_3)$ denote the relative coordinates and $m$ the reduced mass of the system.	We assume that the magnetic field is Coulomb-like, i.e.,
\begin{equation}
\label{eq:4.0}
B_i = g\frac{x_i}{4\pi r^3}~~~.
\end{equation}

 Then the conventional Lorentz force equation, for a particle with an electric charge $q=-e$, reads
\begin{equation}
\label{eq:4.1}
m \ddot{x}_{i} = n \frac{1}{r^{3}} \varepsilon_{ijk} x_{j} \dot{x}_{k}~~~.
\end{equation}
Here $r\equiv |{\bf x}| $ is the radial coordinate, $\varepsilon_{ijk}$ is the Levi-Civita
symbol, $4 \pi n$ is the product $eg$ of the electric and magnetic charges
$e$ and $g$, and dots denote time differentiation.

The equation (\ref{eq:4.1}) reveals a remarkable structure when written in terms of
radial and angular variables. Let
\begin{equation}
\label{eq:4.2}
x_{i} = r \hat{x}_{i}~~,~~~\sum_{i} \hat{x}_{i}^{2} = 1~~.
\end{equation}
Then Eq.(\ref{eq:4.1}) is equivalent to
\begin{eqnarray}
 &  & \ddot{r} = r \sum_{i} \dot{\hat{x}}_{i}^{2}~~, \label{eq:4.3}\\
 &  & \frac{d}{dt}[m\varepsilon_{ijk} x_{j}\dot{x}_{k} + n \hat{x}_{i}] =
0~~.\label{eq:4.4}
\end{eqnarray}
The radial equation (\ref{eq:4.3}) has the same form as for a non-relativistic free particle. But from
Eq.(\ref{eq:4.4}), the conserved angular momentum
\begin{equation}
\label{eq:4.5}
J_{i} = m\varepsilon_{ijk} x_{j}  \dot{x}_{k} + n \hat{x}_{i}
\end{equation}
has an additional piece $n \hat{x}_{i}$ as compared to that of the free
particle. It can be interpreted as contributing a helicity
\begin{equation}
\label{eq:4.6}
\hat{x}_{i} J_{i} = n
\end{equation}
along the line joining the particle and the monopole.

\subsection{The Hamiltonian Formalism}
\label{section_magnetic_monopoles_2}
It is much easier to write down a Hamiltonian description of this system
than it is to write a Lagrangian description. We describe the former in this
section.
 
Let $B = \{ {\bf x} \in {\bf R} |~ r\equiv |{\bf x}| \neq 0 \}$ denote the configuration
space. Note that we have excluded the origin $r = 0$ from $B$. Thus
the electric charge and the magnetic monopole are forbidden to occupy the same space-time point.
The phase space $T^{\star}B$ can be chosen to have coordinates $({\bf x},
{\bf v})$, where ${\bf v} = (v_{1}, v_{2}, v_{3})$ denotes the relative
velocity of the electric charge and the magnetic monopole.

The equation of motion Eq.(\ref{eq:4.1}) is readily verified to be produced by the
Hamiltonian
\begin{eqnarray}
\label{eq:4.7}
H & = & \frac{1}{2} m  \sum_{i} v_{i}^{2}   \nonumber \\
~ & \equiv & \frac{1}{2} m v^{2}~~~,
\end{eqnarray}
provided the Poisson brackets (PB's) are chosen as follows:
\begin{eqnarray}
& & \{ x_{i}, x_{j} \} = 0~~, \label{eq:4.8}\\
& & \{x_{i}, v_{j} \} = \delta_{ij}/m~~, \label{eq:4.9}\\
& & \{v_{i}, v_{j} \} = - \frac{n}{m^{2}} \varepsilon_{ijk} \frac{x_{k}}{r^{3}} \label{eq:4.10}
~~~~~.
\end{eqnarray}
Note that since the right-hand side of Eq.(\ref{eq:4.1}) is proportional to the magnetic field, the PB Eq.(\ref{eq:4.10}) is conventional for \underline{velocities} in the presence of a magnetic field.

      As was the case for the spinning non-relativistic  particle, a global Lagrangian can be found
if a canonical system of coordinates $(Q,P)$ for $T^{\star}B$ can be found. It
may again be shown, however, that no such global system of coordinates exists \cite{Friedman_79}.
Thus, it is not possible to construct a global Lagrangian by application of a
simple Legendre transformation.

\subsection{The Lagrangian Formalism}
\label{section_magnetic_monopoles_3}

      The global Lagrangian can be constructed by enlarging the configuration space
$B$ appropriate to the Hamiltonian to a $U(1)$ bundle $E$ over $B$. This
Lagrangian exhibits weak gauge invariance under time dependent $U(1)$
transformations. As a consequence, the equations of motion are defined entirely
on $B$. The structure of the Lagrangian formalism bears a strong resemblance
to the one for the non-relativistic spinning system, although there are important
points of difference as well.

      Let $\{s\}$ denote the set of all two-by-two unitary unimodular matrices, i.e., elements of the $SU(2)$ group in the defining representation. The space
$E$ is
\begin{equation}
\label{eq:4.11}
E = R^{1} \times SU(2) \equiv \{ (r,s) \}~~~.
\end{equation}
Here $r$ is the radial variable with the restriction $r>0$. So the electric charge and
the magnetic monopole are again forbidden to occupy the same spacetime point. The
relation of $s$ to the relative coordinates $x_{i}$ is given by
\begin{eqnarray}
\label{eq:4.12}
\hat{X}  =  \sigma_{i} \hat{x}_{i}   =  s \sigma_{3} s^{-1}~~~.
\end{eqnarray}

In the Lagrangian below, the basic dynamical variables are $r$ and $s$. So,
wherever $x_{i}$ occurs, it is to be regarded as written in terms of $r$ and
$s$.
 
The Lagrangian is
\begin{eqnarray}
L & = & \frac{1}{2} m \sum_{i} \dot{x}_{i}^{2} + in\mbox{Tr}~ \sigma_{3} s^{-1}
\dot{s} \label{eq:4.13} \\
~ & = & \frac{1}{2} m \dot{r}^{2} + \frac{1}{4} m r^{2}~\mbox{Tr}~
\dot{\hat{X}}^{2} + in\mbox{Tr}~ \sigma_{3} s^{-1} \dot{s}~~~,
\label{eq:4.14}
\end{eqnarray}
 In writing Eq.(\ref{eq:4.14}) the identity $\mbox{Tr}~ \hat{X} \dot{\hat{X}} = 0$
has been used.
Variation of $r$ in Eq.(\ref{eq:4.14}) leads directly to Eq.(\ref{eq:4.3}). The most general
variation of $s$ is
\begin{equation}
 \label{eq:4.15}
\delta s = i \varepsilon_{i} \sigma_{i} s~~,~~~\varepsilon_{i}~\mbox{real}.
\end{equation}
Hence
\begin{equation}
 \label{eq:4.16}
\delta \hat{X} = i [ \varepsilon \cdot \sigma,\hat{X}]~~,~~~\varepsilon \cdot
\sigma = \varepsilon_{i} \sigma_{i}~,
\end{equation}
and
\begin{equation}
 \label{eq:4.17}
\delta ~\mbox{Tr}~ \sigma_{3} s^{-1} \dot{s} = i~\mbox{Tr}~ \dot{\varepsilon}
\cdot \sigma \hat{X}~~.
\end{equation}
The variation of $s$ in Eq.(\ref{eq:4.14}) thus leads to
\begin{equation}
 \label{eq:4.18}
\mbox{Tr}~ \varepsilon \cdot \sigma \frac{d}{dt} \{ - \frac{1}{2} [\hat{X}\, \, ,
mr^{2} \dot{\hat{X}}] + n \hat{X} \} = 0\, ,
\end{equation}
where we have used the identity Eq.(\ref{eq:3_33}). The bracketed expression in Eq.(\ref{eq:4.18})
is a linear combination of Pauli matrices and $\varepsilon_{i}$ is arbitrary.
Therefore,
\begin{equation}
\label{eq:4.19}
\frac{d}{dt} \{ - \frac{i}{2} [\hat{X}, mr \dot{\hat{X}}] + n \hat{X} \} = 0\, \, ,
\end{equation}
which is equivalent to
\begin{equation}
\label{eq:4.20}
\frac{dJ_{i}}{dt} = 0~~~,
\end{equation}
that is, to Eq.(\ref{eq:4.4}).

Thus $L$ leads to both the equations of motion Eq.(\ref{eq:4.3})  and Eq.(\ref{eq:4.4}).

\subsection{Gauge Properties of $L$}
\label{section_magnetic_monopoles_4}

    The Lagrangian $L$ shows a weak gauge invariance under gauge transformations
associated with the $U(1)$ group
\begin{equation}
\label{eq:4.21}
U(1) = \left\{ g = e^{i \sigma_{3}\theta/2 } \right\} ~~~~~.
\end{equation}
This is similar to the weak gauge invariance of the Lagrangian for the
spinning systems. Under the gauge transformation
\begin{equation}
\label{eq:4.22}
s \rightarrow s e^{i \sigma_{3}\theta/2 }~~~~,~~~~\theta =
\theta(t)~~~~,
\end{equation}
we have the weak gauge invariance
\begin{equation}
\label{eq:4.23}
L \rightarrow L - n \dot{\theta}~~~.
\end{equation}

As for the spinning system, associated with $L$, there is the fibre bundle
structure
\begin{equation}
\label{eq:4.24}
U(1) \rightarrow S^{3} \rightarrow S^{2}~~~.
\end{equation}

Again, it is impossible to fix a gauge globally so as to eliminate the $U(1)$
gauge degree of freedom. This is because $L$ is only weakly gauge invariant,
and $S^{3} \neq S^{2} \times U(1)$. Thus there does not exist an $s(\hat{X})
\in SU(2)$ which is continuous for all $\hat{X}$ such that
\begin{equation}
\label{eq:4.25}
\hat{X} = s(\hat{X}) \sigma_{3} s (\hat{X})^{-1}~~.
\end{equation}

It is of course possible to find an $s(\hat{X})$ 
which fails to be
continuous only at one point, say the south pole $[\hat{x} =(0,0,-1)]$. Such an
$s(\hat{X})$ is
\begin{eqnarray}
& & s(\hat{X}) = \frac{1}{2} \{ \alpha {\bf 1} - \frac{1}{\alpha} [\sigma_{3},
\hat{X}] \}~~~,  \nonumber \\
& & \alpha =  [ 2 (1 + \hat{x}_{3})]^{1/2} ~~~.
\label{eq:4.26}
\end{eqnarray}
It is easy to verify that $s(\hat{X})$ appearing in Eq.(\ref{eq:4.26}) is a
unimodular unitary matrix and fulfills  Eq.(\ref{eq:4.25}). Note that $s(\hat{X})$ in Eq.(\ref{eq:4.26}) is, however, 
not differentiable at the south pole. Substitution of Eq.(\ref{eq:4.26}) into the
interaction term appearing in  Eq.(\ref{eq:4.14}) yields
\begin{equation}
\label{eq:4.27}
in\mbox{Tr}~ \sigma_{3} s(\hat{X})^{-1} \dot{s} (\hat{X}) = n \varepsilon_{3ij}
\hat{x}_{i} \dot{\hat{x}}_{j} / (1 + \hat{x}_{3})
\end{equation}
which is a conventional form of the interaction Lagrangian with a string
singularity along the $x_{3}$ axis.

Alternatively, we can cover the two sphere $S^{2} = \{ \hat{X} \}$ by two
coordinate patches $U_{1}$ and $U_{2}$ and find group elements $s_{i}
(\hat{X})$ which are defined and continuous in $U_{1}$ and $U_{2}$ and which fulfill Eq.(\ref{eq:4.25}).
Substitution of $s_{i} (\hat{X})$ for $s$ in Eq.(\ref{eq:4.13}) leads to Lagrangians
$L_{i}$ defined on $U_{i}$. In the intersection region $U_{1} \cap U_{2}$, in
view of Eq.(\ref{eq:4.25}),
\begin{equation}
\label{eq:4.28}
[s_{1} (\hat{X})^{-1}s_{2} (\hat{X})] \sigma_{3} [s_{1}(\hat{X})^{-1} s_{2}
(\hat{X})]^{-1} = \sigma_{3}~~~.
\end{equation}
This means that $s_{i}$ differ from each other in the overlapping region
by a gauge transformation,
\begin{equation}
\label{eq:4.29}
s_{1} (\hat{X}) = s_{2} (\hat{X}) e^{i \frac{\sigma_{3}}{2} \theta}
\end{equation}
for some $\theta = \theta(t)$. Hence $L_{1}$ and $L_{2}$ differ by the total
time derivative of a function in $U_{1} \cap U_{2}$:
\label{eq:4.30}
\begin{equation}
L_{1} = L_{2} - n \dot{\theta}~~~.
\end{equation}
Such a (singularity free) formulation which works with two local Lagrangians
is the non-relativistic analogue of the work of Wu and Yang \cite{Yang_75}.

\noi 
\newpage

%% file: Relativistic_Particles.tex
%
\begin{center}
\section{RELATIVISTIC SPINNING PARTICLES}
\label{relativistic_spin}
\end{center}
\seqnoll
\setcounter{exenum}{1}

{\Huge I}n this chapter, we give the Lagrangian description for relativistic spinning particles,  which is formulated on he
Poincare group manifold \cite{Skagerstam_80}. It describes a particle which,
after quantization,  is associated with any particular irreducible representation of the connected Poincar\'{e} group  ${\cal P}_{+}^{\uparrow}$.	The Lagrangian formalism.can be generalized \cite{Skagerstam_80,Skagerstam_81} to include
couplings with both electro-magnetism and gravity. We recover the usual equations of motion for the two systems, i.e.,  the Bargmann-Michel-Telegdi \cite{Telegdi_59} equations for electro-magnetism
and the Mathisson - Papapetrou  \cite{Papapetrou_51}  equations for gravitation.
The latter equations have been generalized to include coupling to torsion \cite{Hel_71}. Such systems are also recovered from our formalism.

\subsection{The Configuration Space}

The Lagrangian is associated with a non-trivial principal fibre bundle structure  $U(1) \rightarrow {L}_{+}^{\uparrow} \rightarrow {L}_{+}^{\uparrow}/U(1)$,  which is obtained in the following way. The bundle space is the connected component of the Lorentz group ${L}_{+}^{\uparrow}$. The structure group is, as usual, $U(1)$. It acts on ${L}_{+}^{\uparrow} =\{ \Lambda \}$ by right multiplication,  i.e.,
\begin{equation}
\label{eq:5.1}
\Lambda	\rightarrow	\Lambda g	\, \,,\,\,  g	\in  U(1)\, \, .
\end{equation}
Thus the base space is the space of left cosets ${L}_{+}^{\uparrow}/ U(1)$ . As in previous sections, we can infer from connectivity arguments that ${L}_{+}^{\uparrow}\neq ({L}_{+}^{\uparrow}/U(1))\times U(1)$. Thus the bundle is non-trivial.

The configuration space $Q$ for the Lagrangian is the connected Poincar\'{e} group, i.e.,
\begin{equation}
\label{eq:5.2}
{\cal P}_{+}^{\uparrow} = \{z, \Lambda) \,\,  | \,\,   z = ( z^0, z^1, z^2, z^3)  \in   R^4 \, \, , \, \, \Lambda =[\Lambda^a_{\,\,\,b}]		\in {L}_{+}^{\uparrow} \}\, \, .
\end{equation}
Here $z^a$ is interpreted as the components of the space-time coordinate of the particle. The interpretation of $\Lambda $ is as follows. 
If $p_a$ and $S_{ab}$ are the momentum and spin components of the particle, we write
\begin{equation}
\label{eq:5.3}
p_a = m\Lambda_{a0}\,\, ,\,\, m > 0 \, \, ,
\end{equation}
and
\begin{equation}
\frac{1}{2}S_{ab}\sigma^{ab} = \lambda  \Lambda\sigma_{12}\Lambda^{-1} \equiv -iS\, \, ,
\label{eq:5.4}
\end{equation}
where the matrix elements of $\sigma^{ab}$ are given by
\begin{equation}
\label{eq:5.5}
(\sigma^{ab})_{cd}= -i(\delta^a_{c}\delta^b_{d} - \delta^a_{d}\delta^b_{c} )\, \, ,
\end{equation}
and $\lambda$ is a constant. These equations are valid for a time-like four vector $p_a$. The cases where the four vector $p_a$ is not time-like will be treated later. Note that by the definitions above,
\begin{equation}
\label{eq:5.6}
S_{ab}= \lambda (\Lambda_{a1}\Lambda_{b2} - \Lambda_{a2}\Lambda_{b1}  )\, \, ,
\end{equation}
and
\begin{eqnarray}
\label{eq:5.7}
p_0 &=& m\Lambda_{00} > 0\,\, ,\,\, p_ap^a= -m^2 \, \, .
\end{eqnarray}
Therefore we obtain
\begin{equation}
\label{eq:5.8}
\frac{1}{2}S_{ab}S^{ab}= \lambda^2\, \, ,
\end{equation}
and
\begin{equation}
\label{eq:5.9}
p_aS^{ab}= 0\, \, .
\end{equation}
Here the Latin indices are raised and lowered by the Lorentzian
metric
\begin{equation}
\label{eq:5.10}
\eta   =	(-1,1,1,1) \, \, .
\end{equation}

\subsection{The Lagrangian for a Free Spinning Particle}

The Lagrangian for a massive spinning particle is then given by
\begin{equation}
\label{eq:5.11}
L_p  =	p_a\dot{z}^a +i \frac{\lambda}{2} \mbox{Tr}\left[ \sigma_{12}\Lambda^{-1}\dot{\Lambda}\right]\, \, ,
\end{equation}
where $p_a$ is defined in the equation (\ref{eq:5.3}). The dynamical variables $z^a$, $p_a$ and $\Lambda$ in Eq.(\ref{eq:5.11}) are all functions of the parameter $\tau$ which parametrize the space-time trajectory of the particle. The dot indicates differentiation with respect to $\tau$. Note that	the action $\int L_p d\tau $ by construction is invariant under reparametrizations  $\tau \rightarrow f(\tau)$.

Let us first derive the equations of motion. The variation of $z^a$ is standard and leads to
\begin{equation}
\label{eq:5.12}
\dot{p}_a =	0\, \, .
\end{equation}
The most general variation of $\Lambda$ is, 	as usual,
\begin{equation}
\label{eq:5.13}
\delta\Lambda = i \varepsilon \cdot \sigma \Lambda	\, \, ,
\end{equation}
where
\begin{equation}
\label{eq:5.14}
\varepsilon \cdot \sigma =\varepsilon ^{ab} \sigma_{ab}	\, \, .
\end{equation}
This implies
\begin{equation}
\label{eq:5.15}
\delta\Lambda^{-1} = -i \Lambda^{-1}\varepsilon \cdot \sigma 	\, \, .
\end{equation}
Hence
\begin{equation}
\label{eq:5.16}
\delta L_p  =	- i\mbox{Tr}\left[ k \varepsilon \cdot \sigma \right]
+ \frac{i}{2}\mbox{Tr}\left[S\frac{d}{d\tau} (\varepsilon \cdot \sigma)\right] \, \, ,
\end{equation}
where the matrix $k$   is defined by $k_{ab} = \dot{z}_ap_b$. The traces have a conventional  meaning, i.e.,
\begin{equation}
\label{eq:5.17}
\mbox{Tr}\left[ k \varepsilon \cdot \sigma \right] = \sum_{ab}k^{ab}(\varepsilon \cdot \sigma)_{ba} \, \, .
\end{equation}
After a partial integration in the Eq.(\ref{eq:5.16})  and use of
Eq.(\ref{eq:5.12}), we obtain the equation for the conservation  of total angular momentum:
\begin{equation}
\label{eq:5.18}
\frac{d}{d\tau}M^{ab} =0 \, \, ,
\end{equation}
where
\begin{equation}
\label{eq:5.19}
M^{ab} = z^ap^b - z^bp^a + S^{ab}  \, \, .
\end{equation}

The proof that $L_p$ actually describes a particle which is associated with an irreducible representation of the connected
Poincar\'{e} group ${\cal P}_{+}^{\uparrow}$ follows by showing that mass and spin have
definite values. The mass has a definite value due to Eq.(\ref{eq:5.7}).
Note also that the sign of $p_0$   is fixed by Eq.(\ref{eq:5.3}) since $\Lambda \in L_{+}^{\uparrow}$.
Thus $L_p$ does \underline{not} describe a particle which can have  \underline{both} positive and negative energies. Both signs can be obtained by abandoning the condition  that $\Lambda_{00} >0$.

We can show that the spin has a definite value from equations (\ref{eq:5.8}) and (\ref{eq:5.9}).  The latter shows that in the rest frame of the particle, the spin tensor $S^{ab}$ has only space components. The former shows that the magnitude of this spin tensor has a definite numerical value. Thus the spin 3-vector $S_i \equiv \varepsilon_{ijk}S_{jk}/2$ has a definite value in the particle rest frame.  In general, by computing the square of the Pauli-Lubanski vector $W_a$, i.e.,
\begin{equation}
\label{eq:5.20}
W_a =\frac{1}{2} \varepsilon_{abcd}M^{bc}p^d \, \, ,
\end{equation}
where $\varepsilon_{abcd}$ is the usual anti-symmetric tensor with $\varepsilon_{0123}=1$, we find
\begin{equation}
\label{eq:5.21}
W_aW^a=\frac{1}{2} m^2\lambda^2 \, \, .
\end{equation}
It is important to realize that the preceding equations imply
\begin{equation}
\label{eq:5.22}
p_a =\frac{m\dot{z}_a}{\sqrt{-\dot{z}\cdot\dot{z}}}\, \, ,
\end{equation}
and
\begin{equation}
\label{eq:5.23}
\dot{S}_{ab} = 0 \, \, .
\end{equation}
Thus the conventional relation between momentum and velocity is recovered, and Eq.(\ref{eq:5.12}) becomes the usual equation of motion when written in terms of $z_a$. The derivation of these results relies on Eq.(\ref{eq:5.18}) which can be rewritten as
\begin{equation}
\label{eq:5.24}
\dot{z}^ap^b - \dot{z}^bp^a + \dot{S}^{ab} = 0 \, \, .
\end{equation}
in view of Eq.(\ref{eq:5.12}). It also relies on the time derivative of Eq.(\ref{eq:5.9}), i.e., 
\begin{equation}
\label{eq:5.25}
p_a\dot{S}^{ab} = 0 \, \, .
\end{equation}
Multiplication of Eq.(\ref{eq:5.24}) by $p_a$ shows that $p_a$ and $\dot{z}_a$ are, in fact, parallel. The constant of proportionality  can be determined by using the normalization condition $p_ap^a=-m^2$ and the condition that $p_0 > 0$. This gives Eq.(\ref{eq:5.22}). Now Eq.(\ref{eq:5.22}) applied to Eq.(\ref{eq:5.24}) yields Eq.(\ref{eq:5.23}).

The canonical quantization of the Lagrangian  Eq.(\ref{eq:5.11}) will be
carried out in Section \ref{canonical_formalism_3}.

\subsection{The Spinning Particle in an Electro-Magnetic Field}
\label{relativistic_spin_3}

We now discuss the coupling of electro-magnetism  to spinning particles \cite{Skagerstam_81}. In order that our system reduces to the standard formulation in the limit of zero spin and electric charge $q=e$, the minimal coupling term $eA_a(z)\dot{z}^a$ must be present in the interaction Lagrangian.  Here $A_a(z)$ is the electro-magnetic  potential. When the spin is non-zero, an additional coupling to the electro-magnetic  field of the form $cF_{ab}(z)S^{ab}$ may be present, where $F_{ab}= \partial_aA_b - \partial_bA_a$ and $c$ is a constant. As we will see below, the constant $c$ is associated with the gyro-magnetic ratio of the particle. One possible choice for the electro-magnetic  interaction Lagrangian is therefore
\begin{equation}
\label{eq:5.26}
L_{EI}= eA_a(z)\dot{z}^a + c\sqrt{-\dot{z}\cdot\dot{z}}F_{ab}(z)S^{ab}\, \, .
\end{equation}
The second term in Eq.(\ref{eq:5.26}) is the generalization  of the interaction term in the Hamiltonian Eq.(\ref{eq:3_6}) for a non-relativistic, spinning particle. The factor $\sqrt{-\dot{z}\cdot\dot{z}}$ in the second  term of Eq.(\ref{eq:5.26})
was inserted in order to retain the invariance under reparametrization  transformations $\tau \rightarrow f(\tau)$. As will be shown later, alternatives to Eq.(\ref{eq:5.26}) are possible.

The equations of motion are obtained by varying $\Lambda$ and $z$ in
the total action
\begin{equation}
\label{eq:5.27}
S = \int d\tau L_P + \int d\tau L_{EI} \, \, .
\end{equation}
Variations in  $\Lambda$ now lead to
\begin{equation}
\label{eq:5.28}
\dot{S}^{ab} + \dot{z}^ap^b - \dot{z}^bp^a = c\sqrt{-\dot{z}\cdot\dot{z}}\left( F^{ac}(z)S_{c}^{~b}-F^{bc}(z)S_c^{~a} \right)\, \, ,
\end{equation}
where $M^{ab}$ is defined in Eq.(\ref{eq:5.19}). Variations of $z^a$
\begin{equation}
\label{eq:5.29}
\dot{p}_a = eF_{ab}(z)\dot{z}^b + c\sqrt{-\dot{z}\cdot\dot{z}}S_{cd}\partial_aF^{cd}(z) +
c\frac{d}{d\tau}\left( \frac{\dot{z}_a}{\sqrt{-\dot{z}\cdot\dot{z}}} S\cdot F(z)\right) \,\, ,
\end{equation}
where we have introduced the notation $S\cdot F(z) \equiv S_{ab} F^{ab}(z)$.
Note that we no longer have the usual relationship between momentum and velocity Eq.(\ref{eq:5.22}).
In general, the velocity and momentum variables are, in fact,  not even parallel. This follows after substitution
of Eq.(\ref{eq:5.28}) and Eq.(\ref{eq:5.29}) into the condition
\begin{equation}
\label{eq:5.30}
\dot{S^{ab}}p_b + S^{ab}\dot{p}_b  =0\, \, .
\end{equation}
We find,
\begin{eqnarray}
\label{eq:5.31}
 p_a = - \frac{1}{p\cdot\dot{z}}\Big( m^2\dot{z}_a  +c\sqrt{-\dot{z}\cdot\dot{z}}
\left( p_bF^{bc}(z)S_{ca} + S_a^{~b}S_{cd}\partial_b F^{cd}(z)\right)\nonumber \\
+ cS_{ab}\frac{d}{d\tau}\left(\frac{\dot{z}^b}{\sqrt{-\dot{z}\cdot\dot{z}}}S\cdot F(z) \right) +eF_{bc}(z)\dot{z}^bS_{ab} \Big) \, \, .~~~~~~~~~~~
\end{eqnarray}
In order to compare this system with that of Bargmann et al. in Ref.\cite{Telegdi_59}, let us examine the weak and homogeneous  field limit. Upon substituting Eqs. (\ref{eq:5.28}) and (\ref{eq:5.31}) into (\ref{eq:5.29}), we then find
the Lorentz equation of motion, i.e.,
\begin{equation}
\label{eq:5.32}
 m\frac{d}{d\tau} \left(\frac{\dot{z}^b}{\sqrt{-\dot{z}\cdot\dot{z}}}\right) =eF^{ab}(z)\dot{z}_b\,\, .
\end{equation}
The equation for the spin precession can be expressed in terms of the
Pauli-Lubanski vector as defined in  Eq.(\ref{eq:5.20}). Substituting Eqs.(\ref{eq:5.28}),  (\ref{eq:5.31}), and  (\ref{eq:5.31}) into Eq.(\ref{eq:5.29}) into the equation
\begin{equation}
\label{eq:5.33}
 \dot{W}_a  = \frac{1}{2} \varepsilon_{abcd}\left( \dot{S^{bc}}p^d + S^{bc}\dot{p}^d \right) \, ,
\end{equation}
and again keeping terms which are at most linear in the homogeneous field we find
\begin{equation}
\label{eq:5.34}
 \dot{W}_a  = -2c\sqrt{-\dot{z}\cdot\dot{z}}F_{ab}W^b - \frac{2c+e/m}{\sqrt{-\dot{z}\cdot\dot{z}}} \left( \dot{z}^cF_{cb}W^b\right)\dot{z}_a \, \, .
\end{equation}
Equations (\ref{eq:5.32}) and (\ref{eq:5.34}) and  are the Bargmann-Michel-Telegdi
equations for a spinning particle \cite{Telegdi_59} with the identification
\begin{equation}
\label{eq:5.35}
 c = - \frac{eg_e}{4m}\,\, ,
\end{equation}
where $g_e$ being the gyro-magnetic ratio.

The field equations for this system are obtained by adding the usual free field action, i.e.,
\begin{equation}
\label{eq:5.36}
S_E= - \frac{1}{4} \int d^4x  F_{ab}(x) F^{ab}(x)\,\, ,
\end{equation}
to Eq.(\ref{eq:5.27}). By varying the electro-magnetic potentials $A_a$ and
integrating by parts, we find
\begin{equation}
\label{eq:5.37}
\partial_aF^{ab}(x)= -  q\int d\tau \delta^4(x- z(\tau)) \dot{z}^b +
2c\int d\tau \sqrt{-\dot{z}\cdot\dot{z}}\, \partial_a\delta^4(x- z(\tau)) S^{ab}\,\, ,
\end{equation}
where, as above, we use the notation $\partial_a \equiv \partial/\partial x^a$. The second
term on the right hand side of Eq.(\ref{eq:5.37}) represents the \underline{dipole}
contribution to the field in the sense of Papapetrou \cite{Papapetrou_51,Hel_71} (see
also in this context the work by Bailyn and Ragusa \cite{Ragusa_77} and
references therein).

As was stated above the interaction Lagrangian Eq.(\ref{eq:5.26}) is
not uniquely determined (for a related discussion see Ref.\cite{vanDam_1980}).
For instance, we can replace the second term in Eq.(\ref{eq:5.26}) by \cite{Kunzle_72}
\begin{equation}
\label{eq:5.38}
-\frac{c}{m}p_a\dot{z}^aS_{bc}F^{bc}(z)\,\, .
\end{equation}
This term preserve all the symmetries of the previous system, yet it gives a different set of equations of motion. In the
limit of a weak homogeneous field the two systems can, however, be shown to be equivalent. Note that the term in 
Eq.(\ref{eq:5.38}) can be absorbed in the first term in $L_P $ in Eq.(\ref{eq:5.11}), through a "renormalization"
of the mass $m$:
\begin{equation}
\label{eq:5.39}
m \rightarrow M(\alpha) = m +  \frac{eg_e}{4m} \, \, ; \, \, \alpha = F_{ab}S^{ab}\,\, .
\end{equation}
In fact, if we no longer restrict ourselves to Lagrangians which are 1inear in $F_{ab}S^{ab}$, we can consider the case where the mass $M(\alpha)$ is an arbitrary function of $\alpha$ (which may be relevant when one
is considering particles with an anomalous magnetic moment \cite{Wu_73}).
In this case the total particle Lagrangian would be
\begin{equation}
\label{eq:5.40}
L_P + L_{EI} = L_p  =	M(\alpha)\Lambda_{a0}\dot{z}^a + 
i \frac{\lambda}{2} \mbox{Tr}\left[ \sigma_{12}\Lambda^{-1}\dot{\Lambda}\right] + eA_a (z)\dot{z}^a \, \, , \,\, .
\end{equation}
The resulting equations of motion are
\begin{equation}
\label{eq:5.41}
\dot{S}^{ab} + \dot{z}^ap^b - \dot{z}^bp^a = \frac{d\ln M(\alpha)}{d\alpha}p_c \dot{z}^c\left( F^{ac}(z)S_{c}^{~b}-F^{bc}(z)S_c^{~a} \right) \,\, ,
\end{equation}
where we have corrected for a printing error in Eq.(31) of Ref.\cite{Skagerstam_81}, and
\begin{equation}
\label{eq:5.42}
 \dot{p}_a = eF_{ab}(z)\dot{z}^b +\frac{d}{d\alpha}(\ln M(\alpha))p_c\dot{z}^c\partial_a F_{bd}(z)S^{bd} \,\, .
\end{equation}
These equations have also been considered by Dixon \cite{Dixon_65}. Even though the equations Eq.(\ref{eq:5.41}) and Eq.(\ref{eq:5.42}) correspond to a large class of systems, depending on the choice of $M(\alpha)$, they all lead to the Bergmann-Michel-Telegdi equations in the weak and homogeneous field limit. Here the identification of the particle~s
mass and the gyro-magnetic ratio $g_e$ are given by
\begin{equation}
\label{eq:5.43}
m = M(0)  \,\, ,
\end{equation}
and
\begin{equation}
\label{eq:5.44}
g_e = \frac{4}{e}\frac{dM(\alpha)}{d\alpha}|_{\alpha = 0}  \,\, ,
\end{equation}
which are, of course, consistent with the specific choice Eq.(\ref{eq:5.35}).

\subsection{The Spinning Particle in a Gravitational Field}

It is rather straightforward to generalize $L_P$ to include gravitational effects. It is then convenient to regard the gravitational field as a gauge field \cite{Utiyama_56}, i.e., the Poincar\'{e} group is regarded as a local symmetry group. Let $h = h^a_{\mu}$ be the vierbein fields and $A^{ab}_{\mu}$ the corresponding Yang-Mills potentials for the Lorentz group. Our notation is as follows. A Latin index like $a$ is a tangent space index and a Greek index like $\mu$ is a curved space index. The metric tensor is $g_{\mu\nu} =\eta_{ab} h^{a}_{\mu}h^{b}_{\nu}$ and $\delta^{\mu}_{\nu}= h_{a}^{\mu}h^{a}_{\mu}$. The action now is
\begin{equation}
\label{eq:5.45}
S = \int d\tau L_P + \int d^4xL_F  \,\, ,
\end{equation}
where 
\begin{equation}
\label{eq:5.46}
L_P = m\Lambda_{ao}h^a_{\mu}\dot{z}^{\mu} + i\frac{\lambda}{2} \mbox{Tr}[\sigma_{12}\Lambda^{-1}D_{\tau}\Lambda]\,\, ,
\end{equation}
and
\begin{equation}
\label{eq:5.47}
L_F = -\frac{1}{16\pi G}F^{ab}_{\mu\nu}h^{\mu}_{a}h^{\nu}_{b} \mbox{det}(h) \,\, .
\end{equation}
Here the Yang-Mills the components of the field strength $F^{ab}_{\mu\nu}$ are given by 
\begin{equation}
\label{eq:5.48}
F_{\mu\nu} \equiv  \frac{i}{2}F^{ab}_{\mu\nu}\sigma_{ab} = [D_{\mu}, D_{\nu}] \,\, .
\end{equation}
where $D_{\mu}$ is the covariant derivative 
\begin{equation}
\label{eq:5.49}
D_{\mu} = \partial_{\mu} + \frac{i}{2}A^{ab}_{\mu\nu}\sigma_{ab}\,\, ,
\end{equation}
and 
\begin{equation}
\label{eq:5.50}
D_{\tau}\Lambda = \dot{\Lambda}+ \dot{z}^{\mu}A_{\mu}\Lambda \,\, ,
\end{equation}
with $A_{\mu} \equiv A_{\mu}^{ab}\sigma_{ab}$. Furthermore, in Eq.(\ref{eq:5.47}), $G$ is Newton's constant.

The equations of motion are found as follows. If we vary $\Lambda$ as in Eq.(\ref{eq:5.13}),  we see that Eq.(\ref{eq:5.16}) is replaced by 
\begin{equation}
\label{eq:5.51}
\delta L_P = -i  \mbox{Tr}\big[ J\varepsilon \cdot \sigma \big]+ \frac{i}{2}\mbox{Tr} \big[ S\frac{d}{d\tau}(\varepsilon \cdot \sigma) - [\dot{z}^{\mu}A_{\mu},S] \varepsilon \cdot \sigma \big]\,\, ,
\end{equation}
where
\begin{equation}
\label{eq:5.52}
J^{ab} = h^{a\mu}\dot{z}_{\mu}m\Lambda^{b0}\,\, .
\end{equation}
We thus find the equation for spin precession \cite{Weinberg_72}
\begin{equation}
\label{eq:5.53}
J^{ab} -J^{ba} +(D_{\tau}S)^{ab} = 0\, \, ,
\end{equation}
where
\begin{equation}
\label{eq:5.54}
(D_{\tau}S) = \frac{dS}{d\tau} + [\dot{z}_{\mu}A^{\mu},S]\,\, .
\end{equation}
Variation of the coordinate $z_{\mu}$ leads to the Mathisson-Papapetrou equation in the presence of torsion \cite{Papapetrou_51,Hel_71}. We find
\begin{equation}
\label{eq:5.55}
\delta L_P = -\delta z_{\mu} \big(  \dot{p}_{\mu} - \dot{z}_{\lambda}\partial_{\mu}h^{a\lambda}p_a \big)+ \delta\dot{z}_{\mu}\frac{1}{2}\mbox{Tr}\big[ SA^{\mu} \big]+ \delta z_{\mu}\frac{1}{2}\mbox{Tr} \big[S\partial_{\mu}A^{\lambda} \big]\dot{z}_{\lambda}\,\, ,
\end{equation}
where $p_ {\mu} = h^{a}_{\mu}p_a $. Partial integration in the second term and substitution from Eq.(\ref{eq:5.53}) leads to
\begin{equation}
\label{eq:5.56}
 \dot{p}_{\mu} - \dot{z}_{\lambda}\partial_{\mu}h^{a\lambda}p_a + \dot{z}_aA_{\mu}^{ab}p_b -\frac{1}{2} \mbox{Tr} \big[SF_{\mu\nu} \big]\dot{z}^{\nu} = 0\,\, .
\end{equation}
This is actually the same equation as the Mathisson-Papapetrou equation in the presence of torsion, i.e.,
\begin{equation}
\label{eq:5.57}
 (D_{\tau}p)_a  - h^{\mu}_a\dot{z}^{\nu}\big( (D_{\mu}h_{\nu})^b  - (D_{\nu}h_{\mu})^b \big)p_b  -\frac{1}{2}h^{\mu}_{a} \mbox{Tr} \big[SF_{\mu\nu} \big]\dot{z}^{\nu} = 0\,\, .
\end{equation}
In the Equation (\ref{eq:5.57}) we make use of the notation
\begin{equation}
\label{eq:5.58}
 (D_{\tau}p)_a  \equiv \frac{dp_a}{d\tau} + \eta_{ab}\dot{z}^{\lambda}A_{\lambda}^{bc}p_c\,\, ,
\end{equation}
and
\begin{equation}
\label{eq:5.59}
(D_{\mu}h_{\nu})^b \equiv \partial_{\mu}h^{b}_{\nu} + A^{bc}_{\mu}h_{c\nu}\,\, .
\end{equation}
For a discussion of the field equations, we refer to Ref.\cite{Skagerstam_80}.

\subsection{ General Irreducible Representations of ${\cal P}_{+}^{\uparrow}$}

To find Lagrangian descriptions for other unitary irreducible representations of the Poincare group ${\cal P}_{+}^{\uparrow}$ it is sufficient to alter the definitions of $p_a$ and $S_{ab}$. For example, in order to describe a massless particle, like the photon or a massless neutrino, we may replace $m\Lambda_{a0}$ in Eq.(\ref{eq:5.7}) by
\begin{equation}
\label{eq:5.60}
(\Lambda_{a0}+ \Lambda_{a3})\omega\,\, ,
\end{equation}
where $\omega$ corresponds to the angular "frequency" of the massless particle.
Equation (\ref{eq:5.60}) ensures that
\begin{equation}
\label{eq:5.61}
p_ap^a = 0\,\, .
\end{equation}
For a massless particle, the Pauli-Lubanski vector $W_a$, as given
by the Equation (\ref{eq:5.20}), obeys the condition that $W_aW^a = 0$, and a since $p_aW^a = 0$, it is easy to show the following identity
\begin{equation}
\label{eq:5.62}
W_a = \lambda p_a\,\, .
\end{equation}
It follows that for $\lambda = 1/2$ and $\lambda = 1$, we get a "neutrino" and a "photon" of definite helicity. The sign of the helicity can, of course, be reversed by reversing the sign of $\lambda$.

The tachyonic representations are obtained by choosing
\begin{equation}
\label{eq:5.63}
 p_a = \rho\Lambda_{a3}\,\, .
\end{equation}
Different values of $\rho$ and $\lambda$ give different irreducible representations as may be seen from the values of the invariants
$p_ap^a$ and $W_aW^a$:
\begin{equation}
\label{eq:5.64}
p_ap^a = \rho^2\,\, ,
\end{equation}
and 
\begin{equation}
\label{eq:5.65}
W_aW^a= -\rho^2 \lambda^2 \,\, .
\end{equation}

For the irreducible representations with zero four-momentum, we set
\begin{equation}
\label{eq:5.66}
p_a = 0 \,\, ,
\end{equation}
and choose the Lagrangian to be
\begin{equation}
\label{eq:5.67}
L_P = \frac{i}{2} \mbox{Tr} \big[K\Lambda^{-1}\dot{\Lambda} \big]\,\, ,
\end{equation}
where $K = K_{ab}\sigma^{ab}$ is a \underline{fixed} element of the Lie algebra. From
the variation Eq.(\ref{eq:5.13}), we find the equation of motion
\begin{equation}
\label{eq:5.68}
\frac{d}{d\tau}S_{ab} = 0 \,\, ,
\end{equation}
where
\begin{equation}
\label{eq:5.69}
\frac{1}{2}S_{ab}\sigma^{ab} = \Lambda K \Lambda^{-1} \,\, .
\end{equation}
Thus $S$ is the spin angular momentum. The irreducible representations are characterized by the two invariants \cite{Naimark_64}, i.e.,  $S_{ab}S^{ab}/2$ and $S_{ab}^{*}S^{ab}/2$. Since
\begin{equation}
\label{eq:5.70}
\frac{1}{2}S_{ab}S^{ab} = 2K_{ab}K^{ab}  \,\, ,
\end{equation}
and 
\begin{equation}
\label{eq:5.71}
\frac{1}{2}S_{ab}^{* }S^{ab} = 2K_{ab}^{*}K^{ab} \,\, ,
\end{equation}
we can, classically, get any real values for these invariants by choosing valuer for  $K_{ab}$ appropriately. To quantize the system
(see Chapter \ref{canonical_formalism}), we are, as usual, obliged to give them values which are appropriate for unitary irreducible representations \cite{Naimark_64}.

\subsection{Relation Between the Charge-Monopole System and the Massless Spinning Particle System}

In this section we point out some striking analogies between the charge-monopole system and the system of a massless particle
of fixed helicity. The similarities of the two systems become evident when the roles of coordinates and velocities are interchanged.

The analogies are as follows:

 $1)$ The angular momentum of a charged particle in the field of monopole contains a helicity $n$ (see Chapter \ref{section_magnetic_monopoles} and  Eq.(\ref{eq:4.6})) along the direction joining the monopole and the charge. The angular momentum of a massless particle of spin $\lambda$ contains helicity $\lambda$ along the direction of the momentum of the particle.

$2)$ The components of the position vector of the charge-monopole system commute, but the components of the velocity vector do
not (at least not for finite charge-monopole separation). Thus the system cannot be localized in velocity space. Furthermore, there is no globally defined momentum vector, consequently a  globally defined momentum space wave function cannot be defined. For a massless particle, on the other hand, the components of
momenta commute, i.e., 
\begin{equation}
\label{eq:5.72}
[p_i, p_j] = 0\,\, .
\end{equation}
But the components of position do not
\begin{equation}
\label{eq:5.73}
[x_i, x_j] =  -i\lambda\varepsilon_{ijk}\frac{p_k}{p^3}\,\, .
\end{equation}
Using Eqs. (\ref{eq:5.72}) and (\ref{eq:5.73}), along with the canonical commutation relation Eq.(\ref{eq:5.76}) as given below,  we can verify that $J_i = \varepsilon_{ijk}x_jp_k + \lambda p_i/p$ generates rotations for this system. Equation (\ref{eq:5.73}) is analogous
to the commutation relation for the components of velocity for the charge-monopole system Eq.(\ref{eq:4.10}). It is consistent with the
fact that the photon cannot be localized \cite{Pauli_64}. With the Hamiltonian $H = |\mathbf{p}|$ , we are lead to the equations of motion
\begin{equation}
\label{eq:5.74}
[x_i, x_j] =  -i\lambda\varepsilon_{ijk}\frac{p_k}{p^3}\,\, .
\end{equation}
and
\begin{equation}
\label{eq:5.75}
\frac{d}{dt} \left(  \varepsilon_{ijk}x_jp_k + \lambda p_i/p \right) \,\, ,
\end{equation}
if we supplement the commutation relations Eqs.(\ref{eq:5.72}) and (\ref{eq:5.73})
with the canonical commutation relation
\begin{equation}
\label{eq:5.76}
[x_i, p_j] = i\delta_{ij}\,\, .
\end{equation}

 $3)$ The non-trivial topology of the charge-monopole system depends on the fact that their relative spatial separation
cannot become zero. As a consequence, the configuration space has the topology $R\times S^1$.  The unusual topological features of
the charge-monopole system can be characterized in terms of this bundle. If the relative coordinate is allowed to vanish as well, the configuration space becomes $R^3$, which does not admit non-trivial $U(1)$ bundles.

In contrast, since for a massless particle its three momentum cannot be transformed to zero by Lorentz transformations, the origin in momentum space should be excluded. The topology of ${ \mathbf{p} }$  is thus $R^1 \times S^1$. For a non-zero helicity, its Lagrangian
description is facilitated by making use of the $U(1)$  bundle $R^1 \times S^3$ over $R^1 \times S^2$. In the photon Lagrangian, the entire Lorentz group appears to play the role of the bundle space. Consider, however, the translation group $T_2$ as generated by
\begin{equation}
\label{eq:5.77}
\Pi_1 = M_{10}+M_{13} \,\, ,
\end{equation}
and
\begin{equation}
\label{eq:5.78}
\Pi_2 = M_{20}+M_{23}\,\, .
\end{equation}
The photon Lagrangian is invariant under the transformations
\begin{equation}
\label{eq:5.79}
\Lambda \rightarrow \Lambda \dot \exp (i\alpha^a (x)\Pi_a)\,\, .
\end{equation}
Thus it can be globally written on $L_+/T_2 = R^1\times S^3$ by factoring these gauge degrees of freedom. The Euclidean group generated
by $\sigma_3$, $\Pi_1$, and $\Pi_2$ is the familiar stability group of the four momentum $( 1 , 0 , 0, 1)$.

 $4)$ From the expression for the conserved angular momentum, we see that under a parity transformation, $\lambda \rightarrow -\lambda$ for both systems under consideration. Thus a charge-monopole system with a fixed value of $e$ and $g$  ($e (g)$ being the electric (magnetic) charge), or a massless particle of fixed helicity, is incompatible with parity invariance.

 $5)$ There are no bound states in the charge-monopole system. For large times, the motion approaches that of two free particles, i.e.,
 as $t \rightarrow \infty$,   ,
\begin{equation}
\label{eq:5.80}
\mathbf{x}(t) \rightarrow \mathbf{v}t+ \mathbf{x}_0 +{\cal O}(t^{-2}) \,\, .
\end{equation}
where $\mathbf{x}(t)$ is the trajectory in the relative coordinate, and $\mathbf{x}_0$  and $\mathbf{v}$ are constant vectors. It follows that as $t \rightarrow \infty$ the conserved angular momentum $J_i = \varepsilon_{ijk}x_kp_k + \lambda \hat{\mathbf{x}}_i$, where
$\mathbf{p} = m \mathbf{v}$, approaches the value
\begin{equation}
\label{eq:5.81}
J_i = \varepsilon_{ijk}(\mathbf{x}_0)_jx_kp_k + \lambda \hat{\mathbf{p}}_i\,\, ,
\end{equation}
which has the same form as that for a massless particle. In
Ref.\cite{BGV_1978}, the preceding limit for the charge-monopole system was discussed in detail. It was shown that the commutation
relations of $\mathbf{x}_0$ and $\mathbf{p}$ are the same as those in the Eqs.(\ref{eq:5.72}), (\ref{eq:5.73}), and (\ref{eq:5.76}).

A canonical formalism was, furthermore, developed in Ref.\cite{BGV_1978} for a free, non-relativistic particle with no internal
degrees of freedom. This formalism was unusual in that upon quantization, the angular momentum contained a helicity $\lambda$ in the direction of the three momentum $\mathbf{p}$. If $\lambda$ is chosen as half integral, the system thus becomes "fermionic". Such a
system resembles a massless particle or the large time limit of the charge-monopole system. In Ref.\cite{ BGV_1978} no Lagrangian formulation of the system was given. We may notice here that it is just the non-relativistic analogue of the photon Lagrangian
i.e.,
\begin{equation}
\label{eq:5.82}
L = p\hat{p}_k \dot{x}_k  - i\mbox{Tr}\big[ \sigma_3s^{-1}\dot{s}\big]\,\, .
\end{equation}
Here $s \in  SU(2)$, the momentum is $p_k = p\hat{p}_k$  and $\sigma_kp_kk = s\sigma_3s^{-1}$. Thus $p$ is not an independent variable, but is defined in terms of the dynamical group element $s$.

%% file: Yang_Mills_Particles.tex
%
\newpage
\vspace{-3cm}
\begin{center}
\section{YANG-MILLS PARTICLES}
\label{yang_mills_particles}
\end{center}
\seqnoll
\setcounter{exenum}{1}

{\Huge T}he classical description of a charged particle in an
electro-magnetic field is well-known. The motion of the particle is described by the Lorentz force equation, while the
dynamics of the field is described by the Maxwell equations. The non-Abelian generalization of these equations is due to
Wong \cite{Wong_1970}. Instead of an electric charge, the corresponding Yang-Mills particle carries a spin-like variable ${\bf I}$ which
transforms under the adjoint representation of the internal
symmetry group. The Wong equations provide a coarser level
of description than a non-Abelian gauge field theory since
they treat the sources only as particles. Hence they may be more tractable than a gauge field theory and may also
reveal important features of the latter. For such reasons,
there is currently a growing interest in the Wong equations.
Below, we first recall the Wong equations. Then the
Hamiltonian and Lagrangian descriptions of these equations
are discussed. The Lagrangian description in our approach \cite{Bal_78,BSSW_1977} requires the use of non-trivial fiber bundles.

\subsection{The Wong Equations}
The Wong equations, with a gauge coupling $e$, are
\begin{equation}
\label{eq:6_1}
m\frac{d}{d\tau}\left[\frac{\dot{z}_a}{\sqrt{-\dot{z}^2}} \right] = -eF_{ab}^{\alpha}(z)I_{\alpha}\dot{z}^b\,\, ,
\end{equation}
and
\begin{equation}
\label{eq:6_2}
(D^a)_{\alpha\beta}F_{ab}^{\beta}(x) = e\int d\tau \delta^4 (x- z(\tau))\dot{z}_b(\tau) I_{\alpha}(\tau) \,\, .
\end{equation}
Here, $z^a = z^a(\tau)$ denotes the particle trajectory in Minkowski space, while 
$F_{ab}^{\beta} \equiv \partial_aA_b^{\beta} -\partial_bA_a^{\beta} + ec_{\beta\alpha\gamma}A_a^{\alpha}A_b^{\gamma}$
and  $(D_a)_{\alpha\beta} \equiv \delta_{\alpha\beta} \partial_a +ie[T(\gamma)]_{\alpha\beta}A^{\gamma}_a$ with the adjoint representation $[T(\gamma)]_{\alpha\beta} = -ic_{\alpha\beta\gamma}$, are the usual Yang-Mills tensor and co­variant derivative, respectively. 
The range of the indices $\alpha$,$\beta$ and $\gamma$ is equal to the dimension $n$ of the internal symmetry group $G$. The vector ${\bf I}= {\bf I}(\tau)$ transforms under the adjoint representation of $G$. From Eq.(\ref{eq:6_2}) and the identity 
\begin{equation}
\label{eq:6_3}
[D^a,D^b)]_{\alpha\beta}F_{ab}^{\beta} = 0\,\, ,
\end{equation}
one finds the following consistency condition
on ${\bf I}$:
\begin{equation}
\label{eq:6_4}
\frac{d}{d\tau}I_{\alpha}(\tau) -e\dot{z}^{a}(\tau) A^{\rho}_{a}(z(\tau))c_{\rho\alpha\beta}I_{\beta}(\tau) = 0\,\, ,
\end{equation}
Here $c_{\rho\alpha\beta}$ are the structure constants of $G$.

It is known \cite{Bal_78} that the spectrum of the Casimir invariants constructed out of ${\bf I}$ determines the irreducible representations
(IRRs) of $G$ which occur in the quantum-mechanical Hilbert space. We want to describe a particle which belongs to a definite IRR of $G$. Thus we impose also the constraint: \underline{Casimir invariants of ${\bf I}$ have definite} \underline{numerical values}. It is easy to show that this constraint is consistent with the time evolution of ${\bf I}$ according to Eq.(\ref{eq:6_4}).

It is instructive to verify that the preceding equations reduce to the Lorentz and Maxwell equations when $G=U(1)$. In this case ${\bf I}$ has only one component, say
 $I_1$. Since, in this case,   $c_{\rho\alpha\beta} = 0$, the component $I_1$ is a constant of motion by Eq.(\ref{eq:6_4}) and it can be assigned a definite numerical value, say $\lambda$. Identifying $e\lambda$ with the electric charge, Eqs.(\ref{eq:6_1}) and (\ref{eq:6_2}) are seen to reduce to the Lorentz and Maxwell equations of motion.
\subsection{The Hamilton Formalism}
The Hamiltonian is a  simple generalization of the electrodynamic Hamiltonian. It is 
\begin{equation}
\label{eq:6_5}
H = H_F + H_P\,\, ,
\end{equation}
where
\begin{equation}
\label{eq:6_6}
H_P = \big[ (p_i- eA^{\alpha}_{i}(z)I_{\alpha} )^2 + m^2c^4\big]^{1/2} + eA^{\alpha}_{0}(z)I_{\alpha}\,\, ,
\end{equation}
and $H_F$ is the Hamiltonian for the Yang-Mills field \cite{abers_1973}. The latter is well-known. In writing Eq.(\ref{eq:6_6}), we have identified
$z_0$ with time $\tau \equiv t$. The Poisson brackets (PBs) involving $z_i$'s and $p_i$'s, $i = 1,2,3$, are conventional. They have also
zero PBs with $I_{\alpha}$. The PBs involving $I_{\alpha}$ alone are
\begin{equation}
\label{eq:6_7}
\{I_{\alpha},I_{\beta} \} = c_{\alpha\beta\gamma}I_{\gamma}  \,\, .
\end{equation}
It is then straightforward to verify that the Hamiltonian Eq.(\ref{eq:6_5}) and the PBs in Eq.(\ref{eq:6_7}) lead to the required equations of motion.

\subsection{The Lagrangian Formalism}
The presence of a spin-like variable ${\bf I}$  whose Casimir invariants are fixed suggests in analogy to previous sections
that a Lagrangian can be found on a configuration space $E$ which contains additional gauge degrees of freedom. This is indeed the case. The space $E$ turns out to be ${\bf R}^3 \otimes G$ where ${\bf R}^3$ is the usual space of spatial coordinates and $G $ is the internal symmetry group.

We assume as usual that $G$ is a compact and connected Lie group with a simple Lie algebra $\underline{G}$. Let $\Gamma = \{ s \}$ be a faithful
unitary representation of $G$.  The associated Lie algebra  $\underline{\Gamma}$  has a basis $T(\rho) (\rho = 1,2, ..., n))$ with $T(\rho)^{\dagger} =T(\rho) $.
More precisely, this is a basis for $i\underline{G}$.   We choose $T(\rho)$ so that the normalization condition
\begin{equation}
\label{eq:6_8}
\mbox{Tr} [ T(\rho)T(\sigma)] \ =\delta_{\rho \sigma}  \,\, ,
\end{equation}
is fulfilled.

The commutation relations of $T(\rho)$ are
\begin{equation}
\label{eq:6_9}
[T(\rho) , T(\sigma)] = ic_{\rho \sigma \lambda} T(\lambda)   \,\, .
\end{equation}

The Lagrangian for the particle dynamics is
\begin{equation}
\label{eq:6_10}
L = -m[-\dot{z}^2]^{1/2}  - i \mbox{Tr}[Ks^{-1}(\tau)D_{\tau}s(\tau)]\,\, .
\end{equation}
Here $s \equiv s(\tau) \in \Gamma$ represents the novel degrees of freedom in $L$. The covariant derivative $D_{\tau}$  is defined by
\begin{equation}
\label{eq:6_11}
D_{\tau} = \frac{d}{d\tau} - ie\dot{z}^a A_a\,\, , \, \, A_a \equiv A_a^{\alpha}(z(\tau))T(\alpha) \, \, ,
\end{equation}
where $A_a^{\alpha}$ are the Yang-Mills potentials. The matrix $K$ is defined by
\begin{equation}
\label{eq:6_12}
K = K_{\rho}T(\rho) \, \, ,
\end{equation}
where $K_{\rho}$ are real valued \underline{constants}. Their specific values determine the IRR of $G$ to which the particle belongs.

The Yang-Mills Lagrangian
\begin{equation}
\label{eq:6_13}
-\frac{1}{4} \int d^3x  F_{ab}^{\alpha}F^{\alpha ab}\, \, ,
\end{equation}
can be added to $L$. We omit it here since the treatment of the Yang-Mills is standard (see, e.g., Ref.\cite{abers_1973}).

The definition of the internal vector $I \equiv {\bf I}$ in terms of $s$ and $ K$ is
\begin{equation}
\label{eq:6_14}
I = I_{\alpha} T(\alpha) = sKs^{-1}\, \, .
\end{equation}
The resemblance of equations (\ref{eq:6_10}) and (\ref{eq:6_14}) to the corresponding
equations in the previous sections should be noted.

Let us now derive the equations of motion. The general variation of $s$ is, as usual,
\begin{equation}
\label{eq:6_15}
\delta s = i\varepsilon \cdot T s\, \, , \, \, \varepsilon \cdot T =  \varepsilon_{\alpha} T(\alpha)   \, \,  .
\end{equation}
For this variation,
\begin{equation}
\label{eq:6_16}
\delta L = \mbox{Tr}[Ks^{-1}(\dot{\varepsilon}\cdot T + ie[\varepsilon \cdot T,\dot{z}A_a])s]  \, \,  .
\end{equation}
This becomes after a partial integration
\begin{equation}
\label{eq:6_17}
\delta L = -\mbox{Tr}[\varepsilon \cdot T ( D_{\tau}I ) ]\, \,  ,
\end{equation}
where
\begin{equation}
\label{eq:6_18}
D_{\tau}I \equiv \frac{dI}{d\tau} - ie[\dot{z}^a A_a,I]\,\, .
\end{equation}
Since $ D_{\tau}I \in \underline{\Gamma}$ and $\varepsilon_{\alpha}$ are arbitrary, the variation of $s$
leads to Eq.(\ref{eq:6_4}), i.e.,
\begin{equation}
\label{eq:6_19}
D_{\tau}I = 0\,\, .
\end{equation}

The Euler-Lagrange equation for the variation of $z^a$ can be obtained from
\begin{eqnarray}
\label{eq:6_20}
\frac{d}{d\tau}\frac{\partial L}{\partial \dot{z}_a} &=& m \frac{d}{d\tau}\left[ 
\frac{\dot{z}a}{(-\dot{z}^2)^{1/2}}\right] - \frac{d}{d\tau}\mbox{Tr}[iIA^a] \nonumber \\
&=&\frac{\partial L}{\partial z_a} = - e\mbox{Tr}[I\partial^aA^b]\dot{z}_b\,\, .
\end{eqnarray}

In view of Eq.(\ref{eq:6_19}), we thus find Eq.(\ref{eq:6_1}), i.e
\begin{equation}
\label{eq:6_21}
m\frac{d}{d\tau}\left[\frac{\dot{z}^a}{(-\dot{z}^2)^{1/2}} \right] = -e \mbox{Tr}[ IF^{ab}]\dot{z}_b  \,\, ,
\end{equation}
where $F_{ab} \equiv F_{ab}^{\alpha}T(\alpha)$. 

The variation of $A_a$ gives Eq.(\ref{eq:6_2}) in a standard way. Note that for this variation, the relevant term in the interaction
has the conventional form
\begin{equation}
\label{eq:6_22}
-e\dot{z}^aI_{\alpha}A^{\alpha}_ a(z)\,\, .
\end{equation}

It is again helpful to understand the form of $L$ when the
gauge group is $U(1)$, i.e., $s= \exp(i\psi)$ where $\psi$ is a real-valued function of $\tau$. The we can treat $K$ as a constant number and $L$ differs from the usual electro-magnetic Lagrangian by a term proportional to $d\psi/d\tau$. Since the latter is a  time-derivative of a function, we thus see that for the  $U(1)$ gauge group, $L$ is equivalent to the usual Lagrangian.

\subsection{Gauge Properties of $L$}
\label{yang_mills_particles_4}
The Lagrangian $L$ is invariant under the usual Yang-Mills gauge  transformation. Thus if $h(x) \in \Gamma $, it is invariant
under
\begin{eqnarray}
\label{eq:6_23}
A_a(x)&\rightarrow&  h(x)A_a(x)h(x)^{-1} + \frac{i}{e} h(x) \partial_a h(x)^{-1}\nonumber \\
s(\tau) &\rightarrow& h[z(\tau)]s(\tau)  \,\, .
\end{eqnarray}

It is also weakly invariant under a novel gauge group. The latter acts only on $s$ and not on $A_a$. It depends in
general on the nature of $K$. We thus explain it under two  headings: ($A$) the generic case and ($B$) the non-generic case.
In the discussion which follows we assume that $K \neq 0$.

\noindent $A.$\hspace{5mm}  \underline{The Generic Case}

Let $H = \{g\}$ denote all elements in $\Gamma$ with the property
\begin{equation}
\label{eq:6_24}
gKg^{-1} = K\,\, .
\end{equation}
Thus $H$ is the stability group of $K$ under the adjoint action. 

In the generic case, the Lie algebra $\underline{H}$ corresponding to $H$ is just the Cartan subalgebra containing $K$. If $\underline{C}$ is
an a priori chosen Cartan subalgebra, then in this case, there is a $t \in \Gamma$ such that
\begin{equation}
\label{eq:6_25}
tHt^{-1} = C\,\, .
\end{equation}
For example, if $G = SU(2)$, $\Gamma$ is its two-dimensional irreducible representation and $K = \sigma_3$, then $H = U(1) = \{\exp[i\sigma_3 \theta /2]\}$.
On the other hand, if $\Gamma$ is the adjoint representation of $SU(3)$, so that $\Gamma = SU(3)/Z_3$, and $K=I_3$, then $\underline{H}$ is spanned by $I_3$ and $Y$
(with a standard $SU(3)$ notation).

It can be shown that "most" $K$ are of this sort. The closure of the set of such $K$ is all of the Lie algebra $\underline{\Gamma}$ \cite{Michel_1980}. Note that for the generic case the group $H$ and the Lie algebra $\underline{H}$ are Abelian.

Under the gauge transformation
\begin{equation}
\label{eq:6_26}
s\rightarrow sg \,\, , \,\, g\in H \,\, ,
\end{equation}
with $s$ and $g$ $\tau$-dependent, we find
\begin{equation}
\label{eq:6_27}
L \rightarrow L - i \mbox{Tr}[Kg^{-1}\dot{g}]   \,\, .
\end{equation}
The extra term is the time derivative of a function since $H$ is Abelian. For instance, we can choose a basis
$K$, $L_{\alpha}$, where  $\alpha = 1,2, ...,k$ for $\underline{H}$ such that
\begin{equation}
\label{eq:6_28}
\mbox{Tr}[KL_{\alpha}] =0  \,\, .
\end{equation}
Then we can write
\begin{equation}
\label{eq:6_29}
g=e^{\displaystyle{\small i\Theta K}} e^{\displaystyle{i\Theta_{\alpha}L_{\alpha}}} \,\, .
\end{equation}
For this form of $g$,
\begin{equation}
\label{eq:6_30}
- i \mbox{Tr}[Kg^{-1}\dot{g}] = \mbox{Tr}[K^2\dot{\Theta}]  \,\, ,
\end{equation}
in view of Eq.(\ref{eq:6_28}). Thus $L$ is weakly invariant under $H$. The
principal fiber bundle structure relevant to $L$ is
\begin{equation}
\label{eq:6_31}
H \rightarrow \Gamma \rightarrow  \Gamma/H\,\, ,
\end{equation}
where $\Gamma/H = \{ sH \}$ is the space of left cosets. Thus $\Gamma$ and $\Gamma/H$ are the bundle and base spaces and $H$ is the structure
group.

These principal fiber bundles are never trivial. For instance, if $\Gamma$ is the defining representation of $SU(2)$ and
$H = U(1) = \exp(i\Theta\sigma_{3}/2)$ we get the Hopf fibration of the two sphere. The non-triviality of the bundle can also be seen
in general. Since $H$, being Abelian is the product of $U(1)$s (modulo perhaps a discrete group), it is infinitely connected. But $\Gamma$, being the representation of a simple compact Lie group, is finitely connected. Thus $\Gamma \neq \Gamma/H\times H$.

It follows that it is impossible to fix the gauge globally in this problem. However $L$ is \underline{invariant} under gauge transformations of the form $\{\exp(i\Theta_{\alpha}L_{\alpha})\}$ [Cf. equations (\ref{eq:6_29}) and (\ref{eq:6_30}). Thus the corresponding gauge degrees of freedom can be eliminated and $L$ can be written in terms of a configuration space $\Gamma/\{\exp(i\Theta_{\alpha}L_{\alpha})\}$. Since $L$ is only weakly invariant if $\Theta \neq 0$, the gauge degree of freedom for the $U(1)$ gauge group $\{ \exp(iK\Theta) \}$ cannot be so eliminated.

We note here the possibility of a topological problem which can prevent the elimination of the gauge degrees of
freedom associated with $\Gamma/\{\exp(i\Theta_{\alpha}L_{\alpha})\}$. It can occur that the
ratios of the eigenvalues of $K$ are not all rational. Then $\{\exp(i\Theta K)\}$ is isomorphic to the non-compact group of translations
on $R^1$ . The topology of the latter is incompatible with the topology of the compact $H$. Thus, in this case, the decomposition
$g = \exp(i\Theta K)\exp(i\Theta_{\alpha} L_{\alpha})$ is incompatible with the topology of $H$ and we cannot eliminate these gauge degrees of freedom in
a smooth way.

Note that since $\Gamma$ is a faithful representation of $G$
we can replace $\Gamma$ by $G$ · in much of the preceding  discussion.

\noindent $B.$\hspace{5mm}  \underline{The Non-Generic Case}

The non-zero elements in the complement of the generic $K$'s  in $\underline{\Gamma}$ constitute the non-generic $K$'s \cite{Bal_78}. The stability
group
\begin{equation}
\label{eq:6_32}
H \{ g\in \Gamma | gKg^{-1} = K \}\,\, ,
\end{equation}
for a non-generic $K$ is larger than that generated by the Cartan subalgebra containing $K$.
For example, if $G = SU(3)$ and $K = Y$, then $H = U(2)$. A basis for $H$ is $I_1, I_2, I_3, Y$  There are no non-generic
elements for $SU(2)$.

Let $K$ be non-generic with stability group $H$.  We can still choose a basis $K, L_{\alpha}$, where  $\alpha = 1,2, ...,k$  for $\underline{H}$ with the
property (\ref{eq:6_28}). Now
\begin{equation}
\label{eq:6_33}
\mbox{Tr} [K[L_{\alpha}, L_{\beta}]] = \mbox{Tr}[L_{\alpha}[L_{\beta},K]] = 0\,\, .
\end{equation}
If we write
\begin{equation}
\label{eq:6_34}
L_{\alpha}, L_{\beta}] = d_{\alpha \beta \gamma}L_{\gamma} + \xi K \,\, .
\end{equation}
it follows that
\begin{equation}
\label{eq:6_35}
\mbox{Tr} [K^2] = 0\,\, .
\end{equation}
But $\mbox{Tr}[K^2] = \mbox{Tr}[K K^{\dagger}] > 0$ 
Thus  $\xi = 0$. The conclusion is that $\underline{H}$ is the direct sum of two Lie algebras:
\begin{equation}
\label{eq:6_36}
\underline{H} = \underline{H}_0+ \underline{H}_1\,\, .
\end{equation}
The algebra $\underline{H}_0$ is one dimensional and is spanned by $K$. The algebra $\underline{H}_1$ has a basis $L_{\alpha}$, where  $\alpha = 1,2, ...,k$. For the $SU(3)$ example above, $\underline{H}_0 = \underline{U(1)}$ with basis $Y$ and $\underline{H}_1=\underline{SU(2)}$ with basis
$I_1, I_2, I_3$.

A general gauge transformation is of the form
\begin{equation}
\label{eq:6_37}
g=e^{\displaystyle{\small i\Theta K}} e^{\displaystyle{i\Theta_{\alpha}L_{\alpha}}} \,\, .
\end{equation}
Under $s\rightarrow sg$
\begin{eqnarray}
\label{eq:6_38}
L &\rightarrow& L - i \mbox{Tr}[Kg^{-1}\dot{g}] \nonumber \\
&=& L  + \mbox{Tr}[K^2\dot{\Theta}] + \mbox{Tr}[e^{\displaystyle{-i\Theta_{\alpha}L_{\alpha}}}\frac{d}{d\tau}e^{\displaystyle{i\Theta_{\alpha}L_{\alpha}}}] \,\, .
\end{eqnarray}

Since $\underline{H}_1$ is a Lie algebra, the term within the parentheses in Eq.(\ref{eq:6_38})
is in $\underline{H}_1$. Hence the last term is zero by Eq.(\ref{eq:6_28}). It
follows that $L$ is  weakly invariant under gauge transformations due to $H$. The gauge group in this case is in general
non-Abelian.

The principal fiber bundle structure is
\begin{equation}
\label{eq:6_39}
H \rightarrow \Gamma \rightarrow  \Gamma/H\,\, ,
\end{equation}
as in the generic case. It is non-trivial because $H$ is infinitely connected. The latter statement is proved as
follows: Let $H_0$ and $H_1$ be the groups associated with $\underline{H}_0$  and
$\underline{H}_1$. Then the group for $\underline{H}_0 + \underline{H}_1$  is $H = H_0 \otimes H_1$, possibly
modulo some discrete finite group. Thus $H$ is infinitely connected. For example, if $\Gamma = SU(3)$ and $K = Y$, then $H$ is $U(2)$
which is infinitely connected. Notice that $H$ is then not $SU(2) \otimes U(1)$, but $SU(2) \otimes U(1)/Z_2$. 

As in the generic case, $L$ is invariant under $\underline{H}_1$. Thus the $\underline{H}_1$ gauge freedom can be eliminated and $L$ can be written
as a function on $\Gamma/ \underline{H}_1$. After this partial elimination of gauge freedom, there still remains the $\underline{H}_0$ gauge freedom
and the principal fiber bundle structure
\begin{equation}
\label{eq:6_40}
H_0 \rightarrow \Gamma/H_1 \rightarrow  \Gamma/H\,\, .
\end{equation}
This remaining gauge freedom cannot be eliminated.

The gauge group can thus be reduced to $U(1)$ in $L$ both the generic and non-generic case with the possible exception as noted in the section  on the elimination of  some gauge degrees of freedom in the generic case. In fact, in almost all our examples from particle mechanics, the gauge group
is either $U(l$) or can be reduced to $U(l)$ by a process similar to the one above. In Chapter \ref{section_globalformulation} we prove a general theorem
which shows that under certain assumptions nothing more involved than $U(l)$ bundles need appear in mechanics. That is, we show that a global Lagrangian can always be found by enlarging the space of degrees of freedom appropriate to the equations of motion to at most a $U (1)$ bundle on this
space.

%
%
${\scriptsize{ \mid}}$
\subsection{An Application: Scattering off 't Hooft-Polyakov \newline Monopole}
 As an aside, we now illustrate how one can apply the preceding formalism to probe a specific Yang-Mills field configuration. The field configuration of interest is that of the  
't Hooft-Polyakov monopole solution \cite{Polyakov_1974}. When placed at large distances from the monopoles center, the Yang-Mills particle is known to behave similarly to that of an electric charge in a Dirac monopole field \cite{Schechter_1976}. This follows quite simply through the use of the Lagrangian Eq.(\ref{eq:6_10}), as is shown below.

In this example $G = SU(2)$ and we may set $T(\alpha) = \sigma_{\alpha}/\sqrt{2}$.
At large distances from the  center of the 
't Hooft-Polyakov monopole the gauge potentials $A_a$ take the form
\begin{eqnarray}
\label{eq:6_41}
A_i(x) &=& \frac{1}{2e|x|^2} 
\varepsilon_{\alpha ij}x_j\sigma_{\alpha} \, \, ,\nonumber \\
A_0(x) &=& 0 \, \, ,  \,\,  |x|^2 = x_ix_i \,\, , \,\, i =1,2,3 \, \, .
\end{eqnarray}

Here we shall restrict the discussion to non-relativistic particles. Upon substituting into $L$ we obtain
\begin{eqnarray}
\label{eq:6_42}
L &=& \frac{1}{2}m \dot{z}_i^2 - i\mbox{Tr}[Ks^{-1}\dot{s}] - \frac{i}{4}\mbox{Tr}[sKs^{-1}[\hat{z}, \frac{d}{d\tau}\hat{z}]]  \, \, ,\nonumber \\
\hat{z} &=& \frac{z_i\sigma_i}{r}  \,\, , \,\, r = |z| \, \, .
\end{eqnarray}
In analogy with Chapter \ref{section_magnetic_monopoles}, let us write
\begin{equation}
\label{eq:6_43}
\hat{z} = t\sigma_3 t^{-1} \,\, ,
\end{equation}
where
where $t \in \Gamma$ will be regarded as a dynamical variable defining $\hat{z}$. Notice that the dynamics of this system will not
be altered if we make the replacement
\begin{equation}
\label{eq:6_44}
s=tu  \,\, , \, \,  u\in \Gamma \,\, , 
\end{equation}
in Eq.(\ref{eq:6_42}). Variations of $s$ can be implemented through variations
of $u$. They can also be implemented through variations of $t$, which will simultaneously rotate the particle and its isospin $\bf{I}$.
Clearly, the above two variations are equivalent to varying $s$ and $t$ independently.

 Thus an equivalent Lagrangian for this system is
\begin{eqnarray}
\label{eq:6_45}
L &=& \frac{1}{2}m \dot{z}_i^2 - i\mbox{Tr}[K(tu)^{-1}\frac{d}{d\tau}{(tu)}] - \frac{i}{4}\mbox{Tr}[tuK(tu)^{-1}[\hat{z}, \frac{d}{d\tau}\hat{z}]]  \, \,  \\
\label {eq:6_46}
&=& \frac{1}{2}m \dot{r}^2 + \frac{1}{4}mr^2 \mbox{Tr}[(\frac{d}{d\tau}\hat{z})^2] -\frac{1}{2}\mbox{Tr}[I_t\sigma_3]\mbox{Tr}[\sigma_3t^{-1}\dot{t}]- i\mbox{Tr}[Ku^{-1}\dot{u}] \, ,
\end{eqnarray}
using the fact that $\mbox{Tr}[\hat{z}d\hat{z}/d\tau]=0$, and where we have defined $I_t\equiv t^{-1}It$. Let us now take up the equations of motion. Variations of the coordinate $z_i$ will be performed through variations of $r$ and $t$ [cf. Chapter \ref{section_magnetic_monopoles}]. Variations of $u$ yield
\begin{eqnarray}
\label{eq:6_47}
\dot{I}_t - \frac{1}{2}\mbox{Tr}[\sigma_3t^{-1}\dot{t}][I_t,\sigma_3] = 0 \, .
\end{eqnarray}
The isospin vector $I_t$ thus precesses around the third direction in internal space. The precessional frequency depends on
the position of the particle through the variable $t$. By taking the trace of Eq.(\ref{eq:6_47}) with $\sigma_3$,  we find
\begin{eqnarray}
\label{eq:6_48}
\mbox{Tr}[\sigma_3I_t] = -2n = \mbox{constant}\, .
\end{eqnarray}
The remaining equations of motion are obtained from variations in the first three terms in Eq.(\ref{eq:6_46}). Notice that the first
three terms are identical to the Lagrangian Eq.(\ref{eq:4.14}) describing a charged particle in a Dirac monopole field  with the
assignment Eq.(\ref{eq:6_48}). Thus the Yang-Mills particle behaves as a charged particle in a Dirac monopole field with additional
internal dynamics given by Eq.(\ref{eq:6_47}). Here the correspondence is 
\begin{eqnarray}
\label{eq:6_49}
\frac{eg}{2\pi} \leftrightarrow - \mbox{Tr}[\sigma_3I_t]\, .
\end{eqnarray}
Unlike the charge-monopole system of Chapter \ref{section_magnetic_monopoles}, $n$ is not a fixed number in the Lagrangian, but rather a dynamical quantity which obeys the inequality:
\begin{eqnarray}
\label{eq:6_50}
n^2 \leq \frac{1}{2}\mbox{Tr}[I_t^2] = \frac{1}{2}\mbox{Tr}[I^2] \, .
\end{eqnarray}
We thus expect that in the quantum mechanical system for the particle, a spectrum in $n$ will appear, consistent with the
inequality (6.49)\cite{{Schechter_1976}}.

\noi 
\newpage

%% file: Kaluza_Klein_Theory.tex
%
\vspace{-3cm}
\begin{center}
\section{KALUZA-KLEIN THEORY}
\label{kaluza_klein_theory}
\end{center}
\seqnoll
\setcounter{exenum}{1}

{\Huge T}he unified field theory of Kaluza and Klein \cite{Klauza_Klein_1921} has been experiencing a revival of interest since the development of gauge
field theories in elementary particle physics. Here the dynamical
fields, denoted collectively by $\psi$, depend both on a space-time coordinate $x$ and a group element $s$, i.e.,
$\psi  \equiv \psi(x,s)$. If $M^4 = \{ x \}$  is the Minkowski space and $G = \{s\}$ denotes
the internal symmetry group, the fields are thus defined
on the principal fibre bundle $M^4 \times G$ \cite{DeWitt_1965,Cho_1975}.

In this section, we will discuss such theories in the context of particle mechanics. For extended objects \cite{Bal_Al_Bo_79}, the
Kaluza-Klein formalism can be generalized in a straightforward manner \cite{Nielsen_80}.

\subsection{Kaluza-Klein Description of Point Particles}
\label{kaluza_klein_theory_1}
The conventional description of the Kaluza-Klein formalism
is as follows. Let $x^a$ denote the space-time coordinate of the
particle. Let $G={s}$ be a semi-simple, compact Lie group represented
by unitary matrices. Here we wish to use $G$ to describe
the internal degrees of freedom of the particle. The natural
metric to be used on $M^4 \times G$ is a combination of the invariant
line element on $M^4$ and the left invariant metric on $G$ \cite{DeWitt_1965}.
The Lagrangian is chosen to be
\begin{equation}
\label{eq:7.1}
L = -m \left( -\dot{x}^2 - \lambda\mbox{Tr}[ s^{-1}\dot{s} s^{-1}\dot{s} ] \right)^{1/2} \, \, .
\end{equation}
Here $m$ and $\lambda$ are constants, and $x^a \equiv x^a(\tau)$, $s \equiv s(\tau)$. Geometrically, this Lagrangian has the following meaning. Let
us enlarge the Minkowski $M^4$ to $M^4 \times G$ and  regard the latter as the configuration space. Recall that the Lagrangian for a
free particle possessing no internal symmetries is proportional
to the invariant length in $M^4$. Similarly, the Lagrangian (\ref{eq:7.1}) is proportional to the invariant length on $M^4 \times G$.

The system given by Eq.(\ref{eq:7.1}) has the following properties:

$i)$ The states in the quantum system described by $L$ belong to a
reducible representation of $G$. This differs from the quantum
system for the Yang-Mills particle described in Chapter \ref{yang_mills_particles} (also Cf. Sec.\ref{canonical_formalism_4}).

$ii)$ The square of the momentum, $p^2$, depends on the quadratic
Casimir operator. This leads to a mass spectrum for the particle.

Regarding $i)$ we note that quantum mechanical Hilbert space
carries the regular representation (see Sec.\ref{canonical_formalism_5}). Thus the
multiplicity of an irreducible representation is equal to its
dimension by the theorem of Peter and Weyl \cite{Talman_68}.

We can show $ii)$ by computing the operator $p_a$ which generates
translations, and the internal generators $I_{\alpha}$ from $L$, and showing
that they are algebraically related. Thus we find
\begin{equation}
\label{eq:7.2}
p_a = \frac{\partial L}{\partial \dot{x}^a} = \frac{m^2\dot{x}_a}{L} \, \, .
\end{equation}
The generators $I_{\alpha}$ can be found by examine the variation
\begin{equation}
\label{eq:7.3}
\delta s = i\varepsilon_{\alpha} \mbox{T}(\alpha)s\,\, ,
\end{equation}
where $\mbox{T}(\alpha)$ are the hermitian generators of $G$, which fulfill
\begin{equation}
\label{eq:7.4}
\mbox{Tr}[\mbox{T}(\alpha)\mbox{T}(\beta)] =\delta_{\alpha \beta} \, \, .
\end{equation}
It follows that
\begin{equation}
\label{eq:7.5}
\delta L = -i\dot{\varepsilon}_{\alpha} \frac{m^2\lambda}{L} \mbox{Tr}[\mbox{T}(\alpha)\dot{s}s^{-1}]\, \, .
\end{equation}
Thus the quantities
\begin{equation}
\label{eq:7.6}
I_{\alpha} =  i\frac{m^2\lambda}{L} \mbox{Tr}[\mbox{T}\left(\alpha)\dot{s}s^{-1}\right]\, \, , 
\end{equation}
are conserved. In Section \ref{canonical_formalism_5} they will be shown to generate internal
symmetry transformations. 
Now
\begin{equation}
\label{eq:7.7}
p_a p^a =  \frac{m^4}{L^2}\dot{x}^2 \, \, , 
\end{equation}
and furthermore,
\begin{equation}
\label{eq:7.8}
I_{\alpha}I_{\alpha} = -\frac{m^4\lambda^4}{L^2} \mbox{Tr}\left[s^{-1}\dot{s}s^{-1}\dot{s}\right]\, \, ,
\end{equation}
where we have used the completeness of the generators, i.e .
\begin{equation}
\label{eq:7.9}
\mbox{T}(\alpha)\mbox{Tr}\left [\mbox{T}(\alpha)s^{-1}\dot{s}\right] = s^{-1}\dot{s} \, \, .
\end{equation}
Hence we obtain
\begin{equation}
\label{eq:7.10}
p_a p^a - \frac{1}{\lambda}I_{\alpha}I_{\alpha} = -m^2 \, \, .
\end{equation}
By defining the mass $M$ as $p^2 = M^2$, we can rewrite Eq.(\ref{eq:7.10})
as follows
\begin{equation}
\label{eq:7.11}
M^2 = m^2 - \frac{1}{\lambda}I_{\alpha}I_{\alpha}\, .
\end{equation} 
If $\lambda $ is less than zero, the $M^2$-spectrum increases with the
quadratic Casimir operator. If $\lambda $ is larger than zero, the
mass $M$ becomes imaginary for some value of $I_{\alpha}I_{\alpha}$, $L$ becomes
complex and the system is inconsistent.

\subsection{Reformulation of the Kaluza-Klein Theory}
\label{kaluza_klein_theory_2}
We shall now formulate the Kaluza-Klein Lagrangian in a different way. Although the classical equations of motion for this new system are identical to those discussed in the previous section, the corresponding quantum theories differ. Unlike in the previous section, the quantum mechanical Hilbert space derived
from the following Lagrangian carries an irreducible representation of the group $G$.

The idea here is a simple generalization of the Lagrangian formalism used to describe the relativistic point particle
as discussed in detail in Chapter \ref{relativistic_spin}. For the latter, if the mass is $m$ and the spin is zero, the Lagrangian has the form
\begin{equation}
\label{eq:7.12}
L = p_a\dot{x}^a\, ,
\end{equation} 
where $p_ap^a= -m^2$. We can generalize Eq.(\ref{eq:7.12}) to
\begin{equation}
\label{eq:7.13}
L = p_a\dot{x}^a + i\mbox{Tr}\left [Ks^{-1}\dot{s}\right]\, ,
\end{equation} 
where $p_a$ is now defined by 
\begin{equation}
\label{eq:7.14}
p_a = \left( m^2 + \frac{1}{\lambda}\mbox{Tr}\left [K^2\right]\right)^{1/2} \Lambda_{a0}\, .
\end{equation} 
If $K=K_{\alpha}T(\alpha)$ is treated as a dynamical variable we recover precisely the system discussed in the previous section, where the quantum mechanical Hilbert space carries the left regular representation of $G$. Here, however, $K$ will be treated as a constant.

The equivalence of Eqs.(\ref{eq:7.14})  and (\ref{eq:7.1}) at the classical level is now shown by proving that for the Lagrangian Eq.(\ref{eq:7.13}), $\mbox{Tr}\left [K^2\right]$ is the quadratic Casimir operator of the generators of $G$. Now consider the variation Eq.(\ref{eq:7.3}) of $s$ for which
\begin{equation}
\label{eq:7.15}
\delta\left(i\mbox{Tr}\left [Ks^{-1}\dot{s}\right] \right) = -\mbox{Tr}\left [sKs^{-1}\mbox{T}(\alpha)\dot{\varepsilon}_{\alpha}\right] \, .
\end{equation} 
Consequently, the following charges $J_{\alpha}$ are conserved
\begin{equation}
\label{eq:7.16}
J_{\alpha} \equiv \mbox{Tr}\left[\mbox{T}(\alpha)sKs^{-1}\right]  \, .
\end{equation} 
$J_{\alpha}$ actually form the generators of $G$ on the quantum mechanical Hilbert space. The desired result
\begin{equation}
\label{eq:7.17}
p_ap^a =  -m^2 - \frac{1}{\lambda}J_{\alpha}J_{\alpha}  \, ,
\end{equation} 
then follows from Eq.(\ref{eq:7.14}).

Note that the system described here is equivalent to the description of a free Wong article  (Cf.~Chap.\ref{yang_mills_particles}), except
for the constraint (\ref{eq:7.17}), which resulted from the redefinition of momentum. The constraint (\ref{eq:7.17}) is rather arbitrary. In fact,
we can easily arrange for any mass-internal symmetry relation by changing the function appearing in front of $\Lambda_{a0}$ in Eq.(\ref{eq:7.14})). In
order to see this we notice that if $C_n(J)$, $n=1,2,..., , \mbox{rank} (G)$, denotes the Casimir invariants of $G$, then by Eq.(\ref{eq:7.16}), $C_n(J)=C_n(K)$.
It follows that by setting
\begin{equation}
\label{eq:7.18}
p_a = f(C_1(K),C_1(K), ...)\Lambda_{a0}\, ,
\end{equation} 
for a suitable function $f$, we can get any mass spectrum. For a conventional formulation of theories of this kind we refer the reader to the work by N. Mukunda et al. \cite{Mukunda_80}. Note that the procedure of redefining momenta  (or actually the mass) of a particle was also found to be useful in introducing an anomalous magnetic moment for a spinning particle (Cf. Sec.\ref{relativistic_spin_3}).

\subsection{Interaction with External Fields}
\label{kaluza_klein_theory_3}
Above we have considered a non-interacting particle with internal degrees of freedom. The incorporation of external fields
is straightforward and as a result we can obtain the Wong equations \cite{Wong_1970}. In order to achieve this result we replace
the time derivatives of the group element $s(t)$ in Eq.(\ref{eq:7.1}) by the corresponding covariant derivative Eq.(\ref{eq:6_11}), i.e., we consider the
Lagrangian \cite{Nielsen_80}
\begin{equation}
\label{eq:7.19}
L = -m\left( -\dot{x}^2 -\lambda\mbox{Tr}\left [s^{-1}D_{\tau}ss^{-1}D_{\tau}s\right]
\right)^{1/2} \, .
\end{equation} 
The equation of motion for the non-Abelian charges
\begin{equation}
\label{eq:7.20}
I_{\alpha}= i\frac{m^2\lambda}{L}\mbox{Tr}\left [\mbox{T}(\alpha)(D_{\tau}s)s^{-1}\right] \, ,
\end{equation} 
is, as before, obtained by considering the variation Eq.(\ref{eq:7.3}), i.e., $\delta s = i\varepsilon_{\alpha}\mbox{T}(\alpha)s$. The analogue of the Eq.(\ref{eq:7.5}) is then
\begin{eqnarray}
\label{eq:7.21}
\delta L = -i\dot{\varepsilon}_{\alpha} \frac{m^2\lambda}{L} \mbox{Tr}[\mbox{T}(\alpha)(D_{\tau}s)s^{-1}] \nonumber \\
+ \varepsilon_{\alpha} \frac{m^2\lambda}{L} \mbox{Tr}\left[[\dot{x}^aA_a, \mbox{T}(\alpha)(D_{\tau}s)s^{-1}]\right]
\, \, .
\end{eqnarray}
We obtain the equation of motion 
\begin{equation}
\label{eq:7.22}
\frac{dI_{\alpha}(\tau)}{d\tau}= i[\dot{x}^a(\tau)A_a(x(\tau)), I_{\alpha}(\tau)]\, ,
\end{equation} 
i.e., the Eq.(\ref{eq:6_19}).

The Euler-Lagrange equation for $x^a(\tau)$ is
\begin{eqnarray}
\label{eq:7.23}
\frac{d}{d\tau}\left( \frac{\partial L}{\partial \dot{x}_a}\right)  &= &m^2\frac{d}{d\tau}\left( \frac{\dot{x}^a}{L }\right)  -\frac{d}{d\tau}\left( IA^a\right)  \nonumber \\
&=& -e\mbox{Tr}\left[I\partial^a A_b \right]\dot{x}_b\, \, ,
\end{eqnarray}
as in the derivation of Eq.(\ref{eq:6_20}). By choosing the parameter $\tau$ in such a way that $L = m$, we obtain the Lorentz-Maxwell-Wong
equation (\ref{eq:6_1}) in the proper time gauge.

Finally, we notice that in the presence of an external field
the mass-internal symmetry relation Eq.(\ref{eq:7.10}) is changed to
\begin{equation}
\label{eq:7.24}
\left( p_a + eI_{\alpha}A_a^{\alpha}\right)\left( p^a + eI_{\alpha}A^{a\alpha}\right) - \frac{1}{\lambda}I_{\alpha}I_{\alpha} = - m^2 \, .
\end{equation} 
\noi 
\newpage

%% file: Canonical_Quantization.tex
%
\vspace{-3cm}
\begin{center}
\section{THE CANONICAL FORMALISM AND QUANTIZATION}
\label{canonical_formalism}
\end{center}
\seqnoll
\setcounter{exenum}{1}
{\Huge I}n this section we carry out the canonical quantization for the various systems discussed in the previous chapters.
Since all the Lagrangians presented here are singular, i.e., there exist constraints amongst the corresponding phase-space
variables, we will rely on Dirac's quantization procedure. For
extensive reviews on this procedure, see  Ref.\cite{dirac_64}.

A common feature of all the systems presented here is
that elements of a group $G$ appear as dynamical variables. A
method of treating group elements for setting up the canonical
formalism  was given in Refs.\cite{dirac_64} and \cite{Bal_78}. We recall it below.

Let $s \in G$ be parametrized by a set of variables (local coordinates) $\xi = (\xi_{1}, \xi_{2}, \ldots , \xi_{n})$ so that $s = s(\xi)$, $n$ being the
dimension of $G$. The functional form of $s(\xi)$ will not be important for us. We can then regard the Lagrangian as a function of $\xi$
and $\dot{\xi}$ as well as of any other configuration space variables
present in the system and of their velocities.

We first note a preliminary identity. Let us define a set of functions
$f(\varepsilon) = (f_{1} (\varepsilon), f_{2} (\varepsilon), \ldots ,f_{n}
(\varepsilon))$, $\varepsilon = (\varepsilon_{1}, \varepsilon_{2}, \ldots ,
\varepsilon_{n})$, by
\begin{equation}
\label{eq:8_1}
e^{iT(\alpha)\varepsilon_{\alpha}} s(\xi) = s [f(\varepsilon)],~~f(0) = \xi
~~~,
\end{equation}
where $T(\alpha)$'s form a basis for the Lie algebra of the group with
\begin{equation}
\label{eq:8_2}
[T(\alpha),T(\beta)] = i~ c_{\alpha \beta \gamma} T (\gamma)~~~.
\end{equation}
Differentiating Eq.(\ref{eq:8_1}) with respect to $\varepsilon_{\alpha}$ and setting
$\varepsilon = 0$, we find
\begin{eqnarray}
\label{eq:8_3}
& & i T(\alpha) s(\xi) = \frac{\partial s (\xi)}{\partial \xi_{\beta}} N_{\beta
\alpha}  (\xi)~~~,
\end{eqnarray}
where
\begin{eqnarray}
\label{eq:8_4}
& & N_{\beta \alpha} (\xi) = \frac{\partial f_{\beta} (\varepsilon)}{\partial
\varepsilon_{\alpha}}|_{\varepsilon = 0}~~.
\end{eqnarray}
Here $\det N \neq 0$, for if not,  there exist $\chi_{\alpha}$, not all zero,
such that $N_{\rho \sigma} \chi_{\sigma} = 0$. By Eq.(\ref{eq:8_3}), $\chi_{\sigma}
T(\sigma)s(\xi) = 0$, and hence $\chi_{\sigma}T (\sigma) = 0$. But this
contradicts the linear independence of the $T(\alpha)$'s.

Now the coordinates $\xi_{\alpha}$ and their conjugate momenta $\pi_{\alpha}$
fulfill the Poisson bracket (PB) relations
\begin{eqnarray}
\label{eq:8_5}
\{ \xi_{\alpha}, \xi_{\beta} \} & = & \{ \pi_{\alpha}, \pi_{\beta} \} = 0~~~,
\nonumber \\
\{ \xi_{\alpha},\pi_{\beta} \} & = & \delta_{\alpha \beta}~~~~.
\end{eqnarray}
Since $N$ is nonsingular, we can replace the phase space variables
$\pi_{\alpha}$ by $t_{\alpha}$ where
\begin{equation}
\label{eq:8_6}
t_{\alpha} = - \pi_{\beta} N_{\beta \alpha}~~~.
\end{equation}
From Eqs.(\ref{eq:8_3}) and (\ref{eq:8_5}) it follows that
\begin{eqnarray}
\label{eq:8_7}
& & \{t_{\alpha}, s\} = i ~T(\alpha) s~~,\\
\label{eq:8_8}
& & \{t_{\alpha}, s^{-1}\} = -i~ s^{-1} T(\alpha)~~,\\
\label{eq:8_9}
& & \{t_{\alpha}, t_{\beta}\} = c_{\alpha \beta \gamma}~ t_{\gamma}~~~.
\end{eqnarray}
To prove Eq.(\ref{eq:8_9}) note that from the Jacobi identity and Eq.(\ref{eq:8_5}) it follows that
\begin{eqnarray}
\label{eq:8_10}
& & \{ \{t_{\alpha}, t_{\beta} \}, s\} = - \{ \{t_{\beta}, s\}, t_{\alpha} \}
- \{ \{s, t_{\alpha} \}, t_{\beta}\}  \nonumber \\
& & = i~ c_{\alpha \beta \gamma} T(\gamma) s~~~.
\end{eqnarray}
Thus
\begin{equation}
\label{eq:8_11}
\{t_{\alpha} , t_{\beta}\} = c_{\alpha \beta \gamma} t_{\gamma} + F~~~,
\end{equation}
where $\{ F, s(\xi) \} = 0$. Consequently $F$ is independent of the $\pi$'s.
Substituting $\pi_{\alpha} = 0$ in Eq.(\ref{eq:8_11}), we find $F=0$. This proves
Eq.(\ref{eq:8_9}). It also follows from a direct calculation using Eq.(\ref{eq:8_3}) and Eq.(\ref{eq:8_6}).

The PB's Eqs.(\ref{eq:8_7}), (\ref{eq:8_8}), and  (\ref{eq:8_9}) involving $t_{\alpha}$ and $s$ are simple and do
not require a particular parameterization for $s(\xi)$. We therefore find it
convenient to use these variables in canonically quantizing the systems below.

\subsection{Non-Relativistic Spinning Particles~}
\label{canonical_formalism_1}
Here we show how the Hamiltonian description for a spinning
particle (Eqs.(\ref{eq:3_1})-(\ref{eq:3_6})) is obtained from the Lagrangian Eq.(\ref{eq:3_19})
[Eqs.(\ref{eq:3_26})]. Now $G$ is $SU(2)= \{s \}$ and $T(i) = \sigma_i/2$. The phase-space
coordinates are $x_i$,  $p_i$, $s$  and $t_i$, where $p_i$ is canonically
conjugate to $x_i$ . From Eq.(\ref{eq:3_19}) [(\ref{eq:3_26})] we obtain the following
primary constraint:
\begin{equation}
\label{eq:8_12}
\phi_{i} = t_{i} - S_{i} \approx 0~~~,
\end{equation}
where $S_{i}$ is defined in Eq.(\ref{eq:3_18}). From  Eqs.(\ref{eq:8_7}), (\ref{eq:8_8}), and  (\ref{eq:8_9}),
\begin{equation}
\label{eq:8_13}
\{\phi_{i}, \phi_{j} \} = \varepsilon_{ijk} (\phi_{k} - S_{k})~~~.
\end{equation}
Applying Dirac's procedure, the following Hamiltonian is obtained from the Lagrangian Eq.(\ref{eq:3_19}):
\begin{equation}
\label{eq:8_14}
H = \frac{p^2}{2m} + \phi_{i}\eta_i~~~\, ,
\end{equation}
where $\eta_i$ are Lagrange multiples. From the requirement that
\begin{equation}
\label{eq:8_15}
\{\phi_{i}, H \}=0 \, ,
\end{equation}
on the reduced phase space, we find that there exist no secondary constraints. Instead, we obtain conditions on $\eta_i$, i.e.,
\begin{equation}
\label{eq:8_16}
\varepsilon_{ibc}\eta_bt_c = 0\, .
\end{equation}

Since those $\eta_i = \eta_i^{(t)}$ in a direction parallel to $t_i$ are left arbitrary, only those variables which have a weakly zero PB's have a well defined time-evolution. Only such variables are of  physical interest. We will call them observables. Of course, $x_i$ and $p_i$ are observables. In addition, so are $t_i$ and  $S_i$. This follows from
\begin{equation}
\label{eq:8_17}
\{t_{i}, \phi_{j} \} = \varepsilon_{ijk}\phi_k\, ,
\end{equation}
and
\begin{equation}
\label{eq:8_18}
\{s, \eta_i^{(t)}\phi_i \} = -\frac{i}{2}\eta_i^{(t)}\sigma_is  = \frac{i}{2}s\sigma_3\, .
\end{equation}

Eq.(\ref{eq:8_18}), which is weakly valid,  corresponds to an infinitesimal version of the $U(1)$ gauge transformation discussed
in Chap.\ref{section_nonrelativistic_spin}. Hence only those functions of s which are invariant
under gauge transformations (\ref{eq:3_36}) are also observables. But these are precisely $S_i$ or functions thereof. However, we can
eliminate $S_i$ by applying the constraints. Thus a complete set of observables on the reduced phase space are
\begin{equation}
\label{eq:8_19}
x_i\, , p_i\, ~~~\mbox{and}~~~t_i\, ,
\end{equation}
since $S_i$ can be eliminated via the constraints. In so doing note that
\begin{equation}
\label{eq:8_20}
t_it_i = \lambda^2\, .
\end{equation}
It remains to compute the Dirac Bracket (DB's) for the variables (\ref{eq:8_20}). But these are identical to the corresponding
PB's since all variables  (\ref{eq:8_20}) have weakly zero PB with the constraints (Cf. Eq.(\ref{eq:8_17}). Consequently, we have recovered
the Hamiltonian description for a non-relativistic spinning
particle Eqs.(\ref{eq:3_1})-(\ref{eq:3_5}). Note that instead of eliminating $S_i$ via the constraints, we could have eliminated $t_i$. In this
case the DB's involving $S_i$ do differ from the corresponding PB's. It can be shown that DB's for two  $S_i$'s  are given by Eq.(\ref{eq:3_4})
Consequently, both procedures are equivalent. It is straightforward to repeat the above analysis in the case where a
spinning particle with magnetic moment $\mu$ is placed in an external magnetic field.

In passing to the quantum mechanical system, as usual, we replace the Poisson bracket by $-i$ times the commutator bracket. Now the particular representation which occurs in the quantum theory is determined by $\lambda$ (Cf. Eq.(\ref{eq:8_20})). This
implies that $i)$ only one irreducible representation (IRR)
appears in the theory, and $ii)$ quantization is possible only if $\lambda^2$ is restricted to having the values
\begin{equation}
\label{eq:8_21}
t^2 = l(l+1)~~~,~~~ l=0,\frac{1}{2}, 1, ....\, .
\end{equation}
Note $ii)$ is similar to the Dirac charge quantization condition which occurs in magnetic monopole theory.
\subsection{Magnetic Monopoles~}
\label{canonical_formalism_2}
The canonical quantization for the magnetic monopole theory proceeds in a similar fashion to the proceeding section.
The essential difference is due to equation Eq.(\ref{eq:4.13}), which constrains
the configuration space variables for the monopole. Consequently, the independent phase space coordinates now consist of $r$, $p_r$,  $s$ and $t_i$, where $p_r$ is canonically conjugate to $r$. From  Eq.(\ref{eq:4.14}) we find only one primary constraint,
\begin{equation}
\label{eq:8_22}
\phi \equiv  \hat{x}_it_i - n \approx 0\, ,
\end{equation}
where $\hat{x}_i$  is defined in Eq.(\ref{eq:4.12}). Computing the Hamiltonian
~
\begin{equation}
\label{eq:8_23}
H = \frac{p_r^2}{2m} + \frac{1}{2mr^2} \left( t_it_i -n^2 \right) + \eta\phi\, .
\end{equation}
Here $\eta$ is a  Lagrange multiplier. The constraint Eq.(\ref{eq:8_22}) is
rotationally invariant, i.e.,  $\{\phi, t_i \} = 0$. Hence the requirement that $\{H, \phi\} = 0$ on the reduced phase space leads to no secondary constraints.

As before, observables are those variables which have zero PB's with  $\phi$. Among them are
\begin{equation}
\label{eq:8_24}
x_i\,,\,~p_i\, , \, t_i~~~\mbox{and}~~~\hat{x}_i\, .
\end{equation}
The latter follows from
\begin{equation}
\label{eq:8_25}
\{\phi, s\} = \frac{i}{2}s\sigma_3\, ,
\end{equation}
which is analogous to (\ref{eq:8_18}). As before, this corresponds to a $U(1)$ gauge transformation, and only those functions of $s$
which are invariant under sue~ transformations are observables. But these are precisely $\hat{x}_i$ or functions thereof, so Eq.(\ref{eq:8_24}) corresponds to a complete set of observables subject to the constraint (\ref{eq:8_22}).

A representation for the quantum theory can be constructed as follows. Let us regard the wave functions as functions of
$r$ and $s$:
\begin{equation}
\label{eq:8_26}
\psi \equiv \psi(r,s)\, .
\end{equation}
The position coordinates are diagonal in this representation in view of Eq.(\ref{eq:4.12}). The momentum $p_r$ acts as the usual differential
operator on $\psi$. The operators $t_i$  are the differential operators which represent the elements $\sigma_i/2$ in the left regular representation of $SU(2)$, i.e.,
\begin{equation}
\label{eq:8_27}
\left[ \exp \left(i\theta_kt_k \right)\psi  \right](r,s) = \psi (r,\exp (-\frac{i}{2}\theta_kt_k )s)\, .
\end{equation}
The constraint (\ref{eq:8_22}) is taken into account by imposing the condition
\begin{equation}
\label{eq:8_28}
\hat{x}_it_i \psi = n\psi\, ,
\end{equation}
on the wave functions. In view of Eqs.(\ref{eq:8_27}) and (\ref{eq:4.13}), this means
\begin{equation}
\label{eq:8_29}
 \psi (r,\exp (-\frac{i}{2}\theta_kt_k )s) = \psi (r,s)\exp (i\theta n)\, ,
\end{equation}

The scalar product of wave functions is
\begin{equation}
\label{eq:8_30}
 (\psi,\chi) \equiv \int_0^{\infty} dr r^2 \int_{SU(2)}d\mu(s)\psi^{*}(r,s)\chi(r,s)\, ,
\end{equation}
where $d\mu$ is the invariant Haar measure on $SU(2)$.

Let $\left\{ D^j(s) \right\}$ be the representation of $SU(2)$ with angular momentum $j$. Wave functions $\psi$ with finite norm have the expansion
\cite{Talman_68}
\begin{equation}
\label{eq:8_31}
 \psi(r,s) =\sum_j \sum_{\rho,\sigma} \alpha^j_{\rho\sigma} D^j_{\rho\sigma}(s)\, .
\end{equation}
Here $D^j_{\rho\sigma}(s)$ re the matrix elements of $\{ D^j(s)\}$ in the conventional basis with the third component of angular momentum diagonal.

The constraint (\ref{eq:8_29}) means that in Eq.(\ref{eq:8_31}), only those $\alpha^j_{\rho\sigma}$ with $\sigma = -n$ are non-zero. Thus
\begin{equation}
\label{eq:8_32}
\left\{ D^j_{\rho,-n} \right\}~~~,~~~\mbox{fixed}\,  n\,\,  ,
\end{equation}
is a basis for expansions of the form (\ref{eq:8_31}). Since $\sigma$ is necessarily integral or half-integral, we have the Dirac
quantization condition
\begin{equation}
\label{eq:8_33}
2n = \mbox{integer}\, .
\end{equation}
In (\ref{eq:8_32}), $j$ and $\rho$ are half-integral if $2n$ is odd and integral if $2n$ is even.

The quantum mechanics outlined here is essentially equivalent to conventional treatments.
\subsection{Relativistic Spinning Particles~}
\label{canonical_formalism_3}
In this  section we shall only be concerned with free relativistic
spinning particles. The group $G$ is now the connected component of the Lorentz group ${L}_{+}^{\uparrow} =\{ \Lambda^a_{\,\,\, b}\}$ with generators $\sigma_{ab}$ with matrix elements $(\sigma_{ab})_{cd}$ obtained from Eq.(\ref{eq:5.5}). Here Eq.(\ref{eq:8_2}) reads
\begin{equation}
\label{eq:8_34}
\left[\sigma_{ab},\sigma_{cd} \right] = i\left( -\eta_{bc}\sigma_{ad} + \eta_{bd}\sigma_{ac} +\eta_{ac}\sigma_{bd} -\eta_{da}\sigma_{bc}\right)\, .
\end{equation}
In addition Eqs.(\ref{eq:8_9})  and (\ref{eq:8_7}) are replaced by
\begin{equation}
\label{eq:8_35}
\left[t_{ab},t_{cd} \right] = \left( -\eta_{bc}t_{ad} + \eta_{bd}t_{ac} +\eta_{ac}t_{bd} -\eta_{da}t_{bc}\right)\, \, ,
\end{equation}
and
\begin{equation}
\label{eq:8_36}
\{t_{ab}, \Lambda \} = i\sigma_{ab}\Lambda\, .
\end{equation}
\newline
\noindent \underline{{A. Spinnless Particles~}}
\newline\newline
\noindent
For simplicity, we begin with the case where the spin is absent, i.e., $\lambda = 0$ in Eq.(\ref{eq:5.11}). The phase space coordinates
are given by $z_a$, $\pi_a$, $\Lambda$, and $t_{abd}$,  where $\pi_a$ is canonically conjugate to $z_a$. The primary constraints are
\begin{equation}
\label{eq:8_37}
\phi_{ab} = t_{ab}\approx 0 \, ,
\end{equation}
and
\begin{equation}
\label{eq:8_38}
\theta_{a} = p_a - \pi_a \approx 0\, .
\end{equation}
where $p_a$ is defined in Eq.(\ref{eq:5.3}). The equation (\ref{eq:5.37}) follows because there are no time-derivatives of $\Lambda$ appearing in the Lagrangian.
The constraints obey the PB algebra:
\begin{equation}
\label{eq:8_39}
\{\theta_{a}, \theta_{b} \} = 0\, ,
\end{equation}
and
\begin{equation}
\label{eq:8_40}
\{\phi_{ab}, \theta_{c} \} = i(\sigma_{ab})_{cd}p^d\, ,
\end{equation}
along with Eq.(\ref{eq:8_35}).

Because of the reparametrization symmetry of the Lagrangian, the Hamiltonian consists solely of the constraints (for a discussion of this issue, see, for example, Ref.\cite{dirac_64}), i.e.,
\begin{equation}
\label{eq:8_41}
H = \rho^{ab}\phi_{ab} + \kappa^a\theta_a\, ,
\end{equation}
where $\rho^{ab}$ and $\kappa^a$ are Lagrange multipliers. Once again, there are no secondary constraints. Instead $\rho^{ab}$ and $\kappa^a$ are restricted by
\begin{equation}
\label{eq:8_42}
\rho^{ab}p_b = 0\, ,
\end{equation}
and
\begin{equation}
\label{eq:8_43}
\kappa_ap_b-\kappa_b p_a = 0\, ,
\end{equation}
on the reduced phase space. In deriving Eqs.(\ref{eq:8_42}) and (\ref{eq:8_43}) we have used the representation for $\sigma_{ab}$given by Eq.(\ref{eq:5.5}). Eqs.(\ref{eq:8_42}) and (\ref{eq:8_43}) imply that
\begin{equation}
\label{eq:8_44}
\rho_{ia} = \varepsilon_{ijk}r_j\Lambda^{-1}_{ka}\,\,\,,\,\,\, i,j,k=1,2,,3 \, ,
\end{equation}
and
\begin{equation}
\label{eq:8_45}
\kappa_a = kp_a \, ,
\end{equation}
where $r_i$, $i=1,2,3$ and $k$ are undetermined constants. This, in turn, implies that four linearly independent combinations of Eqs.(\ref{eq:8_37}) and (\ref{eq:8_38}) form first class constraints, namely
\begin{equation}
\label{eq:8_46}
\phi_{i} \equiv \varepsilon_{ijk}\phi_j\Lambda^{-1}_{ka}\, ,
\end{equation}
and
\begin{equation}
\label{eq:8_47}
\phi_{0} \equiv \theta_a \Lambda^{a}_{\,\,0} \, .
\end{equation}
Observables, by definition, have zero PB's with $\phi_a$. Among them are
\begin{equation}
\label{eq:8_48}
\pi_{a} \,\, \mbox{and}\,\, J_{ab} = z_a\pi_b - z_b\pi_a \, ,
\end{equation}
where we have applied the constraints. Additional observables can be formed from $p_a$ and $t_{ab}$, however, these degrees of freedom can be eliminated via the constraints Eqs.(\ref{eq:8_37}) and (\ref{eq:8_38}). There exist six independent observables amongst the remaining ten degrees of freedom for the system. The former are exactly given by Eqs.(\ref{eq:8_48}), since four constraints now exist on the variables $\pi_a$ and  $J_{ab}$,
\begin{equation}
\label{eq:8_49}
\pi_{a}\pi^a = -m^2  \, ,
\end{equation}
and 
\begin{equation}
\label{eq:8_50}
W^{a} = 0 \,\, ,\,\, W^a \equiv \frac{1}{2}\varepsilon^{abcd}\pi_bJ_{cd} \,\, .
\end{equation}
Note that Eq.(\ref{eq:8_50}) yields three relations since $\pi_a W^a$ is identically zero. Equations (\ref{eq:8_49}) and ((\ref{eq:8_50}) indicate that the above system describes a particle of mass $m$ and spin zero.

It remains to compute the DB's for the variables (\ref{eq:8_48}). We first define $J^*_{ab}$:
\begin{equation}
\label{eq:8_51}
J^*_{ab} = J_{ab} + \phi_{abd}\, ,
\end{equation}
which, along with $\pi_a$, form a complete set of first class variables. Consequently, all DB's involving $J^*_{ab}$ and $\pi_a$ are identical to the corresponding PB's. Equivalently, we can define a DB with
$J_{ab}$ according to
\begin{equation}
\label{eq:8_52}
\{J_{ab}, \cdot \}^* \equiv \{J_{ab}^* , \cdot \}\, \, .
\end{equation}
Using Eq.(\ref{eq:8_52}), we obtain the usual Poincare algebra for$\pi_a$ and $J_{ab}$, 
\begin{eqnarray}
\label{eq:8_53}
\{\pi_{a}, \pi_{a} \}^* &=& 0\, , \nonumber \\
\{J_{ab}, \pi_{c} \}^* &=& \eta_{ac}\pi_b - \eta_{bc}\pi_a \, , \\
\{J_{ab}, J_{cd}\}^* &=& \eta_{ac}J_{bd}+ \eta_{bd}J_{ac} + \eta_{ad}J_{cb} +\eta_{bc}J_{da}\,\,  . \nonumber
\end{eqnarray}
Note that the equations (\ref{eq:8_49}) and (\ref{eq:8_50}) lie in the center of the algebra generated by $\pi_a$ and  $J_{ab}$. So if desired, one can eliminate redundant variables from $\pi_a$ and  $J_{ab}$, by hand, without conflict with their DB's (\ref{eq:8_53}).

Since the Hamiltonian is simply a linear combination of the constraints (\ref{eq:8_46}) and (\ref{eq:8_47}), it generates no time-evolution for  $\pi_a$ and  $J_{ab}$.  So if desired, we can declare that $\pi_0$ 
generates time-translations. Also we can identify
\begin{equation}
\label{eq:8_54}
x_i = \frac{1}{\pi_0}J_{i0}\,\, , \, \, i=1,2,3 \, ,
\end{equation}
as the space coordinate of the particle. It fulfills
\begin{eqnarray}
\label{eq:8_55}
\{x_i, \pi_j\}^* = \delta_{ij}\, ,
\end{eqnarray}
and 
\begin{eqnarray}
\label{eq:8_56} \{x_i, x_j\}^* =0 \, .
\end{eqnarray}
In proving Eq.(\ref{eq:8_56}), a direct computation yields
\begin{equation}
\label{eq:8_57}
\{x_i, x_j\}^* = \frac{1}{\pi_0^2} \left( - J_{ij}+ \frac{1}{\pi_0} (J_{i0}\pi_j - J_{j0}\pi_i) \right) \, .
\end{equation}
The result is then obtained after applying the definition for $J_{ab}$ (Cf. Eq.(\ref{eq:8_48})).
\newline\newline
\noindent \underline{{B. Spinning Particles~}}
\newline\newline
\noindent In this case, we consider non-zero values for $\lambda$ in Eq.(\ref{eq:5.11}).
Here, Eq.(\ref{eq:8_37}) s replaced by
\begin{equation}
\label{eq:8_58}
\phi_{ab} = t_{ab}- S_{ab} \approx 0 \, ,
\end{equation}
where $S_{ab}$ is given by Eq.(\ref{eq:5.4}). Equation (\ref{eq:8_58}), along with
(\ref{eq:8_38}), form the primary constraints for this system. Their PB' are given by (\ref{eq:8_39}), (\ref{eq:8_40}) and
\begin{eqnarray}
\label{eq:8_59}
\{\phi_{ab}, \phi_{cd}\}^* &=& \eta_{bc}(\phi_{da}- S_{da})-\eta_{ad}(\phi_{bc}- S_{bc}) \nonumber \\
&+& \eta_{ac}(\phi_{bd}- S_{bd}) -\eta_{dc}(\phi_{ca}- S_{ca})\, . 
\end{eqnarray}
The Hamiltonian, once again, consists solely of the constraints, i.e., Eq.(\ref{eq:8_41}).  Again there are no secondary constraints, and, instead, the Lagrange multipliers are restricted by (\ref{eq:8_42}) and
\begin{equation}
\label{eq:8_60}
\kappa_{a}p_b - \kappa_{b}p_a = 2\left( S_{ac}\rho^c_{\,\,b}- S_{bc}\rho^c_{\,\,a} \right)\, ,
\end{equation}
on the reduced phase space. After applying the definitions for $S_{ab}$ and $p_a$ (Cf. Eqs.(\ref{eq:5.3}) and (\ref{eq:5.4})), we find
\begin{equation}
\label{eq:8_61}
\frac{m}{2\lambda} \Big( \tilde{\kappa}_a\eta_{b0} +\eta_{a1}\tilde{\rho}_{b2} - \eta_{a2}\tilde{\rho}_{b1} - \tilde{\kappa}_b\eta_{a0} - \eta_{b1}\tilde{\rho}_{a2} + \eta_{b2}\tilde{\rho}_{a1}\Big) = 0\, ,
\end{equation}
%
%
where $\tilde{\kappa}\equiv \Lambda^{-1}\rho$ and $\tilde{\rho}\equiv \Lambda^{-1}\rho\Lambda$. Eqs.(\ref{eq:8_42}) and (\ref{eq:8_61}) along with 
\begin{equation}
\label{eq:8_62}
\tilde{\rho}_{ab} = -\tilde{\rho}_{ba} \,\, ,
\end{equation}
imply that all components of $\tilde{\kappa}$ and $\tilde{\rho}$ vanish  except for  $\tilde{\kappa_0}$ and $\tilde{\rho}_{12}$. Consequently, there are two first class constraints: Eq.(\ref{eq:8_47}) and
\begin{equation}
\label{eq:8_63}
\phi_{ab}\Lambda^{a1}\Lambda^{b2} \approx 0 \,\, .
\end{equation}

Once again $p_a$ and $J_{ab}$ (Cf. Eq.(\ref{eq:8_48})) are observables for the system. However, they no longer form a complete set of observables. Since now only two first class constraints can be found, there exist a total of eight observables for the system. But there are only six independent degrees of freedom in
$p_a$ and $J_{ab}$. Additional observables for this system are $S_{ab}$. Note that there are four constraining equations on $S_{ab}$:
\begin{eqnarray}
\label{eq:8_64}
S_{ab}\pi^b &= &0 \,\, ,  \\
\label{eq:8_65}
\frac{1}{2}S_{ab}S^{ab} &=& \lambda^2\,\, . 
\end{eqnarray}
Eq.(\ref{eq:8_64}), which holds on the reduced phase space, contains a total of three constraints since $S_{ab}\pi^a\pi^b$ vanishes identically. Thus two independent degrees of freedom remain in $S_{ab}$, i.e., $\Pi_a$, $J_{ab}$ and$ S_{ab}$ form a complete set of observables.

Alternatively, the five independent degrees of freedom in $J_{ab}$ and$ S_{ab}$ can be expressed more compactly by
\begin{equation}
\label{eq:8_66}
M_{ab} \equiv J_{ab} + S_{ab} \,\, .
\end{equation}
Now $M_{ab}$ contains all five degrees of freedom since
\begin{equation}
\label{eq:8_67}
W_{a}W^{a} =m^2\lambda^2 \,\, ,
\end{equation}
where
\begin{equation}
\label{eq:8_68}
W^{a} = \frac{1}{2}\varepsilon^{abcd}\pi_bM_{cd}\,\, ,
\end{equation}
is the only constraining equation on $M_{ab}$.

Equation (\ref{eq:8_67}) indicates that particle has a fixed spin $\lambda$. It, along with (\ref{eq:8_49}), can be used to eliminate, by hand, the redundant degrees of freedom from $\pi_a$ and $M_{ab}$. This follows because (\ref{eq:8_49}) and (\ref{eq:8_67}) lie in the center of the algebra generated $\pi_a$ and $M_{ab}$.  remains to be shown that this algebra is, once again, the Poincar\'{e} algebra.

With this in mind, we define
\begin{equation}
\label{eq:8_69}
M_{ab}^* \equiv M_{ab} + \phi_{ab} = J_{ab} + t_{ab} \,\, ,
\end{equation}
which, along with $\pi_a$, form a complete set of first class a variables. DB's involving $M_{ab}$ can the be defined by 
\begin{equation}
\label{eq:8_70}
\{M_{ab}\,,\, \cdot \}^*  \equiv \{M_{ab}^*\,,\, \cdot \} \,\, ,
\end{equation}
while DB's involving $\pi_a$ are equivalent to the corresponding PB's.  Using Eq.(\ref{eq:8_70}) we then verify that $\pi_a$ and  $M_{ab}$ generate the Poincar\'{e} algebra.
the Poincare algebra .

Again, if desired, we can declare that time-translations are generated by $\pi_0$. The standard canonical DB's (\ref{eq:8_55}) and (\ref{eq:8_56}) are obtained after defining the space coordinate $x_i$ according to (see Sudarshan and Mukunda, Ref.\cite{dirac_64}, p.439-454):
\begin{equation}
\label{eq:8_71}
x_i = \frac{1}{\pi_0} \left( M_{i0} - \frac{\varepsilon_{ijk}\pi_jW_k}{m(m-\pi_0)} \right) \,\, .
\end{equation}
Note that Eq.(\ref{eq:8_71}) does reduce to Eq.(\ref{eq:8_54}) in the limit of zero spin. In addition,spin $3$-vectors $\tilde{S}_{i}$ with the usual brackets
\begin{equation}
\label{eq:8_72}
\{\tilde{S}_{i}\,,\, \tilde{S}_{j} \}^* =\varepsilon_{ijk} \tilde{S}_{k} \,\, ,
\end{equation}
can be defined in terms of the Poincar\'{e} generators \cite{dirac_64}:
\begin{equation}
\label{eq:8_73}
\tilde{S}_{i} = - \frac{1}{m} \left( W_{i} - \frac{W_j\pi_j\pi_i}{\pi_0(m-\pi_0)} \right) \,\, .
\end{equation}
The variables $\tilde{S}_{i}$ differs from $S_i = \varepsilon_{ijk}S_{jk}/2$ which can be reconstructed in terms of the Poincar\'{e} generators. The latter variables do not satisfy Eq. (\ref{eq:8_72}).

In conclusion, the eight degrees of freedom in $\pi_a$  and $M_{ab}$ can be expressed in terms of $x_i$, $\pi_i$, and $\tilde{S}_{i}$, which have standard bracket relations. The Hamiltonian for $\pi_0$ is
\begin{equation}
\label{eq:8_74}
H = \sqrt{ \pi_i^2 + m^2} \,\, ,
\end{equation}
where we have chosen the positive root in eliminating the constraint (\ref{eq:8_49}). In terms of the variables  $x_i$, $\pi_i$, and $\tilde{S}_{i}$ the constraint (\ref{eq:8_67}) translates to
\begin{equation}
\label{eq:8_75}
\tilde{S}_{i}\tilde{S}_{i} = \lambda^2 \,\, .
\end{equation}
This system represents the obvious generalizations of the non-relativistic spinning particle system described in Sections \ref{section_nonrelativistic_spin_1} and \ref{canonical_formalism_1}. As before, we find that only one IRR appears in the quantum theory, and quantization is possible only if $\lambda^2$ is
restricted to having the values given in Eq.(\ref{eq:8_21}).

\subsection{Yang-Mills Particles~}
\label{canonical_formalism_4}

For simplicity we shall specialize to the case of non-relativistic particles. Consequently, we replace the first term in Eq.(\ref{eq:6_10}) by $m\dot{x}_i^2/2$. Now the phase space coordinates are  $x_i$, $p_i$, $s$, and $t_{\alpha}$. Here $s\in \Gamma$, where $\Gamma$ being a faithful unitary representation of an arbitrary compact connected Lie group $G$.

The Hamiltonian for this system is
\begin{equation}
\label{eq:8_76}
H = \frac{1}{2m}\left( p_i -  eA_i^{\alpha}t_{\alpha}\right)^2 +eA_0^{\alpha}t_{\alpha}+ \eta_{\alpha}\phi_{\alpha} \,\, ,
\end{equation}
where the primary constraints are given by
\begin{equation}
\label{eq:8_77}
\phi_{\alpha} = I_{\alpha} -  t_{\alpha} \approx 0\,\, .
\end{equation}
In deriving Eq.(\ref{eq:8_76}) we have used the constraints to rearrange terms. As usual, there are no secondary constraints and the $\eta$'s are restricted by
\begin{equation}
\label{eq:8_78}
c_{\sigma\rho\lambda}\eta_{\rho}t_{\alpha} = 0\,\, ,
\end{equation}
on the constrained surface. Let there be $k$ independent vectors $\{ \eta_{\rho} = \eta_{\rho}^{(A)} , A=1,2,...,k \}$ satisfying Eq.(\ref{eq:8_78}). The first class constraints of the theory are
\begin{equation}
\label{eq:8_79}
\phi^A = \eta_{\rho}^{(A)}\phi_{\rho}\,\, .
\end{equation}
Observables have zero PB's with $\phi^A$. They consist of
\begin{equation}
\label{eq:8_80}
x_i \,\, , \, \, p_i \,\, , \,\, t_{\alpha}\,\, .
\end{equation}
Note that the $I_{\alpha}$'s are also observables. This follows from
\begin{equation}
\label{eq:8_81}
{\ s, \phi^A } = - \eta(A)s\,\, ,
\end{equation}
where $\eta(A)= \eta_{\alpha}^AT(\alpha)$ generate the stability group of $t=t_{\alpha}T(\alpha)$ under the adjoint action. This group is isomorphic to the group $H$ (Cf. Sec.\ref{yang_mills_particles_4}). From Eq.(\ref{eq:8_81}), only those functions of $s$ which are invariant under the action of the little group of $t$ are of interest. These must be functions of $I$. However, the $I$'s can be eliminated via the constraint. Thus we are left with variables (\ref{eq:8_80}). Since they all have weakly zero PB with $\phi_{\alpha}$, all DB's involving these variables are identical to the corresponding PB's.

As in Section \ref{canonical_formalism_1}, not all the $t_{\alpha}$'s are independent. From Eqs.(\ref{eq:6_14})
and (\ref{eq:8_77}), $t$ is constrained to lie on a certain orbit in $\underline{\Gamma}$. These orbits are labeled by the constants $K_{\alpha}$. Using Eq.(\ref{eq:6_14}) any function of the $t_{\alpha}$ which is a constant on the orbits can be written as a function of the $K_{\alpha}$'s. in particular, the Casimir invariants can be expressed in terms of $K$. For the case of $G=SU(2)$, we are left with one constraint, which is analogous to (\ref{eq:8_20}).

The particular representation which occurs in the quantum theory is determined by the Casimir invariants, which in turn are determined by the $K_{\alpha}$'s. Once again, only one IRR appears in the quantum theory and quantization is possible only if the Casimir invariants formed out of the $K_{\alpha}$'s are restricted to a certain discrete set.

\subsection{Kaluza-Klein Formulation~}
\label{canonical_formalism_5}
As was true in several other cases the Lagrangian here (Cf. Eq.(\ref{eq:7.1})) contains a re-parametrization symmetry. Here we shall remove it by fixing $x^0 = \tau$. The Hamiltonian for this system is
\begin{equation}
\label{eq:8_82}
H = \sqrt{ m^2 + p_i^2 - \frac{1}{\lambda}t^2_{\alpha} }\,\, ,
\end{equation}
where $x_i$, $p_i$, and $t_{\alpha}$ are the usual phase space variables. Unlike the previously discussed systems, there are no constraints on the phase space variables. This is due to the fact that $t_{\alpha}$
 can be  expressed in terms or $\dot{s}s^{-1}$ (Cf. Eq.(\ref{eq:7.6}). Here $I_{\alpha}= t_{\alpha}$).

In the previous section $t_{\alpha}T(\alpha)$ was constrained to lie on certain orbits in the Lie algebra. These orbits determined which IRR was to appear in the quantum theory. Now there are no constraints on the variables $t_{\alpha}$ and, consequently, all IRR's appear in the quantum theory.

In setting up the quantum theory, we can write down wave-functions which are functions of $s$ as well as $x_i$:
\begin{equation}
\label{eq:8_83}
\psi = \psi(s,x)\,\, .
\end{equation}
This follows since all components $s_{\alpha\beta}$ can be simultaneously diagonalized. Then the $t_{\alpha}$ 's are differential operators which represent the generators $T(\alpha)$ in the left regular representation of the group. In particular,
\begin{equation}
\label{eq:8_84}
(\exp\left( i\theta_{\alpha}L_{\alpha}\right)\psi)(s,x) = \psi(\exp\left( -i\theta_{\alpha}T({\alpha})\right)s,x) \,\, .
\end{equation}
The scalar product with respect to which the $t_{\alpha}$ 's are Hermitian a is given by
\begin{equation}
\label{eq:8_85}
(\phi,\psi) = \int d\mu(s)d^3x\phi^*(s,x)\psi(s,x) \,\, ,
\end{equation}
where $d\mu(s)$ is the invariant Haar measure of the group. The left regular representation is highly reducible. Every irreducible representation occurs with a multiplicity equal to its own dimension.

If an irreducible representation of the Kaluza-Klein system is desired, we must deal with the formulation given in Section \ref{kaluza_klein_theory_2}. As was noted earlier, the system there is identical to that of the Yang-Mill particle with the mass
\begin{equation}
\label{eq:8_86}
\sqrt{m^2 + \frac{1}{\lambda}\mbox{Tr}[K^2]} \,\, .
\end{equation}
Excluding this additional requirement, the quantization of such a system .is identical to that discussed in Section \ref{canonical_formalism_4}.

\noi 
\newpage

%% file: Pseudo_Classical.tex
%
\vspace{-3cm}
\begin{center}
\section{PSEUDO-CLASSICAL DESCRIPTION}
\label{pseudo_classical}
\end{center}
\seqnoll
\setcounter{exenum}{1}

{\Huge I}n the previous sections, we have seen how to describe the spin and isospin degrees of freedom of a particle in terms of dynamical group elements $g(\xi)\in G$.  It has been pointed out however,
that Grassmann variables, i.e., anti-commuting $c$-numbers, can be utilized for the same purpose. Such a formulation is usually, referred to as pseudo-classical mechanics. Upon quantization the
anti-commuting $c$-numbers leads to certain irreducible representations of some symmetry group $G$. In the case of spin degrees of freedom it was discussed by Volkov, Peletminskii \cite{Volkov_59}  and Martin \cite{Martin_59} that the classical Grassmann variables, are replaced by Pauli matrices after quantization. These considerations for point particles (and extended objects) have recently been discussed
in much detail in the literature \cite{Grassmanns_1960,Teilteboim_1980}.  They have also been applied to internal degrees of freedom \cite{BSSW_1977, Casalbuoni_1976,{Casalbuoni_1977}}.

The algebra of the anti-commuting Grassmann variables can be used to extend the notion of Lie algebras to graded Lie-algebras \cite{Berezin_1970}. The notation of graded Lie algebras, usually referred
to as super-symmetry, has been extended to field theory
leading to global and local (i.e., super-gravity) super-symmetric field theories. For a review of this very dynamic field of research and for further references see, e.g., Ref.\cite{Fayet_1977}.
In this Chapter we apply some of these concepts to the
description of the systems discussed in the previous chapters.

\subsection{Non-Relativistic Spinning Particles~}
\label{pseudo_classical_1}
The Lagrangian for a free, non-relativistic spinning particle
involving dynamical anti-commuting Grassmannian variables, $f_a(\tau)$, is \cite{Teilteboim_1980,Casalbuoni_1976}
\begin{equation}
\label{eq:9_1}
L_0 = \frac{1}{2}\dot{x}_i^2  + \frac{1}{2}f_a\dot{f}_a\, \, .
\end{equation}
The equations of motion derived from Eq.(\ref{eq:9_1}) are
\begin{equation}
\label{eq:9_2}
m\ddot{x}_a = 0 \, \, ,\, \, \dot{f}_a = 0 \, \, .
\end{equation}
The orbital angular momentum $L_a = \varepsilon_{abc}x_ap_b$ and spin $S_a$, as defined by
\begin{equation}
\label{eq:9_3}
S_a =  -\frac{i}{2}\varepsilon_{abc}f_bf_a\, \, .
\end{equation}
Note that the Lagrangian (\ref{eq:9_1}) is weakly invariant under the transformations
\begin{eqnarray}
\label{eq:9_4}
x_a \rightarrow x_a -i\epsilon \frac{f_a}{\sqrt{m}}  \, \, , \,\,
f_a \rightarrow f_a + \epsilon \sqrt{m}\dot{x}_a  \, \, ,
\end{eqnarray}
where $\epsilon$ is a c-number Grassmann parameter. We define (\ref{eq:9_4}) to be a "super-symmetry" transformation. Under (\ref{eq:9_4}),
\begin{eqnarray}
\label{eq:9_5}
L_0 &\rightarrow&  L_0 + i\frac{\sqrt{m}}{2}\frac{d}{dt}\Big(f_a\dot{x}_a  \epsilon\Big)  \, \, ,
\end{eqnarray}
Thus the action $\int d\tau L_0$ is invariant.

Interactions can be added to the Lagrangian (\ref{eq:9_1}) in a straightforward manner, although in general, they will not be invariant under (\ref{eq:9_4}). For example, consider the interaction
of a particle having magnetic moment $\mu$ (and no charge) with an external electro-magnetic field $\mathbf{B}$. The form of the Lagrangian is then the same as in Eq.(\ref{eq:3_26}), i.e.,
\begin{eqnarray}
\label{eq:9_6}
L=   L_0  -\mu S_aB_a \, \, .
\end{eqnarray}
A variation of the coordinate $x_a$ leads to the equation of motion Eq.(\ref{eq:3_31}). A variation of $f_a$ leads to
\begin{eqnarray}
\label{eq:9_7}
\dot{f}_a + \mu\varepsilon_{abc}f_bB_c  = 0\, \, .
\end{eqnarray}
Using the definition (\ref{eq:9_3}) of the spin angular momentum, we obtain
the spin precession equation Eq.(\ref{eq:3_34}), i.e., 
\begin{eqnarray}
\label{eq:9_8}
\dot{S}_a =  \mu\varepsilon_{abc}B_bS_c \, \, .
\end{eqnarray}
The interaction given in Eq.(\ref{eq:9_3}) is not invariant under the super-symmetry
transformation (\ref{eq:9_4}), since it transforms according to
\begin{eqnarray}
\label{eq:9_9}
-\mu\dot{S}_aB_a \rightarrow  -\mu\dot{S}_aB_a  +i\mu\sqrt{m}\epsilon\varepsilon_{abc}B_a\dot{x}_bf_c  + i\frac{\mu \epsilon}{\sqrt{m}}S_af_b\partial_bB_a\, \, .
\end{eqnarray}
On the other hand, if we add the term $qA_a\dot{x}_a$ for a particle with charge $q=-e$, which transforms according to 
\begin{eqnarray}
\label{eq:9_10}
-eA_a\dot{x}_a \rightarrow  -eA_a\dot{x}_a  + i\frac{e}{\sqrt{m}}\partial_b A_a\epsilon f_b\dot{x}_a
  + i\frac{e}{\sqrt{m}}\dot{f}_aA_a\, \, ,
\end{eqnarray}
to the interaction Lagrangian (\ref{eq:9_6}), and set
\begin{eqnarray}
\label{eq:9_11}
\mu =   \frac{e}{m} \, \, ,
\end{eqnarray}
the weak invariance under Eq.(\ref{eq:9_4}) is restored. Here
\begin{eqnarray}
\label{eq:9_12}
F_{ab} =   \varepsilon_{abc}B_c = \partial_aA_b -\partial_bA_c\, \, .
\end{eqnarray}

Next we give a super-field formulation \cite{Salomonson_1978,Salam_1975} of the above systems. It provides us with a systematic method for constructing Lagrangians which are invariant under the super-symmetry transformations as in Eq.(\ref{eq:9_4}). We define $X_a(t,\Theta)$ to be a "super-coordinate",
i.e., it depends an super-space parameters $t$ and $\Theta$, the latter being a Grassmann parameter. We then identify the coefficients of the Taylor expansion of $X_a(t,\Theta)$ in $\Theta$ with $x_a$ and $f_a/\sqrt{m}$, i.e.,
\begin{eqnarray}
\label{eq:9_13}
X_a(t,\Theta) = x_a(t) + i\frac{\Theta}{\sqrt{m}}f_a(t) \, \, .
\end{eqnarray}
Now consider the following "super-charge" operator 
\begin{eqnarray}
\label{eq:9_14}
Q \equiv  = \Big( i\frac{\partial}{\partial \Theta} - \Theta \frac{\partial}{\partial t} \Big) \, \, .
\end{eqnarray}
It follows that
\begin{eqnarray}
\label{eq:9_15}
[Q,Q]_+ = - 2i\frac{\partial}{\partial t} \, \, ,
\end{eqnarray}
i.e., the anti-commutator of two super-charges  yields the "energy operator". Equation (\ref{eq:9_15}) thereby expresses a general property of super-symmmetry algebras \cite{Fayet_1977}. Furthermore, it can be easily verified that the super-charge $Q$ induce translations in super-space $(t,\Theta)$ according to
\begin{eqnarray}
\label{eq:9_16}
\delta X_a(t,\Theta) \equiv i\epsilon QX_a(t,\Theta) = X_a(t+i\epsilon \Theta,\Theta- \epsilon) - X_a(t,\Theta)  \, \, .
\end{eqnarray}
The transformation defined by the Equation (\ref{eq:9_16}) applied to the super-coordinate $X_a(t,\Theta)$ is identical to the super-symmetry transformations (\ref{eq:9_16}) applied to $x_a$ and $f_a$. 

For the purpose of constructing weakly invariant Lagrangians, we now note the following:

i) Let $Y = Y(t,\Theta)  + \Theta \eta(t)$ be a super-coordinate, which undergoes the transformation
\begin{eqnarray}
\label{eq:9_17}
\delta Y_a(t,\Theta) = i\epsilon QY_a(t,\Theta)  \, \, .
\end{eqnarray}
Then $\eta(t)$ is invariant under this transformation up to a total time derivative. This is analogous to the transformation properties of the $D$-term in $3+1$ dimensional super-symmetry \cite{Fayet_1977}.

ii) Let $P(t,\Theta)$ be a (fermionic-) bosonic operator which (anti-) commutes with $Q$. Then if the super-field $Y$ transforms according to Eq.(\ref{eq:9_17}), so does $P(t,\Theta)Y$. Examples of first-order
differential operators $P$ fulfilling this property are
\begin{eqnarray}
\label{eq:9_18}
\frac{\partial}{\partial t}  \, \, , \, \, d_{\Theta}\equiv   \frac{\partial}{\partial \Theta}-i\Theta\frac{\partial}{\partial t}  \, \, , 
\end{eqnarray}
the former being bosonic while the latter is fermionic.

iii) If $Y$ and $Z$ transform according to Eq.(\ref{eq:9_17}), then so does the product $YZ$.

Let $L_{0*} = L_{0*}(t,\Theta)$ transform under super-symmetry according to Eq.(\ref{eq:9_17}). Then the coefficient of $L_{0*}$ is invariant up to a time derivative. Since we desire an invariant quantity which is
bosonic, $L_{0*}$ should be fermionic. A choice for $L_{0*}$ which is quadratic in first order derivations of $X_a$ is
\begin{eqnarray}
\label{eq:9_19}
L_{0*} = i\frac{m}{2}
\frac{\partial X_a}{\partial t}d_{\Theta}X_a   \, \, .
\end{eqnarray}
The $\Theta$ coefficient of $L_{0*}(t,\Theta)$ can be extracted by integrating over e
and utilizing the usual rule \cite{Berezin_1966}
\begin{eqnarray}
\label{eq:9_20}
\int d\Theta \Theta = 1 \, \, , \, \, \int d\Theta = 0   \, \, , 
\end{eqnarray}
Applying this to Eq.(\ref{eq:9_19}), we then find
\begin{eqnarray}
\label{eq:9_21}
\int d\Theta L_{0*}(t,\Theta) = L_0(t)   \, \, , 
\end{eqnarray}
where $L_0(t)$ is the free particle Lagrangian (\ref{eq:9_1}).

Next we consider adding an interaction term to ((\ref{eq:9_19}). We first take up the case of a particle interacting with a scalar (bosonic) potential $V = V(X)$. The latter transforms under super-symmetry
according to Eq.(\ref{eq:9_17}). Since $V(X)$ is bosonic it must appear in $L_* = L_{0*} + L_{I*}$ times a fermionic operator. The latter must anti-commute with $Q$. Thus interactions like $L_{I*} = \Theta V(X)$
are excluded since they explicitly break the super-symmetry invariance. On the other hand, interactions like $L_{I*} = d_{\Theta} V(X)$ preserve the super-symmetry invariance. However, integrating with respect to $\Theta$ leaves only a total time derivative so no interaction results. Consequently, it appears difficult to construct  super-symmetric invariant version of a particle interacting with a scalar potential. The latter is possible, however, for other treatments of the super-symmetric point particle (Cf . Ref.\cite{Witten_1981}).

In our formalism we can quite easily write down the interaction of the particle with a vector potential $A_i = A_i(X)$. Here we simply make the replacement
\begin{eqnarray}
\label{eq:9_22}
\frac{\partial}{\partial t}X_a &\rightarrow&    \frac{\partial}{\partial t}X_a - 2\frac{e}{m}A_a(X) \, \, , \nonumber \\
d_{\Theta} X_a &\rightarrow& d_{\Theta} X_a \,\, ,
\end{eqnarray}
in Eq.(\ref{eq:9_19}). Expanding the total Lagrangian in $\Theta$ and performing
the $\Theta$-integral according to the rule Eq.(\ref{eq:9_20}), one obtains
\begin{eqnarray}
\label{eq:9_23}
L(t)  = L_0(t) - e\dot{x}_aA_a(x) - \frac{e}{m}S_aB_a  \, \, .
\end{eqnarray}
This is identical to the Lagrangian (\ref{eq:9_23}) added with (\ref{eq:9_10}), with the restriction that the electric charge $q=-e$ and the magnetic moment $\mu$ are related  according to Eq.(\ref{eq:9_11}).

Here we notice that the Lagrangian in Eq.(\ref{eq:9_23}), or Eq.(\ref{eq:9_11}), leads to the conclusion that the gyro-magnetic ratio of the particle is $2$. This is actually a general feature of super-symmetric point particles (see, e.g., Refs.\cite{last_1980, Skagerstam_81, BSSW_1977}). In super-symmetric field theories, where the super-symmetry is unbroken, this situation corresponds to the anomalous magnetic moment being zero (see, e.g., Ref. \cite{Ferrara_1974}). 

In the above we have used a hermitian Grassmannian variable $\Theta$ to describe the spin degrees of freedom. One can also develop a non-relativistic super-symmetry by making use of a complex Grassmann variable. As was shown by Witten \cite{Witten_1981}  and discussed by other authors \cite{Salomonson_1982,Cooper_1900}  super-symmetric quantum theories can be useful for studying the non-perturbative breaking of super-symmetry.

\subsection{Super-Symmetric Point Particles in the Field of a Magnetic
Monopole}
\label{pseudo_classical_2}
In this Section we will  super-symmetrize \cite{Witten_1981}  the qlobal Lagrangian of Chapter \ref{section_magnetic_monopoles}. This can be achieved by applying the rules given in Section \ref{pseudo_classical_1} for constructing, weakly invariant super-symmetric Lagrangians. In Chapter \ref{section_magnetic_monopoles} a dynamical group element $s(t)$ entered in the construction of the global Lagrangian
(\ref{eq:4.13}). We can write $s(t)$ in the form

\begin{eqnarray}
\label{eq:9_24}
s(t)  = \exp\left(iT(a)\varepsilon_a (t) \right) \, \, , 
\end{eqnarray}
where $T(a) = \sigma_a/2$.
Similarly, we can define a group element $s_{*}$ on the super-space $(t,\Theta)$, according to
\begin{eqnarray}
\label{eq:9_25}
s_*(t)  = \exp\left(iT(a)\eta_a (t,\Theta) \right) \, \, , 
\end{eqnarray}
where $\eta_a (t,\Theta)$ is now a "super-field". If we write $\eta_a (t,\Theta)  = \varepsilon_a (t) -2\Theta \xi_a(t)$, then Eq.(\ref{eq:9_25}) can be expressed by
\begin{eqnarray}
\label{eq:9_26}
s_*(t)  = (1+\Theta\xi)s(t) \, \, , 
\end{eqnarray}
where $\xi =\xi_a\sigma_a$. Note that no conditions have to placed on $\xi_a$other then it being an add Grassmann variable in order that Eq.(\ref{eq:9_26}) be consistent with $s_*^{\dagger}s_* = 1$ and $\mbox{det} (s_*) = 1$.

A natural extension of the Lagrangian (\ref{eq:4.13}) is
\begin{eqnarray}
\label{eq:9_27}
L_*(t)  = L_{0*} - n\mbox{Tr}\left[ \sigma_3 s_*^{\dagger}d_{\theta}s_*\right] \, \, , 
\end{eqnarray}
where $L_{0*}$ is given by the Eq.(\ref{eq:9_19}). We must, furthermore, generalize (\ref{eq:4.12}), i.e., the relation between the relative coordinate $x_a$ and the dynamical group element $s$, to the $(t,\Theta)$-space. This extension can now be easily achieved after constructing
the following polar decomposition of the super-coordinate $\mathbf{X}(t,\theta)$:
\begin{eqnarray}
\label{eq:9_28}
\mathbf{X}(t,\theta)  =\mathbf{x}(t,\theta) +i\Theta\frac{\mathbf{f}}{\sqrt{m}} = R_*(t,\Theta)\hat{\mathbf{X}}_*(t,\theta)\, \, , 
\end{eqnarray}
where
\begin{eqnarray}
\label{eq:9_29}
R_*(t,\Theta)  = r(t) + +i\frac{\Theta}{\sqrt{m}}\hat{\mathbf{x}}(t)\cdot \mathbf{f}(t) \, \, , 
\end{eqnarray}
and
\begin{eqnarray}
\label{eq:9_30}
\hat{\mathbf{X}}_*(t,\theta)  =\hat{\mathbf{x}}(t) +i\frac{\Theta}{r\sqrt{m}}  \big( \mathbf{f}(t)- \hat{\mathbf{x}}(t)(\hat{\mathbf{x}}(t)\cdot\mathbf{f}(t)) \big)\, \, . 
\end{eqnarray}
In Eqs.(\ref{eq:9_29}) and (\ref{eq:9_30}) $\mathbf{x}= r\hat{\mathbf{x}}$. The super-symmetric generalization of the Eq.(\ref{eq:4.12}) then is
\begin{eqnarray}
\label{eq:9_31}
\hat{X}_*  =\hat{X}_{*a}\sigma_a = s_*\sigma_3 s_*^{\dagger}\, \, . 
\end{eqnarray}
The Eqs.(\ref{eq:9_19}), (\ref{eq:9_26}), (\ref{eq:9_27}), and (\ref{eq:9_31}) lead to the following Lagrangian
\begin{eqnarray}
\label{eq:9_32}
L = \int d\Theta L_*(t,\Theta)= \frac{1}{2}m\dot{x}_a^2 + \frac{1}{2}f_a\dot{f}_a  + in\mbox{Tr}\left[\sigma_3 s^{\dagger}\dot{s} \right] + 2ni\varepsilon_{abc}\hat{x}_a\xi_b\xi_c\, \, . 
\end{eqnarray}
By making use of the constraint Eq.(\ref{eq:9_31})  and the explicit form of $s_*(t,\Theta)$, as given by Eq.(\ref{eq:9_26}), we obtain
\begin{eqnarray}
\label{eq:9_33}
\hat{X}_*(t,\Theta) = \hat{x}(t) + \Theta[\xi(t),\hat{x}(t)]\, \, ,
\end{eqnarray}
where $\hat{x} = s\sigma_3s^{\dagger}$. Eqs.(\ref{eq:9_30}) and (\ref{eq:9_33})  then lead to
the following relationship
\begin{eqnarray}
\label{eq:9_34}
\varepsilon_{abc}\hat{x}_af_bf_c = 4mr^2\varepsilon_{abc}\hat{x}_a\xi_b\xi_c\, \, ,
\end{eqnarray}
i.e., the  Lagrangian (\ref{eq:9_32}) can now be written in the following form
\begin{eqnarray}
\label{eq:9_35}
L = \frac{1}{2}m\left (\dot{r}^2 +r^2\dot{\hat{x}}^2_a \right) + \frac{1}{2}f_a\dot{f}_a  + in\mbox{Tr}\left[\sigma_3 s^{\dagger}\dot{s} \right] - \frac{e}{m}S_aB_a\, \, . 
\end{eqnarray}
Here $\mathbf{B}$ is the magnetic field of the monopole,  i.e.,
\begin{eqnarray}
\label{eq:9_36}
B_a  = \frac{g}{4\pi} \frac{x_a}{r^2} \, \, ,
\end{eqnarray}
and $4\pi n = eg$ (Cf. with Section \ref{section_magnetic_monopoles_1}). In the expression
(\ref{eq:9_35}) $\hat{x}_a$ is to be regarded as a function of $s$ (Cf. Eq.(\ref{eq:4.12})).
For $f_a = 0$, (\ref{eq:9_35}) becomes the Lagrangian (\ref{eq:4.14}).

In order to obtain the equations of motion we consider variations of the dynamical variables $r$, $f_a$,  and $s$. The variation of $r$ in Eq.(\ref{eq:9_35}) gives
\begin{eqnarray}
\label{eq:9_37}
m\ddot{r} = r\dot{\hat{x}}^2_a + \frac{2n}{mr^3}\mathbf{S}\cdot\hat{\mathbf{x}} \, \, . 
\end{eqnarray}
A variation of the Grassmann variables $f_a$ leads to a spin precession equation (Cf. Eq.(\ref{eq:9_8}))
\begin{eqnarray}
\label{eq:9_38}
\dot{S}_a =  \frac{e}{m}\varepsilon_{abc}B_bS_c \, \, .
\end{eqnarray}
Again for variations in $s$, we take (Cf. Eqs.(\ref{eq:3_21}) and (\ref{eq:3_21}))
\begin{eqnarray}
\label{eq:9_39}
\delta S =   i\epsilon_k\sigma_ks \, \, .
\end{eqnarray}
By the Eq.(\ref{eq:4.12}),  (\ref{eq:9_39}) will induce an infinitesimal rotation of the unit vector $\hat{\mathbf{x}}$ as given by Eq.(\ref{eq:4.16}), i.e.,
\begin{eqnarray}
\label{eq:9_40}
\delta\hat{{x}}_a =   -2\varepsilon_{abc}\epsilon_b\hat{{x}}_c \, \, .
\end{eqnarray}
We therefore obtain the following result due to the variation of  (\ref{eq:9_39}):
\begin{eqnarray}
\label{eq:9_41}
\delta L  =   2 \epsilon_a\left( \frac{d}{dt}(L_a + n\hat{x}_a) + \frac{e}{m}\varepsilon_{abc}S_bB_c  \right)\, \, .
\end{eqnarray}
Eqs.(\ref{eq:9_38}) and (\ref{eq:9_41}) can now be combined to give
\begin{eqnarray}
\label{eq:9_42}
 \frac{d}{dt}\left(L_a + n\hat{x}_a + S_a \right) = 0\, \, ,
\end{eqnarray}
i.e., angular momentum conservation. After a long but straightforward
calculation, the Eqs.(\ref{eq:9_37}),  (\ref{eq:9_38}), and (\ref{eq:9_42}) can be combined to yield the following equation (Cf. Eq.(\ref{eq:3_11}):
\begin{eqnarray}
\label{eq:9_43}
m\ddot{\mathbf{x}} = -e \dot{\mathbf{x}}\times\mathbf{B} - \frac{e}{m}\nabla (\mathbf{S}\cdot\mathbf{B})\, \, ,
\end{eqnarray}
which is the equation of motion for a spinning particle in a non-homogeneous magnetic field (Cf. Section \ref{section_nonrelativistic_spin_2} and Ref.\cite{Jackson75}) The equation (\ref{eq:9_43}) can also be obtained directly from the Lagrangian (\ref{eq:9_35}) by considering simultaneous variations
of $s$ and $r$. Here one makes use of the relation
\begin{eqnarray}
\label{eq:9_44}
2n\dot{\hat{x}}_a\epsilon_a = -e(\dot{\mathbf{x}}\times\mathbf{B})\cdot\delta \mathbf{x}\, \, ,
\end{eqnarray}
where $\delta\mathbf{x}= -2\mathbf{\epsilon}\times\mathbf{x} + \delta r\hat{\mathbf{x}}$. Equation (\ref{eq:9_35}) follows from (\ref{eq:9_40}) and the explicit form of the magnetic monopole field \ref{eq:9_36}. As expected (Cf. Section \ref{pseudo_classical_1}) the gyro-magnetic ratio of the particle is $2$ according to the Eqn.(\ref{eq:9_38}). Here we also notice that although $\mathbf{L} + n\hat{\mathbf{x}}$ is not conserved, its projection along the ${\mathbf{x}}$-direction is. In fact, the latter is just $n$. This fact will turn out to be important when we quantize the system (Cf. Section \ref{pseudo_classical_5}) .

\subsection{The Super-Symmetric Hopf Fibration}
\label{pseudo_classical_3}

In Chapter \ref{section_magnetic_monopoles}  we have seen that the non-trivial $U(1)$ bundle on the two-sphere $S^2$ could be used to find a global Lagrangian description of magnetic monopoles. Let us recall how these bundles, here denoted by $L_M$, were constructed \cite{Steenrod_51,last_1980,Hilton_62}. For some related work see also Refs.\cite{Ryder_80,Minami_79,Quiro_82} . In Section \ref{section_magnetic_monopoles_4} we regarded $SU(2)$ as a $U(1)$ bundle over $S^2$, where the action of the $U(1)$ group corresponded to the gauge transformation Eq.(\ref{eq:4.22}), i.e.,
\begin{equation}
\label{eq:9_45}
s(t) \rightarrow s(t) \exp\big(i \sigma_{3}\alpha(t)/2 \big)~~~~.
\end{equation}
The projection map from the $SU(2)$s bundle to the two-sphere $S^2$ is given by Eq.(\ref{eq:4.12}), i.e.,
\begin{equation}
\label{eq:9_46}
s(t) \rightarrow s(t)\sigma_3s^{\dagger}(t) = \hat{X}(t) ~~~~.
\end{equation}
Now consider the following cyclic subgroup of $SU(2)$
\begin{equation}
\label{eq:9_47}
Z_M =  \{ z_k = \exp\left( i\sigma_3\frac{2\pi k}{M}\right)\,|\, k= 0,1, ..., M_1 \} ~~~~,
\end{equation}
where $M$ is a positive integer. $Z_M$ has an action on $s \in SU(2)$
which commutes with the projection (\ref{eq:9_46}), i.e., if
\begin{equation}
\label{eq:9_48}
s \rightarrow sz_k ~~~~,
\end{equation}
then
\begin{equation}
\label{eq:9_49}
s \rightarrow s\sigma_3s^{\dagger} \rightarrow sz_k\sigma_3(sz_k)^{\dagger} = \hat{X} ~~~~.
\end{equation}
The $U(1)$-bundles over the two-sphere $S^2$ are then generated by the quotient of $SU(2)$ with the group-action (\ref{eq:9_48}) \cite{Comment_70}. A function $f$ on $L_M$ can now be regarded as a function on $SU(2)$ which is $Z_M$ invariant, i.e.,
\begin{equation}
\label{eq:9_50}
f(s_z) = f(s) ~~~~,
\end{equation}
for all $z_k \in Z_M$. In view of the fact that the wave functions the charge-monopole system have the property of being $Z_{|2n|}$ invariant they can be regarded as on $R^1 \times Z_{|2n|}$. In this sense, there is a topological interpretation of the Dirac quantization condition Eq.(\ref{eq:8_33}) (Cf.
Ref.\cite{Quiro_82}).

In the present chapter we have constructed the super-symmetric generalization of the fibrations of $S^3$ as discussed above. Let us here briefly examine the corresponding mathematical
structure. The super-symmetric version $SU(2)_*$ of $SU(2)$ is defined by letting the group parameters become super-fields as indicated by the Eq.(\ref{eq:9_25}). Let
\begin{equation}
\label{eq:9_51}
U(1)_* =  \{ \exp\left(i\sigma_3\gamma \right)  \}~~~~,
\end{equation}
where $\gamma$ is an even Grassmann variable. $U(1)_*$ ha a right-handed
action on $SU(2)_*$, i.e., 
\begin{equation}
\label{eq:9_52}
s_* \rightarrow s_*\exp\left(i\sigma_3\alpha \right)~~~~.
\end{equation}
The projection map (\ref{eq:9_25}) can therefore be generalized to
\begin{equation}
\label{eq:9_53}
s_* \rightarrow s_*\sigma_3s^{\dagger}_* =\sigma_a\hat{x}_{*a} ~~~~,
\end{equation}
where the image of the map (Cf. Eq.(\ref{eq:9_30})) is the super-symmetric version $S^2_*$ of the two-sphere $S^2$. The bundle which describes the spinning charge-monopole system is then
\begin{equation}
\label{eq:9_54}
L_{M*} = SU(2)_*/Z_M ~~~~.
\end{equation}
As will be shown in Section \ref{pseudo_classical_5}, the Dirac quantization condition
Eq.(\ref{eq:8_33}) is full-filled also in this case, i.e., we choose $M = |2n|$.

The $U(1)_*$ gauge transformation will induce a transformation on the $\xi$-variables defined in the  Eq.(\ref{eq:9_26}). In fact, if we write
\begin{equation}
\label{eq:9_55}
 \exp\left(i\sigma_3\gamma(t,\Theta) \right) = \left(1+\Theta \sigma_3\beta (t) \right)\exp\left(i\sigma_3\alpha(t) \right)~~~~,
\end{equation}
then by the projection map (\ref{eq:9_31}) and (\ref{eq:9_26}) $\xi$  will transform according to
\begin{equation}
\label{eq:9_56}
\xi (t) \rightarrow \xi (t) + \beta (t)\hat{x}(t)~~~~,
\end{equation}
i.e.,  $\xi$ undergoes a translation parallel to $\hat{x}$. Since $X_*$  is gauge invariant, it determines $\xi$ only up to a transformation (\ref{eq:9_56}). Eq.(\ref{eq:9_33}) is consistent  with this observation.

\subsection{Super-Symmetric Yang-Mills-Particles}
\label{pseudo_classical_4}

In the present Section we will combine the description of Yang-Mills particles (Cf. Chapter \ref{yang_mills_particles}) with the super-symmetry discussed above. For simplicity, we will restrict ourselves to non-relativistic particles, but the discussion can easily be generalized to the relativistic case \cite{Comment_71}.

The free part of the Lagrangian will again be given by
Eq.(\ref{eq:9_19}). We now extend the minimal coupling prescription Eq.(\ref{eq:9_2}) to the non-Abelian case, where the Yang-Mills vector potential is a matrix (Cf. Eq.(\ref{eq:6_11})). The super-symmetric generalization of the Lagrangian Eq.(\ref{eq:6_10}) is therefore
\begin{equation}
\label{eq:9_57}
L_* = L_{0*} + L_{I*} ~~~~,
\end{equation}
where $L_{0*}$ is given by Eq.(\ref{eq:9_19}) and the minimal coupling term $L_{I*}$ is
\begin{equation}
\label{eq:9_58}
L_{I*} = \mbox{Tr}\left[ Ks^{\dagger}_*(t,\Theta)D_{(t,\Theta)}s_*(t,\Theta) \right] ~~~~.
\end{equation}
Here we have generalized the covariant derivative appearing in Eq.(\ref{eq:6_11}) to
\begin{equation}
\label{eq:9_59}
D_{(t,\Theta)} = d_{\Theta} - ie(d_{\Theta}X(t,\Theta))A_a(X(t,\Theta))~~~.
\end{equation}
Next we expand the dynamical group element $s_*(t,\Theta)$ (Cf. Eq.(\ref{eq:9_26})):
\begin{equation}
\label{eq:9_60}
s_*(t,\Theta) = (1+ \Theta \xi(t))s(t)~~~,~~~\xi(t) = \xi_a(t) T(a)\,\, ,
\end{equation}
and the super-coordinate $X(t,\Theta)$ according to Eq.(\ref{eq:9_13}). We then integrate Eq.(\ref{eq:9_57})
 with respect to $\Theta$, i.e., 
\begin{equation}
\label{eq:9_61}
L(t)  = \int d\Theta L_{*}(t,\Theta) = L_{0}(t) + L_{I}(t) ~~~~.
\end{equation}
The result is that $L_{0}(t)$ is given by Eq.(\ref{eq:9_1}) and that
\begin{equation}
\label{eq:9_62}
L_{I} = -\mbox{Tr}\left[ Ks^{\dagger}D_{t}s \right] - i\frac{e}{m}\mbox{Tr}\left[If_af_b\partial_bA_a\right] - \nonumber \\
\mbox{Tr}\left[ I\xi\xi +\frac{e}{\sqrt{m}}[I,\xi] A_af_a\right]~~~~,
\end{equation}
where $I$ is given by Eq.(\ref{eq:6_14}) and $D_t$ is the same as in Eq.(\ref{eq:6_11}). Since the Lagrangian Eq.(\ref{eq:9_61}) does not contain time derivatives of the dynamical variable $\xi$, it plays the role of an auxiliary field (see, e.g., Ref.\cite{Fayet_1977}). The $\xi$-variable in the Lagrangian
is necessary in order that successive super-symmetric transformations, induced by the translations
\begin{equation}
\label{eq:9_63}
t \rightarrow t + i\epsilon\Theta~~~,~~~ \Theta \rightarrow \theta -\epsilon ~~~~,
\end{equation}
close without the use of the equations of motion (see, e.g., Ref.\cite{Salomonson_1978}). We are allowed to substitute the equation of motion for $\xi$, i.e.,
\begin{equation}
\label{eq:9_64}
[\xi,I] = -\frac{e}{\sqrt{m}}f_a[A_a,I] ~~~~,
\end{equation}
back into the Lagrangian Eq.(\ref{eq:9_62}). Equation (\ref{eq:9_64}) leads to
\begin{equation}
\label{eq:9_65}
\mbox{Tr}[I\xi\xi] = \frac{e^2}{2m}f_af_b\mbox{Tr}[I[A_a,A_b]] ~~~~.
\end{equation}
After substituting (\ref{eq:9_64}) and (\ref{eq:9_65}) into Eq.(\ref{eq:9_62}), we then find
\begin{equation}
\label{eq:9_66}
L_I = - \mbox{Tr}[Ks^{\dagger}D_ts] =- \frac{e}{m}\mathbf{S}\cdot\mbox{Tr}[I\mathbf{B}] ~~~~,
\end{equation}
 where $\mathbf{B}$ is the non-Abelian magnetic field strength.

Concerning the equations of motion as derived from the Lagrangian (\ref{eq:9_66}), or (\ref{eq:9_62}), we notice that the spin precession Eq.(\ref{eq:9_38}) will be modified according to
\begin{equation}
\label{eq:9_67}
\dot{S}_a =  \frac{e}{m} \varepsilon_{abc}B_b^{\alpha} I_{\alpha}S_c   ~~~~.
\end{equation}
Thus the gyro-magnetic ratio is $2$ as expected.

\subsection{Canonical Formulation and Quantization of Pseudo-Classical Systems}
\label{pseudo_classical_5} 
In deriving the canonical formalism for the preceding systems, we follow the methods used in Chapter \ref{canonical_formalism}. For treating the fermionic variables, we shall apply the methods of Ref.\cite{Casalbuoni_1976}, which are as follows.

Let $\chi_a$ denote the momenta denote the momenta conjugate to $f_a$. If $C$ and $D$ are
any anti-commuting variables, then PB is defined according to
\begin{equation}
\label{eq:9_68}
\{C,D\} \equiv  -\left(\frac{\partial C}{\partial f_a}\frac{\partial D}{\partial \chi_a} + \frac{\partial C}{\partial \chi_a}\frac{\partial D}{\partial f_a}\right)   ~~~~.
\end{equation}
Hence,
\begin{equation}
\label{eq:9_69}
\{f_a,f_b\} =  \{\chi_a,\chi_b\} = 0~~~,~~~~\{f_a,\chi_b\}= -\delta_{ab}  ~~~~.
\end{equation}
The remaining PB's are defined in the usual way.

For the non-relativistic particle interacting with a magnetic field (Cf. Eqs.(\ref{eq:9_1}) and (\ref{eq:9_23})),
\begin{equation}
\label{eq:9_70}
\chi_a  =  \frac{\partial L}{\partial \dot{f}_a} = -\frac{i}{2}f_a  ~~~~.
\end{equation}
Thus we obtain the primary constraints
\begin{equation}
\label{eq:9_71}
\zeta_a  =  \chi_a + \frac{i}{2}f_a \approx 0~~~~.
\end{equation}
The Hamiltonian is
\begin{equation}
\label{eq:9_72}
H  =   \frac{1}{2m}\left(p_a - eA_a \right)^2 + \frac{e}{m}\mathbf{S}\cdot \mathbf{B} + \lambda_a\zeta_a~~~~,
\end{equation}
where  $\lambda_a$ are Lagrange multipliers. The requirement that $\{\zeta_a, H \}\approx 0$ determines the $\lambda_a$'s, i.e., 
\begin{equation}
\label{eq:9_73}
  \lambda_a = \frac{e}{m}\varepsilon_{abc}B_bf_c~~~~,
\end{equation}
and thus leads to no secondary constraints.

The constraints $\zeta_a$ are second class, since
\begin{equation}
\label{eq:9_74}
\{\zeta_a, \zeta_b \} = -i\delta_{ab}~~~~.
\end{equation}
They many be eliminated by introducing the DB's \cite{Casalbuoni_1976}:
\begin{equation}
\label{eq:9_75}
\{f_a, f_b \}^* = -i\delta_{ab}\,\,,\,\,\{f_a, \chi_b \}^* = -\frac{1}{2}\delta_{ab}~~~~,
\end{equation}
as well as
\begin{equation}
\label{eq:9_76}
\{\chi_a, \chi_b \}^* = \frac{i}{4}\delta_{ab}~~~~.
\end{equation}
The DB's which involve $x_a$ or $p_a$ are all equal to the corresponding PB's. Thus we can replace PB's by DB's and then eliminate $\chi_a$ via Eq.(\ref{eq:9_70}).

The generator of the super-symmetry transformation on the phase space variables is
\begin{equation}
\label{eq:9_77}
Q = \frac{1}{\sqrt{m}}f_a\left( p_a -e A_a\right)~~~~,
\end{equation}
since
\begin{equation}
\label{eq:9_78}
\{f_a, Q\}^* = -\frac{i}{\sqrt{m}}\left( p_a -e A_a\right)~~~,~~~\{\chi_a, Q\}^*=\frac{1}{\sqrt{m}}f_a   ~~~~.
\end{equation}
(Cf. with Eq.(\ref{eq:9_4}).) Furthermore, the Hamiltonian Eq.(\ref{eq:9_72})
can be expressed by
\begin{equation}
\label{eq:9_79}
H = \frac{1}{2i}\{Q, Q\}^*   ~~~~.
\end{equation}
In passing to the quantum theory we replace the DB's in Eq.(\ref{eq:9_75}) by $(-i)$ times the anti-commutator brackets (and the remaining DB's by $(-i)$ times the commutator brackets). In particular
\begin{equation}
\label{eq:9_80}
[f_a, f_b]_+ = \delta_{ab}  ~~~~.
\end{equation}
It is known, as a consequence \cite{Casalbuoni_1976}, that an IRR of the $f_a$'s is obtained in the quantum theory by the identification
\begin{equation}
\label{eq:9_81}
f_a= -\frac{1}{\sqrt{2}}\sigma_a   ~~~~,
\end{equation}
with $\sigma_a$'s being the Pauli matrices. Consequently, the spin of the particle is $1/2$. Furthermore, Eq.(\ref{eq:9_79}) becomes
\begin{equation}
\label{eq:9_82}
H= Q^2 ~~~~.
\end{equation}

For the monopole system described in Section \ref{pseudo_classical_2}, we replace the above variables $x_a$ and $p_a$ by $r$, $p_r$ , $s$ and $t_a$ ($p_r$ and $t_a$ are canonically conjugate to $r$ and $x_a$, respectively (Cf. Section \ref{canonical_formalism_2}). The variables $t_{\alpha}$ and $s$ again satisfy the Poisson bracket relations Eqs.(\ref{eq:8_7}),  (\ref{eq:8_8}), and (\ref{eq:8_9}). For this system, in addition to the constraint Eq.(\ref{eq:9_71}) we have Eq.(\ref{eq:8_22}), i.e.,
\begin{equation}
\label{eq:9_83}
\phi \equiv  \hat{x}_it_i - n \approx 0\, \, .
\end{equation}
The Hamiltonian is now
\begin{equation}
\label{eq:9_84}
H = \frac{p_r^2}{2m}+ \frac{1}{2mr^2}\left(t_at_a - n^2 \right) +\frac{e}{m} \mathbf{S}\cdot\mathbf{B} +\lambda_a\zeta_a+ \eta\phi\,\,\, ,
\end{equation}
$\lambda_a$ and $\eta$ being Lagrange multipliers. As before there are no secondary constraints. The constraints $\zeta_a$ are once again second class while $\phi$ is first class. The former are eliminated via
the DB's Eq.(\ref{eq:9_75}, while the gauge symmetry generated by the latter is eliminated by working on the reduced phase space, which is coordinatized by $r$, $p_r$ $t_a$ and $\hat{x}_a$ (Cf. Section \ref{canonical_formalism_2}).

For the monopole system we can express the super-symmetry
generator globally by
\begin{equation}
\label{eq:9_85}
Q = \frac{1}{\sqrt{m}}\left(f_a\hat{x}_ap_r  - \frac{1}{r}\varepsilon_{abc}t_af_b\hat{x}_c\right) \,\,\, ,
\end{equation}
as compared with Eq.(\ref{eq:9_77}). After applying the constraint (\ref{eq:9_83}), we can once again show that Hamiltonian is given by Eq.(\ref{eq:9_79}). 

In passing to the quantum theory we again make the identification Eq.(\ref{eq:9_80}), yielding the spin-half particle. The quantization of the remaining variables is the same as in Section
\ref{canonical_formalism_2}. In particular, quantization is possible on1y if $2n$ = integer.

Next, we take up the canonical quantization of the super-symmetric Yang-Mills particle described in Section \ref{pseudo_classical_4}. We pick up the discussion with the interaction Lagrangian Eq.(\ref{eq:9_66}), where the auxiliary variables $\xi$ have already been eliminated. The corresponding phase space for this system is spanned by $x_a$, $s$, $f_a$ and the canonically conjugate variables $p_a$ , $t_{\alpha}$, and $\chi_a$. The variables $t_{\alpha}$ and $s$ again satisfy the Poisson bracket relations Eqs.(\ref{eq:8_7}),  (\ref{eq:8_8}), and (\ref{eq:8_9}). The bosonic variables are constrained by equation
(\ref{eq:8_77}), i.e.,
\begin{equation}
\label{eq:9_86}
\phi_{\alpha} = I_{\alpha} - t_{\alpha} \approx 0\,\,\, ,
\end{equation}
while the fermionic variables are constrained by  Eq.(\ref{eq:9_71}). The Hamiltonian for this system is
\begin{equation}
\label{eq:9_87}
H = \frac{1}{2m}\left(p_a - e A_a^{\alpha}t_{\alpha} \right)^2 + \frac{e}{m}S_at_{\alpha}B_a^{\alpha} + \lambda_a\zeta_a + \eta_{\alpha}\phi_{\alpha}\,\,\, .
\end{equation}

The treatment of the bosonic constraints and the fermionic constraints have both been previously discussed (the former in Section \ref{canonical_formalism_4}). Here the super-symmetry generator is
\begin{equation}
\label{eq:9_88}
Q = \frac{1}{\sqrt{m}} \left(p_a - eA^{\alpha}_a (x)t_{\alpha} \right)f_a\,\,\, . 
\end{equation}
It can have a non-trivial action on the isospin variables when an external field is present
\begin{equation}
\label{eq:9_89}
\{t_{\alpha}, Q\}^* = -\frac{e}{\sqrt{m}}f_ac_{\alpha\beta\gamma} A^{\beta}_a (x)t_{\gamma} \,\,\, . 
\end{equation}
The quantum theory for the above system describes a particle of spin-half and isospin which is determined by the value of the constants $K_a$.
\noi 
\newpage

%% file: Lagrangians_10.tex
%
\begin{center}
\section{LOCAL AND GLOBAL LAGRANGIANS}
\label{section_globalformulation}
\end{center}
\seqnoll
\setcounter{exenum}{1}

{\Huge I}n the previous chapters, we considered systems which admit a global Hamiltonian description. That is, these systems
have a globally defined Hamiltonian or energy function, and the corresponding symplectic form (or equivalently, the Poisson
bracket) is globally defined. However, these systems do not admit global canonical coordinates. Thus a global Lagrangian
cannot be found in terms of the variables which occur in the Hamiltonian description. Now by a theorem of Darboux \cite{Westenholz_78},
local canonical coordinates always exist. Thus, locally, the Legendre transform can be made and a Lagrangian can be found.
These local Lagrangians are defined on coordinate neighbourhoods and are, in general, not defined globally. In previous Chapters,
in effect, we have constructed global Lagrangians from these local ones by introducing additional gauge degrees of freedom,
that is, a principal fibre bundle structure.

In this Chapter we now give a systematic method for finding the global Lagrangian when the system admits local Lagrangians and a global Hamiltonian description. The analysis presented here is similar to an analysis used in context of geometric quantization.

Three striking results emerge from the analysis:

$i)$\hspace{2mm} The construction in terms of $U(1)$ fibre bundles works only if,
\underline{classically}, a certain "quantization" is fulfilled. For the system of several charges and monopoles, this result has been
proved by Friedman and Sorkin \cite{Friedman_79}.  For that system, the condition is 
\begin{equation}
\label{eq:10_1}
\frac{e_ig_j}{e_kg_l} = \mbox{a rational number}\,\, , 
\end{equation}
where $e_i$ and $g_i$ are electric and magnetic charges. Note that this implies that electric and magnetic charges (and hence their product) are separately quantized.
( Take  $g_j=g_l$ to get the first result. Take $ge_i=e_k$ to get the second result.)
Note also that Eq.(\ref{eq:10_1}) is weaker than Dirac's result \cite{Dirac_31,Dirac_48}
\begin{equation}
\label{eq:10_2}
e_ig_j = 2\pi k \,\,\, , \,\,\,\mbox{$k$ integer}\,\, , 
\end{equation}
the proof of which requires quantum mechanics.

$ii)$\hspace{2mm} Once the quantization condition is fulfilled, a global
Lagrangian can be found by introducing $U(1)$ gauge degrees of freedom, that is a $U(1)$ fibre bundle. It is interesting
that in such a case nothing more involved than a $U(1)$ fibre bundle is required or the construction of the global Lagrangian.

$iii)$\hspace{2mm} Global Lagrangians can be constructed even if the
quantization condition is not fulfilled and hence the fibre bundle approach fails. The fibre bundle construction is a
special case of this more general construction.

In  the proof of these results, we use the language of differential geometry because of its convenience. We have done so sparingly however, so that a reader with a small familiarity
with differential geometry can follow the argument.

\bigskip
 
\bigskip
 
\subsection{The Fibre Bundle Construction}

Before discussing the main result, we first recall the
proof of a theorem due to Weil \cite{Simms_76}.  For our purposes, Weil's
result can be stated as follows: Let $\Omega$ be a closed two-form
on $Q$, i.e.,
\begin{equation}
\label{eq:10_3}
\Omega = \Omega_{ij} dx^{i}\hspace{-1mm}\wedge dx^j \,\, , 
\end{equation}
and
\begin{equation}
\label{eq:10_4}
d\Omega = 0\,\,\,\mbox{or}\,\,\, \partial_i\Omega_{jk}+ \partial_j\Omega_{ki} + \partial_k\Omega_{ij} = 0\,\, . 
\end{equation}
Further, for every closed two-surface $M$ in $Q$, let
\begin{equation}
\label{eq:10_5}
\int_M \Omega = 2\pi\nu\lambda \,\,\, , \,\,\, \nu = 0, \pm 1, \pm 2, ...\,\, .
\end{equation}
Here $\lambda$ is the same for all $M$ and $\nu$ is characteristic of $M$.
Then there exists a $U(1)$ bundle $E$ on $Q$, and a form $\tilde{\Omega}$ on $E$
with the following properties:

$1)$ $\tilde{\Omega}$ is exact, i.e., $\tilde{\Omega}=d\Lambda$. \newline
Here $\Lambda$ is a globally defined one-form on $E$.

$2)$ $\tilde{\Omega}$ is "gauge invariant''   and hence projects down to a
form on $Q$.

$3)$ The latter is precisely $\Omega$.

\noindent Here by gauge invariance we mean the following: Let $\phi$ and $\phi '$
and  be two sections (Cf. Chapter \ref{section_nonrelativistic_spin}) from a coordinate
neighbourhood in $Q$ to $E$. Then the pull backs $\phi^* \tilde{\Omega}$ and $\phi '^* \tilde{\Omega}$
are equal. Stated in another way, let $\pi: E \rightarrow Q$ be the projection
map from the bundle $E$ to the base $Q$,  then $\tilde{\Omega} = \pi^* \Omega$.

In conventional classical mechanics, where global canonical
coordinates exist, the symplectic form
\begin{equation}
\label{eq:10_6}
dp_i \wedge dq^{i} \,\, ,
\end{equation}
is necessarily exact:
\begin{equation}
\label{eq:10_7}
dp_i \wedge dq^{i} = d(p_idqî) \,\, .
\end{equation}
Weil's result gives us conditions under which a non-exact symplectic form can be turned into an exact one. This is
accomplished by introducing gauge  degrees of freedom. Note
in this context the "quantization" of the integrals in Eq.(\ref{eq:10_5}). The origin of the classical quantization condition is this
equation.

If $\lambda = 0$, then $\Omega$ is exact. We shall  also assume hereafter that $\lambda \neq 0$. We shall  also assume that $Q$ is paracompact. Under this technical assumption, $Q$ has a contractible covering $\{ U_{\alpha} \}$
by coordinate neighbourhoods $U_{\alpha}$. In such a covering, each of the sets $U_{\alpha}$, $U_{\alpha}\cap U_{\beta}$, $U_{\alpha}\cap U_{\beta}\cap U_{\gamma}$, ...,  is either empty or can be smoothly contracted to a point. The proof of the converse to the Poincar\'{e} lemma \cite{Westenholz_78} is therefore valid on each of these
sets. It follows from Eq.(\ref{eq:10_3}) that
\begin{equation}
\label{eq:10_8}
\Omega |U_{\alpha} = d\Theta_{\alpha} \,\,\, ,
\end{equation}
where $\Omega |U_{\alpha}$ is the restriction of $\Omega$ to $U_{\alpha}$.  Also, since
$d(\Theta_{\alpha} -\Theta_{\beta}) = 0$  on $U_{\alpha}\cap U_{\beta}$, we have
\begin{equation}
\label{eq:10_9}
\Theta_{\alpha} -\Theta_{\beta} = df_{\alpha\beta}\,\,\, \mbox{on}\,\,\, U_{\alpha}\cap U_{\beta} \,\, ,
\end{equation}
where
\begin{equation}
\label{eq:10_10}
d(f_{\alpha\beta}+f_{\beta\gamma}+f_{\gamma\alpha}) = 0\,\,\, \mbox{on}\,\,\, U_{\alpha}\cap U_{\beta}\cap U_{\gamma} \,\, .
\end{equation}
Equation (\ref{eq:10_10}) states that $f_{\alpha\beta}+f_{\beta\gamma}+f_{\gamma\alpha}$ 
a constant on $U_{\alpha}\cap U_{\beta}\cap U_{\gamma}$. Suppose further that
\begin{equation}
\label{eq:10_11}
f_{\alpha\beta}+f_{\beta\gamma}+f_{\gamma\alpha} = 2\pi n_{\alpha\beta\gamma}\lambda \,\, ,
\end{equation}
where $n_{\alpha\beta\gamma}$ takes integer values. Then the map $F:Q\rightarrow U(1)$ as defined by
\begin{equation}
\label{eq:10_12}
F(f_{\alpha\beta}) \equiv g_{\alpha\beta} = \exp \left( \frac{if_{\alpha\beta}}{\lambda} \right)\,\, ,
\end{equation}
fulfills the cocycle property
\begin{equation}
\label{eq:10_13}
g_{\alpha\beta}g_{\beta\gamma}g_{\gamma\alpha}=1 \,\,\, \mbox{on}\,\,\, U_{\alpha}\cap U_{\beta}\cap U_{\gamma} \,\, .
\end{equation}
The functions $g_{\alpha\beta}$ are defined on $U_{\alpha}\cap U_{\beta}$ and have values in $U(1)$.
Hence they define a $U(1)$ bundle on $Q$.

It may be shown \cite{Simms_76} that Eq.(\ref{eq:10_11}) is equivalent to
Eq.(\ref{eq:10_5}). Thus with Eq.(\ref{eq:10_13}), we have a $U(1)$ bundle on $Q$. It is
defined as follows. Let $x$ and $x'$ be the co-ordinates of the
same point $p$ in $U_{\alpha}\cap U_{\beta}$ for the coordinate systems appropriate to 
$U_{\alpha}$ and $U_{\beta}$. Then $(x,h^{(\alpha )})$ and $(x',h^{(\alpha)}g_{\alpha\beta})$ 
define the same point in the fibre over $p$ in the bundle space $E$. Here $h^{(\alpha)}\in U(1)$. Such a definition of principal fibre bundles is equivalent
to the definition we gave in Chapter \ref{section_nonrelativistic_spin}. 

Let 
\begin{equation}
\label{eq:10_14}
m_{\alpha} = -i\lambda (h^{(\alpha)})^{-1}dh^{(\alpha)} \,\, .
\end{equation}
The form $m_{\alpha}$ is defined on the fibres over $U_{\alpha}$ in the coordinate system appropriate to $U_{\alpha}$. We have,
\begin{equation}
\label{eq:10_15}
m_{\alpha} - m_{\beta}= i\lambda g^{-1}_{\alpha\beta}dg_{\alpha\beta} = - df_{\alpha\beta} \,\,\, ,
\end{equation}
on $U_{\alpha}\cap U_{\beta}$. Comparison of Eq.(\ref{eq:10_9}) and Eq.(\ref{eq:10_15}) shows that
\begin{equation}
\label{eq:10_16}
\Theta_{\alpha} + m_{\alpha}= \Theta_{\beta} + m_{\beta} \,\,\, .
\end{equation}
Thus the one-form
\begin{equation}
\label{eq:10_17}
\Theta = \Theta_{\alpha} + m_{\alpha} \,\, ,
\end{equation}
is globally defined on $E$. Further, since
\begin{equation}
\label{eq:10_18}
dm_{\alpha} = 0\,\,\, ,
\end{equation}
we can write $\Omega = d\Theta$  if we regard $\Omega$ as a form on $E$. (More correctly,
it is the form $\tilde{\Omega}$ in the statement of the theorem). The theorem
is thus proved.

In the statement of our result, we regard the Hamiltonian or energy and the symplectic form as defined in terms of
coordinates and velocities (and not in terms of coordinates
and momenta). We define $Q$ to be the configuration space for a dynamical system. Let $\{U_{\alpha}\}$ be a contractible covering of $Q$
(again assumed to be paracompact) by coordinate neighbourhoods $U_{\alpha}$ and $\mbox{T}U_{\alpha}$
be the tangent bundle (the space of coordinates and velocities)
over $U_{\alpha}$. Suppose now that the following is true:

 $i)$\hspace{2mm} The dynamical system admits local Lagrangians $L^{(\alpha)}$  defined on $\mbox{T}U_{\alpha}$.

$ii)$\hspace{2mm} The energy function $H$ is defined globally  on $\mbox{T}Q=\cup_{\alpha}\mbox{T}U_{\alpha}$.
In local coordinates, this means
\begin{equation}
\label{eq:10_19}
\frac{\partial L^{(\alpha)}}{\partial \dot{x}_i}\dot{x}_i -L^{(\alpha)} =  \frac{\partial L^{(\beta)}}{\partial \dot{x}_i}\dot{x}_i -L^{(\beta)}\,\,\, ,
\end{equation}
on $\mbox{T}U_{\alpha}\cap \mbox{T}U_{\beta}$ (assumed not to be empty). 

$iii)$\hspace{2mm} The symplectic $\omega$ exists globally, that is
\begin{equation}
\label{eq:10_20}
d \left[ \frac{\partial L^{(\alpha)}}{\partial \dot{x}_i}d {x}_i \right] =  d \left[ \frac{\partial L^{(\beta)}}{\partial \dot{x}_i}d {x}_i \right]\,\,\, ,
\end{equation}
on $\mbox{T}U_{\alpha}\cap \mbox{T}U_{\beta}$.

$iv)$\hspace{2mm} The integral of $\omega$ over any closed two-dimensional
surface $M$ in  $Q$ fulfills an analogue of Eq.(\ref{eq:10_5}):
\begin{equation}
\label{eq:10_21}
\int_M \omega = 2\pi\nu\lambda \,\,\, , \,\,\, \nu = 0, \pm 1, \pm 2, ...\,\, .
\end{equation}
Here $\lambda$ is the same for all $\mbox{T}Q$ and $\nu$ is characteristic of $\mbox{T}Q$.
Then there exists a $U(1)$ bundle $E$ on $Q$ and a global Lagrangian on $\mbox{T}E$ for this system.

Both assumptions $ii)$ and $iii)$ are necessary conditions for
the existence of a Hamiltonian description. A system of charges
and monopoles fulfills these conditions. Condition $iv)$ is surprising
in a classical context since it "quantizes" certain
integrals of $\omega$.  We shall show that for a system of charges
and monopoles, it coincides with the Friedman-Sorkin condition
mentioned previously.

To prove our result we can proceed as follows. If $\psi_{\alpha} = (\partial L^{(\alpha)}/\partial \dot{x}_i)dx_i$
then by Eq.(\ref{eq:10_20}), 
\begin{equation}
\label{eq:10_22}
d(\psi_{\alpha}-\psi_{\beta}) =d \left[ \frac{\partial }{\partial \dot{x}_i}\left(  L^{(\alpha)}- L^{(\beta)}\right) d {x}_i\right ] = 0 \,\,\, .
\end{equation}
Hence, $\psi_{\alpha}- \psi_{\beta}$ can be regarded as a closed one-form on $U_{\alpha}\cap U_{\beta}$.  Since $U_{\alpha}\cap U_{\beta}$ is contractible,
\begin{equation}
\label{eq:10_23}
\psi_{\alpha}-\psi_{\beta} =df_{\alpha\beta} \,\,\, \mbox{on} \,\,\, U_{\alpha}\cap U_{\beta} \,\,\,  ,
\end{equation}
where $f_{\alpha\beta}$ fulfills Eq.(\ref{eq:10_11}) by Eq.(\ref{eq:10_21}). As before, we can
construct a $U(1)$ bundle $E$ on $Q$ and forms $m_{\alpha}$ with the property (\ref{eq:10_15}). Hence the form $\chi$
defined by
\begin{equation}
\label{eq:10_24}
\chi = \kappa_{\alpha}  + m_{\alpha} \,\,\,  ,
\end{equation}
exists globally on $E$ and
\begin{equation}
\label{eq:10_25}
d\chi = \omega \,\,\,  ,
\end{equation}
or more precisely $d\chi =\pi^{*}\omega$  where $\pi$ is the projection $\pi: E\rightarrow Q$.

Now by the energy condition Eq.(\ref{eq:10_19}),
\begin{equation}
\label{eq:10_26}
 L^{(\alpha)} -  L^{(\beta)} =  \frac{\partial f_{\alpha\beta}}{\partial x_i}\dot{x_i} \,\, .
\end{equation}
Thus the Lagrangian
\begin{equation}
\label{eq:10_27}
 \tilde{L} =  L^{(\alpha)} - i\lambda(h^{(\alpha)})^{-1}\frac {dh^{(\alpha)}}{dt} \,\,\, ,
\end{equation}
is globally defined on $TE$. Since the last term is (locally) the time-derivative of a function, $L^{(\alpha)}$ and $L^{(\beta)}$ also give the same
equations of motion. The result is thus proved.

Let us understand the result in terms of the charge-monopole system. If
\begin{equation}
\label{eq:10_28}
\{\xi_i, \xi_j \} =  \omega^{ij}(\xi) \,\,\, ,
\end{equation}
are the PB's for a system of coordinates $\xi = (\xi^{1}, xi^{2}, ..., \xi^{2n})$ 
for the phase space, the symplectic form is
\begin{equation}
\label{eq:10_29}
\omega=  \frac{1}{2}\omega_{ij}\,(\xi)d\xi^i\wedge \xi^j\,\,\, ,
\end{equation}
where 
\begin{equation}
\label{eq:10_30}
 \omega_{ij} \omega^{jk} = \delta^{k}_i\,\,\, .
\end{equation}
For a system of one charge and one monopole, the PB's are given in Chapter \ref{section_magnetic_monopoles} by Eqs.(\ref{eq:4.7}) - (\ref{eq:4.9}). With coordinates $(x_{1},x_{2},x_{3},v_{1},v_{2},v_{3})$, they imply that
\begin{equation}
\label{eq:10_31}
 \omega = 2mdv_i\wedge dx_i + \frac{1}{2}F_{ij} dx_i\wedge dx_j \,\,\, ,
\end{equation}
where 
\begin{equation}
\label{eq:10_32}
F_{ij} = n\frac{\varepsilon_{ijk}x_k}{r^3}  \,\,\, ,
\end{equation}
If $M$ is a closed surface in $Q$  not enclosing the monopole, it
follows by Stokes theorem that $\int_M \omega = 0$. If $M$ is a 2-sphere $S^2$
(with outward orientation) which encloses the monopole we get $\int_M \omega = -4\pi n$. Multiple integrations over $S^2$ with different
orientations effectively correspond to different $M$. Thus, in
general,
\begin{equation}
\label{eq:10_33}
\int_M \omega = 4\pi n\nu_n  \,\,\, ,
\end{equation}
where $\nu_n$ is an integer.

In comparing with Eq.(\ref{eq:10_21})  we may set $\nu = \nu_n$  and $\lambda = 2n$. Consequently, the requirement Eq.(\ref{eq:10_21}) imposes no restrictions
on the system and only defines $\lambda$. This, however, is not the
case when more than one monopole is present.

If an additional monopole is introduced to the above system we must add
\begin{equation}
\label{eq:10_34}
-n'  \varepsilon_{ijk}
\frac{x_{k}'}{r^{'3}} \,dx_i' \wedge dx_j' \,\,\, ,
\end{equation}
to Eq.(\ref{eq:10_32}). Here $x'$  corresponds to the distance between the
electric charge and the additional monopole. Now
\begin{equation}
\label{eq:10_35}
\int_M \omega = 4\pi n\nu_n + 4\pi n'\nu_{n'}  \,\,\, ,
\end{equation}
where $\nu_n$  and  $\nu_{n'}$ are integers. Consequently, Eq.(\ref{eq:10_21}) implies
that
\begin{equation}
\label{eq:10_36}
\lambda \nu  = 4\pi n\nu_n + 4\pi n' \nu_{n'} \,\, .
\end{equation}
Since Eq.(\ref{eq:10_36}) holds for any $M$, we can choose it such that
$\nu_{n'}=0$.  It then follows that $\lambda$ equals $n$  times a rational number.
Similarly, by choosing $M$M such that $\nu_{n'} = 0$, we can conclude
that $\lambda$ equals  $n'$ times a rational number. But then
\begin{equation}
\label{eq:10_37}
\frac{n}{n'}  = \mbox{a rational number}\,\,\, ,
\end{equation}
which is consistent with Eq.(\ref{eq:10_1}). Only here $e_i = e_j$e.

The following brief remarks about the consequences of discarding the global energy condition Eq.(\ref{eq:10_19}) may be of interest. If this condition is abandoned, the global nature of symplectic form Eq.(\ref{eq:10_29}) implies only that
\begin{equation}
\label{eq:10_38}
L^{(\alpha)} - L^{(\beta)}  = \frac{\partial f_{\alpha\beta}}{\partial x_i}\dot{x}_i + \rho_{\alpha\beta}\,\,\, ,
\end{equation}
where $\rho_{\alpha\beta}$ does not depend on $\dot{x}_i$.  Further,

$i)$\hspace{2mm} $\rho_{\alpha\beta} = -\rho_{\beta\alpha}$ and $\rho_{\alpha\beta}+\rho_{\beta\gamma}+\rho_{\gamma\alpha}=0$.

$ii)$\hspace{2mm} Since $L^{(\alpha)}$ and $L^{(\beta)}$ give the same equations of motion on $\mbox{T}U_{\alpha}\cap \mbox{T}U_{\beta}$, the $\rho_{\alpha\beta}$'s are actually constants and hence are globally defined.

Now let $\phi_{\alpha}$ be a partition of unity subordinated to the covering $\{U_{\alpha} \}$, i.e., 
\begin{equation}
\label{eq:10_39}
\mbox{Supp}[\phi_{\alpha}]  = U_{\alpha}\,\,\,\,, \,\,\,\,\phi_{\alpha}\ge 0\,\,\,\, , \,\,\,\, \sum_{\alpha} \phi_{\alpha} = 1\,\,\,\, .
\end{equation}
Then the globally defined functions
\begin{equation}
\label{eq:10_43}
\kappa_{\alpha} = \sum_{\lambda}\rho_{\alpha\lambda}\phi_{\lambda}\,\,\, ,
\end{equation}
\begin{equation}
\label{eq:10_44}
\kappa_{\alpha} - \kappa_{\beta}  = \rho_{\alpha\beta}\,\,\, ,
\end{equation}
in view of $ii)$ above.
Thus
\begin{equation}
\label{eq:10_42}
\hat{L} = \hat{L}^{(\alpha)} - \kappa_{\alpha}\,\,\, ,
\end{equation}
fulfill an equation of the form Eq.(\ref{eq:10_26}) and
\begin{equation}
\label{eq:10_43}
\hat{L} = \hat{L}^{} - i\lambda (h^{(\alpha)})^{-1}\frac{dh^{(\alpha)}}{dt}\,\,\, ,
\end{equation}
is globally defined on $TE$. Also $\omega$ has the usual relation to $\hat{L}$.
However, since $\kappa_{\alpha}$ a can depend on $x$, $\hat{L}$ may not give the original
equations of motion.

\subsection{Global Lagrangians without the Quantization Condition}

The discussion which follows is taken from Ref.\cite{Zaccaria_81}.

The variational-principles which follow often involve the phase space as the space $Q = \{ \xi \}$. They are thus often related
to Hamilton's variational principle.

We shall discuss Hamiltonian systems. Thus a globally defined Hamiltonian H and a globally defined symplectic form $\omega$
[Cf. Eq.(\ref{eq:10_29}) ] are assumed to exist. Further $\omega$ is closed and non-degenerate, i.e.,
\begin{equation}
\label{eq:10_44}
d\omega = 0\,\,\, ,
\end{equation}
and
\begin{equation}
\label{eq:10_45}
\mbox{det}\,\omega_{ij} \neq 0 \,\,\, .
\end{equation}
The Hamilton equations of motion for this system are
\begin{equation}
\label{eq:10_46}
\frac{\partial H}{\partial \xi^i} = \omega_{ji}\dot{\xi}^j \,\,\, .
\end{equation}

Suppose now that $\omega$ is exact. By definition, then, there
exists a globally defined one-form $f = f_i(\xi)d\xi^i$ such that
\begin{equation}
\label{eq:10_47}
\omega = df \,\,\, .
\end{equation}
The equations of motion in this case follow from the global Lagrangian
\begin{equation}
\label{eq:10_48}
L =  f_i(\xi)\dot{\xi}^j -H(\xi)\,\,\, .
\end{equation}
In familiar situations where $Q$ admits global canonical coordinates,
we see from Eq.(\ref{eq:10_46}) that the variational principle  associated
with $L$ is just Hamilton's variational principle.

If $\omega$ is not exact as for the charge-monopole system, then
a global $f$ does not exist. Thus we have to modify the above procedure for finding $L$. One such procedure was described in the previous section. We now point out an alternative approach.

The first step in the modification is to change the configuration space from $Q$ to the space of paths $PQ$ over $Q$. It is
defined as follows. Let $\xi_0$ be a fixed reference point in $Q$. This point may be chosen at will.
Then a point of $PQ$ is a path $\gamma$  from $\xi_0$  to some point $\xi$: 
\begin{equation}
\label{eq:10_49}
\gamma = \{\, {\gamma} (\sigma)\, | \,0 \leq \sigma \leq 1~,~ {\gamma} (0) =
{\xi}_{0}\, , \,  {\gamma} (1) = \xi \}~.
\end{equation}
These paths are defined at a given time. We denote the time-dependent paths by
\begin{equation}
\label{eq:10_50}
\gamma (\sigma, t)\,\,\,  \left[ {\gamma} (\sigma, 0) = \xi _0 \right] \,\,\, .
\end{equation}
We now show that we can always write an action principle with configuration space as $PQ$. The procedure, of course, works also when $\omega$ is exact. We illustrate it in this context first.

The Hamiltonian $H$ can first be promoted to a functional $\tilde{H}$ on paths at a given time:
\begin{equation}
\label{eq:10_51}
\int_0^1 d\sigma \tilde{H}[\gamma (\sigma, t)] =  {H}[\gamma (1, t)]\,\,\, .
\end{equation}
Consider next a family of paths $\gamma (\sigma, t)$ with
\begin{equation}
\label{eq:10_52}
\gamma (1, t) = \xi(t)\,\,\, .
\end{equation}
Thus as $\sigma$ and $t$  vary, $\gamma (\sigma, t)$ sweeps out a surface $\Delta$  in $Q$ with the
boundary
\begin{equation}
\label{eq:10_53}
\partial \Delta~  =  \partial \Delta_{1} \cup \partial \Delta_{1} \cup \partial\Delta_{3}\,\,\, ,
\end{equation}
where
\begin{eqnarray}
\label{eq:10_54}
\partial \Delta_{1} & = & \{\, \xi (t) \,\,|\,\,\,t_1 \leq t \leq t_2  \}~~,  \nonumber \\
\partial \Delta_{2} & = & \{ \gamma (\sigma, t_{1}) \,\,|\,\,\,1 \leq \sigma \leq 1 \}~~,  \nonumber \\
\partial \Delta_{3} & = & \{ \gamma (\sigma,t_2) \,\,|\,\,\,1 \leq \sigma \leq 1  \}~~.
\end{eqnarray}
By applying Stokes' theorem, we can write the action $S$ as
\begin{equation}
S = \int_{t_1}^{t_2}dt\left[ f_i(\xi)\dot{\xi}^{i} - H(\xi)\right] = \int_{\partial\Delta_1}\left[ f_i(\xi)d\xi^{i} - H(\xi)dt\right] \,\,\, ,
\end{equation}
as
\begin{eqnarray}
\label{eq:10_55}
S = \int_{\Delta}\left[ \frac{1}{2}\omega_{ij}[\gamma(\sigma,t)]d\gamma^{i}(\sigma,t)\wedge d\gamma^{j}(\sigma,t) - 
\tilde{H}[\gamma(\sigma,t)]d\sigma \wedge dt \right] \,\,\, + \nonumber \\
\left\{ \int_{\partial\Delta_3}f_i[\gamma(\sigma,t)]d\gamma^{i}(\sigma,t) - \int_{\partial\Delta_2}f_i[\gamma(\sigma,t)]d\gamma^{i}(\sigma,t)\right\}\,\,\, .\,\,\,\,\,\,\,\,\,\,\,\,\,\,
\end{eqnarray}
Since we shall not vary the initial and final paths $\gamma(\sigma,t_1)$ and $\gamma(\sigma,t_2)$ in
the variational principle, the expression in the "script bracket'' above will not contribute to the equations of motion.
The action on the space of paths $PQ$ can thus be taken to be
\begin{equation}
\label{eq:10_57}
S = \int_{\Delta}\left[ \omega [\gamma(\sigma,t)] - 
\tilde{H}[\gamma(\sigma,t)]d\sigma \wedge dt \right]  \,\, .
\end{equation}
It involves only the symplectic form $\omega$ and not the one-form $f$.
It appears to define a field theory in one "space" and one time.

The action $S$ as given by Eq.(\ref{eq:10_57}) was derived in the case that  $\omega$ was exact.
However, it involves only  $\omega$ and thus is expected to be valid
even if  $\omega$ is not exact. This expectation is correct as may 
shown by varying
\begin{eqnarray}
\label{eq:10_58}
S = \frac{1}{2}\int_{\Delta}\omega_{ij}\frac{\partial \gamma^i}{\partial\sigma^a} \frac{\partial \gamma^j}{\partial\sigma^b} \varepsilon^{ab}d\sigma \wedge dt - \int_{\partial\Delta_1}Hdt \,\,\, ,
\end{eqnarray}
where $\sigma^0=t$, $\sigma^1= \sigma$, and $\varepsilon^{01}= -\varepsilon^{10} = 1$.
We find, upon using $d\omega = 0$, i.e.,  $\partial_i\omega_{jk}+\partial_j\omega_{ki}+\partial_k\omega_{ij}=0$, and regrouping terms,
\begin{eqnarray}
\label{eq:10_58}
\delta(\frac{1}{2}\int_{\Delta}\omega_{ij}\frac{\partial \gamma^i}{\partial\sigma^a} \frac{\partial \gamma^j}{\partial\sigma^b} \varepsilon^{ab}d\sigma \wedge dt) = - \int_{\Delta}d\left[\omega_{ij}d\gamma^i \delta\gamma^j\right] =  - \int_{\partial\Delta_1}\omega_{ij}d\gamma^i \delta\gamma^j\,\,\, ,
\end{eqnarray}
since $\delta\gamma^j = 0$  on $\partial\Delta_2\cup \partial\Delta_3$.
Also
\begin{eqnarray}
\label{eq:10_59}
\delta(\int_{\partial\Delta_1}Hdt) = \int_{\partial\Delta_1}\frac{\partial H}{\partial\xi^j}\delta\gamma^j dt\,\, .
\end{eqnarray}
Thus the equations of motion (Cf. Eq.(\ref{eq:10_46}) is recovered.

For a charge-monopole system the preceding technique can be directly applied to the conventional local Lagrangian:
\begin{eqnarray}
\label{eq:10_60}
L= L_0 + L_I\,\, ,\,\,L_0 =\frac{1}{2}m\dot{x}^2_i\,\,,\,\,L_I = eA_i\dot{x}_i \,\, ,
\end{eqnarray}
in order to find the global Lagrangian. Here (Cf. Eq.(\ref{eq:9_12})),
\begin{eqnarray}
\label{eq:10_61}
\partial_iA_j -\partial_jA_i = -\varepsilon_{ijk}\frac{gx_k}{4\pi r^3}\,\, ,
\end{eqnarray}
is the monopole magnetic field. The latter is globally defined but, of course, the potential $A_i$ is not.
The space $Q$ is $\mathbf{R}^3-\{0\}= S^2 \times \mathbf{R}_+$ , where
\begin{eqnarray}
\label{eq:10_62}
S^2 = \{ \hat{x}_i =\frac{x_i}{\sqrt{x_ix_i}}\}\,\, ,\,\, \mathbf{R}_+ =\{r =\sqrt{x_ix_i} \,\,\, | \,\, x_ix_i > 0\} \,\, .
\end{eqnarray}
Let the reference point be $\xi_0 = (1, 0, 0)$. Then the space $PQ$ is the space of paths $\gamma$ radiating from  $\xi_0$. The globally defined action and Lagrangian are 
\begin{eqnarray}
\label{eq:10_63}
&&S = \int \tilde{L}(\sigma,t)d\sigma\wedge dt \,\, , \,\, \tilde{L}(\sigma,t) = \tilde{L}_0(\sigma,t)+ \tilde{L}_I(\sigma,t) \,\, , \nonumber \\
&&S = \int \tilde{L}(\sigma,t)d\sigma\wedge dt \,\, , \\
&&\tilde{L}_0(\sigma,t) = -\frac{eg}{8\pi}\varepsilon_{ijk}\varepsilon^{ab}\hat{\gamma}_i(\sigma,t)\frac{\partial \hat{\gamma}_j}{\partial \sigma^a}(\sigma,t)\frac{\partial \hat{\gamma}_k}{\partial \sigma^b}(\sigma,t) \,\, .\nonumber
\end{eqnarray}
Here
\begin{eqnarray}
\label{eq:10_64}
\hat{\gamma}_i(\sigma,t) = \frac{\gamma_i(\sigma,t)}{\sqrt{\gamma_j(\sigma,t)\gamma_j(\sigma,t)}} \,\, .
\end{eqnarray}
and we identify $\gamma_i(1,t)$ with $x_i(t$).

Finally we make contact with the fibre bundle approach as follows. If $\omega$ fulfills a quantization condition of the form Eq.(\ref{eq:10_21}), we know that there is a $U(1)$ bundle $E$ over $Q$ on which $\omega$ becomes exact:
\begin{equation}
\label{eq:10_65}
\omega  = d\chi \,\, \mbox{on} \,\, E \,\,\,  .
\end{equation}
The action $S$ can be thought of as defined on ${PE}$, the path space for $E$. Thus we now regard $\Delta$ as a surface in $E$. Now
\begin{equation}
\label{eq:10_66}
\int_{\Delta}\omega  = \int_{\partial\Delta_1}\chi\,\,  ,
\end{equation}
plus terms which are not varied and may be discarded. In this way, we have a globally defined action on $E$:
\begin{equation}
\label{eq:10_67}
S= \int_{\partial\Delta_1}\left( \chi -Hdt\right)\,\,  .
\end{equation}
Here ,as in the treatment of the kinetic energy term for the charge-monopole system, we regard $H$ as being defined on $E$. Note that this procedure does not work without the quantization condition.

The quantization condition allows us to reduce the path space ${PE}$ to the $U(1)\times U(1)\times ... \times U(1)$ - bundle $E$ over $Q$ with $k$ factors of $U(1)$ when there are $k$ two-cycles $S_1, S_2, ...S_k$, such that
\begin{equation}
\label{eq:10_69}
\int_{S_j}d\omega = a_j \,\,\mbox{and}\,\, \frac{a_i}{a_j} \,\,\mbox{is rational}\,\,  .
\end{equation}
For further details, see Ref.\cite{Zaccaria_81,{bal_1991}}.
